\documentclass[twoside,11pt]{article}

\usepackage{blindtext}
\usepackage{amsmath}
\usepackage{mathtools}
\usepackage{amsthm}
\usepackage{graphicx}
\usepackage{enumerate}
\usepackage{natbib}
\usepackage{url} 
\usepackage{multirow}
\usepackage{tikz}
\usepackage{float}
\usepackage{amssymb}
\usepackage{booktabs}
\usepackage{array}
\usepackage{tabularx}
\usepackage[center]{caption}
\usepackage{longtable}
\usepackage[section]{placeins}
\usepackage{multibib}
\usepackage{anyfontsize}

\newcommand{\independent}{\perp\!\!\!\perp} 
\newcommand{\notindependent}{ \not\!\perp\!\!\!\perp}
\newcommand{\bP}{\mathbb{P}}
\newcolumntype{L}{>{\raggedright\arraybackslash}X}
\newcolumntype{P}[1]{>{\raggedright\arraybackslash}p{#1}}
\newcommand{\tabmech}[2]{\begin{tabular}[t]{@{}l@{}}#1\\[-0.2ex]{\scriptsize #2}\end{tabular}}
\newcolumntype{C}[1]{>{\centering\arraybackslash}p{#1}}
\newcommand{\posmark}{$\checkmark$}
\newcommand{\norestr}{$\ast$}
\newcommand{\pathyes}{\ensuremath{\checkmark}}
\newcommand{\pathno}{\ensuremath{\times}}

\newtheoremstyle{customassumption}
  {}
  {}
  {\itshape}
  {}
  {\bfseries}
  {}
  {.5em}
  {\thmname{#1}\thmnumber{ #2}\textbf{ #3}}

\theoremstyle{customassumption}
\newtheorem*{assumption}{Assumption}
\newtheorem*{ivassumption}{IV Assumptions}

\usepackage[preprint]{jmlr2e}
\allowdisplaybreaks

\begin{document}

\title{Identifiability of the instrumental variable model with the treatment and outcome missing not at random}

\author{\name Shuozhi Zuo \email shuozhi@umich.edu \\
       \addr Department of Statistics\\
       University of Michigan\\
       Ann Arbor, MI 48109, USA
       \AND
       \name Peng Ding$^\dagger$ \email pengdingpku@berkeley.edu \\
       \addr Department of Statistics\\
       University of California\\
       Berkeley, CA 94720, USA
       \AND
       \name Fan Yang$^\dagger$ \email yangfan1987@tsinghua.edu.cn \\
       \addr Yau Mathematical Sciences Center\\
       Tsinghua University\\
       Haidian District, Beijing 100084, China\\[1ex]
       $^\dagger$: Corresponding author
       }


\maketitle

\begin{abstract}
The instrumental variable model of \citet{imbens1994identification} and \citet{angrist1996identification} identifies the local average treatment effect, also known as the complier average causal effect (CACE). In practice, however, the treatment and outcome are often missing, and when they are missing not at random (MNAR), the CACE is generally not identifiable without further assumptions, because the underlying data distribution itself cannot be recovered. We study when the CACE remains identifiable under MNAR. Through an exhaustive search over missingness mechanisms, we characterize all those that identify the CACE without auxiliary information, in two scenarios: (1) missing data in either the treatment or the outcome alone, and (2) missing data in both the treatment and outcome under prospective data collection. Along the way, we unify existing results and establish many new ones, giving a complete picture of identifiability in each case. Our theory suggests that before any practical data analysis under the instrumental variable model, it is important to check whether the CACE is identifiable under the proposed missingness mechanism; moreover, because the true mechanism is typically unknown and untestable, it is more robust to conduct sensitivity analyses across multiple plausible missingness mechanisms.
\end{abstract}

\begin{keywords}
  causal inference, complier average causal effect, local average treatment effect, nonignorable missing data, nonparametric identification, potential outcome
\end{keywords}

\section{Introduction}\label{sec:intro}

Instrumental variable (IV) approaches are frequently adopted to study causal effects in the presence of unmeasured confounders in the treatment and outcome relationship. An IV is a variable that meets the following conditions: $(i)$ it is associated with the treatment, $(ii)$ it does not directly affect the outcome except through the treatment, and $(iii)$ it is uncorrelated with the unmeasured confounders given the observed covariates. In effect, an IV isolates variation in the treatment that is free of unmeasured confounding and uses it to identify the treatment's causal effect on the outcome. Together with the monotonicity assumption that there are no defiers who would take the treatment only when not encouraged by the IV, \citet{imbens1994identification} and \citet{angrist1996identification} showed that the classic two-stage least squares estimator in econometrics identifies the complier average causal effect (CACE), where compliers are subjects who take the treatment only when encouraged by the IV. A separate line of work bounds rather than point-identifies effects: \citet{balke1997bounds} established sharp nonparametric bounds on the average treatment effect for the entire population. We instead follow \citet{imbens1994identification} and \citet{angrist1996identification} and focus on point identification of the CACE.

However, missing data in the treatment and outcome are common in empirical research in the IV setting  \citep{Mealli2004analyzing,lui2010notes,eren2013theeffect,frolich2014treatment,von2016genetic,huber2021plausibility,rai2023efficient}. The difficulty of identification depends on the missingness mechanism. When data are missing completely at random (MCAR), so that missingness is independent of all variables, a simple complete-case analysis identifies the CACE. When data are missing at random (MAR), so that missingness is independent of the unobservables given the observables, standard approaches such as multiple imputation or non-response weighting yield valid inference. When data are missing not at random (MNAR), so that missingness depends on the unobservables even given the observables, identifying the CACE becomes challenging and generally requires additional assumptions. In the IV setting, MNAR arises in two main ways. First, missingness may depend on the latent compliance status, that is, on the unit's compliance behavior with respect to the IV encouragement; for instance, individuals who comply with the IV may be more willing to report their treatment and outcome than those who do not. Second, missingness may depend on the missing values themselves; for instance, individuals may be reluctant to disclose treatment or outcome values that are private or sensitive. Section \ref{sec: examples} gives empirical examples of both.

The existing IV literature has largely studied identification of the CACE when the outcome is MNAR, and the proposed mechanisms fall into two groups. The first group lets missingness depend on the latent compliance status. \citet{frangakis1999addressing} established nonparametric identification under one-sided noncompliance, invoking latent ignorability, namely that outcome missingness is conditionally independent of the outcome given the latent compliance status, treatment, and IV, together with a response exclusion restriction whereby the IV affects missingness only through the treatment. \citet{Zhou2006ITT} and \citet{O'Malley2005likelihood} extended these results to two-sided noncompliance for binary and continuous outcomes, respectively. Under the same mechanism, \citet{yahong2004extend} used parametric models to estimate the CACE for continuous and categorical outcomes, \citet{taylor2009multiple} proposed multiple imputation methods, and \citet{frolich2014treatment} extended the setting to multiple outcome periods. Departing from \citet{frangakis1999addressing}, \citet{Mealli2004analyzing} and \citet{mattei2007application} obtained nonparametric identification under one-sided noncompliance and latent ignorability using an alternative response exclusion restriction, namely that for compliers the outcome missingness is unaffected by the IV. \citet{jo2010handling} provided parametric estimation strategies for both the \citet{frangakis1999addressing} and \citet{Mealli2004analyzing} mechanisms, and \citet{nguyen2024identification} and \citet{huber2021plausibility} examined the role and plausibility of latent ignorability in practice.

The second group lets missingness depend on the outcome value itself. \citet{Chen2009identifiability} gave nonparametric identification results for discrete outcomes; \citet{chen2015semiparametric} extended these to continuous outcomes whose distribution belongs to an exponential family. By incorporating covariates that affect the outcome but not its missingness, \citet{Chen2009identifiability} further allowed missingness to depend on the latent compliance status and/or the IV in addition to the outcome value. \citet{Imai2009statistical} provided nonparametric identification for a binary outcome whose missingness depends on both the missing outcome value and the treatment.

This paper studies nonparametric identification of the CACE without auxiliary information. When missingness occurs only in the outcome, we unify existing results and provide new findings for the MNAR case. Treatment MNAR has received far less attention in the IV literature than outcome MNAR; an exception is \citet{calvi2022late}, who studied identification under treatment MNAR using two proxies. As with missingness in the outcome alone, for missingness in the treatment alone we conduct an exhaustive search over all possible missing data mechanisms, identifying the most general ones that permit nonparametric identification under positivity and conditional dependence conditions and providing counterexamples for those that are not identifiable. We then extend the analysis to the case where both the treatment and outcome are missing. Because the missingness mechanism is typically unknown and untestable, \citet{huber2021plausibility} recommends comparing alternative mechanisms as a robustness check; our identification results supply the theoretical basis for such comparisons across multiple plausible mechanisms.

The rest of the paper is organized as follows. In Section \ref{sec: notation}, we introduce the notation and basic concepts for the IV model. In Section \ref{sec: examples}, we provide real-world IV examples where the treatment and/or outcome may be MNAR. In Section \ref{sec:misoutcome}, we present nonparametric identification results with missingness only in the outcome. In Section \ref{sec: mistreatment}, we provide nonparametric identification results with missingness only in the treatment. In Section \ref{sec: mistreatmentoutcome}, we give nonparametric identification results with missingness in both the treatment and outcome. In Section \ref{sec:numerical}, we illustrate the practical implications of the identification results through a simulation study, and in Section \ref{sec:njcs} we apply the framework to the National Job Corps Study. We conclude with a discussion and provide the proofs of the theorems in the appendix.

\section{Review of the IV model without missing data}\label{sec: notation}
We focus on the setup with a binary IV and a binary treatment. Let $Z$ denote the IV, with $Z=1$ encouraging receipt of the treatment and $Z=0$ otherwise, and let $D$ denote the treatment received, with $D=1$ and $D=0$ for the treatment and control conditions. Noncompliance arises when $Z\neq D$ for some units. Let $Y$ denote the outcome of interest. 
To simplify the presentation, we omit the pretreatment covariates by implicitly conditioning on them, and we consider the setup where $Z$ is randomized within strata of the covariates. All assumptions and identification results in the paper should be read as holding conditional on these covariates. 
We adopt the potential outcomes framework, defining the potential treatments $\{D(1), D(0)\}$ and potential outcomes $\{Y(1), Y(0)\}$ with respect to the IV. The observed values are $D = Z D(1) + (1 - Z) D(0)$ and $Y = Z  Y(1) + (1 - Z) Y(0)$. We index potential outcomes by the IV as $Y(z)$; the alternative notations $Y(d)$ and $Y(z,d)$ also appear in the literature (see Chapter 21.6 of \citet{ding2024first} for a discussion of these notational issues). Following \citet{imbens1994identification} and \citet{angrist1996identification}, we classify units into four latent compliance statuses, denoted $U$, according to the joint potential treatments $\{D(1),D(0)\}$:
\begin{equation*}
U = \begin{cases}
~a, ~~\text{if } D(1)=1 \text{ and } D(0)=1; \\
~c, ~~\text{if } D(1)=1 \text{ and } D(0)=0; \\
~d, ~~\text{if } D(1)=0 \text{ and } D(0)=1; \\
~n, ~~\text{if } D(1)=0 \text{ and } D(0)=0,
\end{cases}
\end{equation*}
where $a$, $c$, $d$, and $n$ denote always-taker, complier, defier, and never-taker, respectively. We call the situation in which units in both the $Z=1$ and $Z=0$ groups fail to fully comply with their IV encouragement two-sided noncompliance. When units in the $Z=0$ group have no access to the treatment, i.e., $D(0) = 0$, the situation is one-sided noncompliance, involving only compliers and never-takers, with no defiers or always-takers. The principal stratum $U=\{D(1),D(0)\}$ captures all unmeasured confounding between $D$ and $Y$, in the sense that given $U$ and $Z$, the treatment $D=ZD(1)+(1-Z)D(0)$ is a deterministic function of $Z$, so that $\{Y(1), Y(0)\}\independent D\mid U, Z$. Conditioning on $U$ therefore suffices to remove all $D$--$Y$ confounding, and we use the same $U$ throughout the DAGs and in $\bP(Z,U,D,Y)$. The causal effect of interest is the CACE, defined as $\mathbb{E}\{Y(1)-Y(0)\mid U=c\}$. We impose the following IV assumptions throughout the paper.

\begin{ivassumption}\label{assiv}
\leavevmode\newline
$(1)$ Randomization: $Z \independent \{D(1),D(0),Y(1),Y(0)\}$;\\
$(2)$ Monotonicity: $D(1) \geq D(0)$ for all units, which implies that there are no defiers;\\
$(3)$ Nonzero average causal effect of $Z$ on $D$: $\mathbb{E}\{D(1)-D(0)\} \neq 0$;\\
$(4)$ Exclusion restriction: $Y(1) = Y(0)$ for always-takers $(U = a)$ and never-takers $(U = n)$.

\end{ivassumption}

Under the IV Assumptions, \citet{imbens1994identification} and \citet{angrist1996identification} showed that the CACE is identified by:
\begin{align}
\mathbb{E}\{Y(1)-Y(0)\mid U=c\}=\frac{\mathbb{E}(Y\mid Z=1)-\mathbb{E}(Y\mid Z=0)}{\bP(D=1\mid Z=1)-\bP(D=1\mid Z=0)}.
\end{align}
The right-hand side of (1) is the ratio of the IV-induced difference in mean outcomes to the IV-induced difference in mean treatment received, both estimable from the sample moments of $Y$ and $D$ across values of the IV. 

Formula (1) applies when no data are missing. The remainder of the paper addresses identification when $D$, $Y$, or both are missing.

\section{Examples of IV studies with missing data}\label{sec: examples}
We review four empirical IV studies where the treatment and/or outcome may be MNAR.

\begin{example}\label{exp1}
\textup{(MNAR in $Y$)} 
\citet{Mealli2004analyzing} examined the effect of an enhanced training course on the practice of breast self-examination (BSE) in Faenza, Italy. In their study, 657 women were randomly assigned to either the standard treatment of receiving only mailed information about BSE $(Z = 0)$ or the enhanced treatment of an additional training course $(Z = 1)$. The binary indicator $D$ records whether a woman received the enhanced treatment, and the outcome $Y$ records whether she practiced BSE one year later. Only 55\% of the women assigned to the enhanced treatment adhered to their assignment, and those assigned to the standard treatment had no access to the enhanced treatment, so the noncompliance is one-sided. The missing rate for $Y$ is 35\%. \citet{Mealli2004analyzing} suggested that missingness may depend on the compliance status $U$ and the treatment assignment $Z$; for example, mothers who complied with their assigned treatment might be more likely to respond to the survey. Beyond $U$ and $Z$, \citet{Small2009discussion} argued that missingness in $Y$ may also depend on $Y$ itself, since mothers who practiced BSE might be more willing to respond than those who did not.
\end{example}

\begin{example}\label{exp2}
\textup{(MNAR in $D$ and MAR in $Y$)}
\citet{rai2023efficient} examined the effect of educational attainment and IQ scores on wages, using another indicator of ability as an IV to account for measurement error in IQ scores. The data come from the 1976 National Longitudinal Survey of Young Men in the United States \citep{card1993using}. Here $Z$ measures scores on a ``Knowledge of the World of Work'' test, $D$ is the IQ score, and $Y$ is the hourly wage in 1976. About 58\% of subjects have both $D$ and $Y$ observed, about 26\% have $Y$ but not $D$, about 11\% have $D$ but not $Y$, and about 5\% have both missing. \citet{rai2023efficient} argued that missingness in $Y$ stems from survey design rather than intentional non-response, so it is reasonable to treat $Y$ as MAR; in contrast, missingness in $D$ could be MNAR if it arises from individuals or schools unwilling to report low IQ scores. 
\end{example}

\begin{example}\label{exp3}
\textup{(MNAR in both $D$ and $Y$)} 
\citet{scholder2014alcohol} studied the effect of prenatal alcohol exposure on children's academic achievement for a cohort born in the Avon area of England, using a validated genetic variant as the IV \citep{zuccolo2009anon}. Here $Z=1$ when the mother carried the rare variant and $Z=0$ otherwise. The treatment is $D=1$ if the mother reported drinking any amount at any time during pregnancy and $D=0$ if she reported not drinking in any of the three trimesters. The outcomes ($Y$) are several measures of children's academic achievement, namely scores on nationally set examinations at different ages, drawn from the National Pupil Database in England. Depending on the outcome measure, the fraction of subjects missing $D$ and/or $Y$ ranges from 40\% to 53\%. Missingness in $Y$ could arise from challenges in data linkage across datasets, children not sitting all (or any) exams, or other factors \citep{jay2019data}; while linkage issues might be MAR, not sitting exams could correlate with lower achievement and thus be MNAR. Similarly, because of the societal stigma around alcohol use during pregnancy, mothers who drank might be less likely to report it, which could make $D$ MNAR. 
\end{example}

\begin{example}\label{exp4}
\textup{(MNAR in both $D$ and $Y$)} 
Using the 1988 National Education Longitudinal Survey, \citet{eren2013theeffect} investigated the impact of noncognitive ability on the earnings of young men, employing an earlier measure of noncognitive ability as the IV to address measurement error. Here $Z$ is the standardized eighth-grade Rosenberg and Rotter scales, $D$ is the tenth-grade noncognitive ability (the standardized average of the Rosenberg and Rotter scales), and $Y$ is log weekly earnings. About 18\% of subjects are missing $D$ and/or $Y$. Because the data were collected by survey, individuals with low noncognitive ability might be less willing to provide such information, and those with high earnings might be less willing to report the amount.
\end{example}

These examples illustrate how, in practice, the plausible missingness DAGs should be chosen on the basis of subject-matter knowledge and study design. Because the missingness mechanism is generally untestable from the observed data, the identification results then determine which of these plausible mechanisms permit recovery of the CACE. In Example \ref{exp1}, the concern that response may depend on compliance status and on the unobserved outcome suggests considering mechanisms in which $R^Y$ depends on $U$ and/or $Y$. In Example \ref{exp2}, reluctance to report low IQ scores suggests considering treatment-missingness mechanisms in which $R^D$ may depend on the missing treatment value $D$. In Example \ref{exp3}, the stigma around drinking during pregnancy and the link between low achievement and not sitting exams suggest combined mechanisms in which $R^D$ depends on $D$ and $R^Y$ on $Y$. In Example \ref{exp4}, reluctance to report low noncognitive ability and high earnings points to the same structures; both fall under the combined mechanisms of Section \ref{sec: mistreatmentoutcome}. This motivates the strategy we adopt in the rest of the paper. Because the mechanism is untestable from the observed data alone, we recommend using subject-matter knowledge and the data-collection design to select a set of plausible mechanisms, estimating the CACE under each identifiable one, and comparing the results: agreement across mechanisms supports the stability of the substantive conclusion, whereas large discrepancies signal sensitivity to the missingness assumptions and should be reported. We illustrate this comparison across identifiable mechanisms in Sections \ref{sec:numerical} and \ref{sec:njcs}. When the plausible set also includes nonidentifiable mechanisms, the comparison can be further supplemented with a formal sensitivity analysis, such as those of \citet{Small2009discussion} and \citet{zuo2024mediation}.

\section{Missingness only in the outcome}
\label{sec:misoutcome}

This section considers missingness only in the outcome. Let $R^Y$ be the response indicator for $Y$, with $R^Y=1$ if $Y$ is observed and $R^Y=0$ otherwise. We describe missingness mechanisms with directed acyclic graphs (DAGs) \citep{mohan2021graphical, fay1986causal, ma2003identification}. The most general mechanism lets $R^Y$ depend on all of $Z$, $U$, $D$, and $Y$, but in general the CACE is not nonparametrically identifiable under this mechanism (see Section \ref{subsec::counterexamples1} of the appendix for counterexamples). Figure \ref{fig: missingness in outcome} displays this most general mechanism along with MCAR, MAR, and four MNAR mechanisms that permit nonparametric identification of the CACE. We label each mechanism by 1 together with the variables $R^Y$ may depend on; for example, Assumption 1$ZD$ describes the mechanism in which $R^Y$ depends on $(Z,D)$ but is conditionally independent of $(U,Y)$.
 
When data are MCAR, i.e., $R^Y\independent (Z, U, D, Y)$ and $\bP(Z,D,Y) = \bP(Z,D,Y\mid R^Y=1)$, the complete-case analysis gives a consistent estimate of the CACE. Besides MCAR, the CACE is nonparametrically identifiable under the MAR mechanism and each of the four MNAR mechanisms shown in Figure \ref{fig: missingness in outcome} (d)-(g), under the conditions stated in Theorem \ref{the1} below. For nonparametric identification of the CACE, Assumptions 1$ZD$ and 1$UD$ require only positivity. The self-censoring mechanisms involving $Y$ (Assumptions 1$DY$, 1$ZY$, and 1$UY$) require a binary outcome; Assumptions 1$DY$ and 1$ZY$ additionally require two-sided noncompliance and a conditional dependence assumption. Table \ref{tab:outcome-missing-summary} summarizes these identifying conditions, and we present the formal statements in Theorem \ref{the1}. These four MNAR mechanisms are the most general that permit nonparametric identification: each lets $R^Y$ depend on two variables, and allowing $R^Y$ to depend on any further variable yields a mechanism under which the CACE is no longer identifiable. We provide counterexamples for the nonidentifiable mechanisms in Section \ref{subsec::counterexamples1} of the appendix, namely the two-variable mechanism $1ZU$ and the three-variable mechanisms. Below, we present the MAR and four MNAR assumptions that allow nonparametric identification of the CACE.

\begin{figure*}[ht]
\centering
\scalebox{0.8}{
\begin{tikzpicture}

    \node (z)  at (2,0) {$Z$};
    \node (d)  at (3.5,0) {$D$};
    \node (ry) at (5,1.5) {$R^Y$};
    \node (y)  at (5,0) {$Y$};
    \node (u)  at (4.25,-0.8) {$U$};
    \node (c)  at (3.5,-1.5) {$(a)$ Most general missing model};

    \path[-latex] (z) edge (d);
    \path[-latex] (d) edge (y);
    \path[-latex] (u) edge (d);
    \path[-latex] (u) edge (y);
    \path[-latex] (z) edge (ry);
    \path[-latex] (d) edge (ry);
    \path[-latex] (u) edge (ry);
    \path[-latex] (y) edge (ry);

    \node (z)  at (7,0) {$Z$};
    \node (d)  at (8.5,0) {$D$};
    \node (ry) at (10,1.5) {$R^Y$};
    \node (y)  at (10,0) {$Y$};
    \node (u)  at (9.25,-0.8) {$U$};
    \node (c)  at (8.5,-1.5) {$(b)$ MCAR};

    \path[-latex] (z) edge (d);
    \path[-latex] (d) edge (y);
    \path[-latex] (u) edge (d);
    \path[-latex] (u) edge (y);

    \node (z)  at (12,0) {$Z$};
    \node (d)  at (13.5,0) {$D$};
    \node (ry) at (15,1.5) {$R^Y$};
    \node (y)  at (15,0) {$Y$};
    \node (u)  at (14.25,-0.8) {$U$};
    \node (c)  at (13.5,-1.5) {$(c)$ 1$ZD$ (MAR)};

    \path[-latex] (z) edge (d);
    \path[-latex] (d) edge (y);
    \path[-latex] (u) edge (d);
    \path[-latex] (u) edge (y);
    \path[-latex] (z) edge (ry);
    \path[-latex] (d) edge (ry);

    \node (z)  at (0.25,-3.8) {$Z$};
    \node (d)  at (1.75,-3.8) {$D$};
    \node (ry) at (3.25,-2.3) {$R^Y$};
    \node (y)  at (3.25,-3.8) {$Y$};
    \node (u)  at (2.5,-4.6) {$U$};
    \node (c)  at (1.75,-5.3) {$(d)$ 1$UD$};

    \path[-latex] (z) edge (d);
    \path[-latex] (d) edge (y);
    \path[-latex] (u) edge (d);
    \path[-latex] (u) edge (y);
    \path[-latex] (d) edge (ry);
    \path[-latex] (u) edge (ry);

    \node (z)  at (4.75,-3.8) {$Z$};
    \node (d)  at (6.25,-3.8) {$D$};
    \node (ry) at (7.75,-2.3) {$R^Y$};
    \node (y)  at (7.75,-3.8) {$Y$};
    \node (u)  at (7,-4.6) {$U$};
    \node (c)  at (6.25,-5.3) {$(e)$ 1$DY$};

    \path[-latex] (z) edge (d);
    \path[-latex] (d) edge (y);
    \path[-latex] (u) edge (d);
    \path[-latex] (u) edge (y);
    \path[-latex] (d) edge (ry);
    \path[-latex] (y) edge (ry);

    \node (z)  at (9.25,-3.8) {$Z$};
    \node (d)  at (10.75,-3.8) {$D$};
    \node (ry) at (12.25,-2.3) {$R^Y$};
    \node (y)  at (12.25,-3.8) {$Y$};
    \node (u)  at (11.5,-4.6) {$U$};
    \node (c)  at (10.75,-5.3) {$(f)$ 1$ZY$};

    \path[-latex] (z) edge (d);
    \path[-latex] (d) edge (y);
    \path[-latex] (u) edge (d);
    \path[-latex] (u) edge (y);
    \path[-latex] (z) edge (ry);
    \path[-latex] (y) edge (ry);

    \node (z)  at (13.75,-3.8) {$Z$};
    \node (d)  at (15.25,-3.8) {$D$};
    \node (ry) at (16.75,-2.3) {$R^Y$};
    \node (y)  at (16.75,-3.8) {$Y$};
    \node (u)  at (16,-4.6) {$U$};
    \node (c)  at (15.25,-5.3) {$(g)$ 1$UY$};

    \path[-latex] (z) edge (d);
    \path[-latex] (d) edge (y);
    \path[-latex] (u) edge (d);
    \path[-latex] (u) edge (y);
    \path[-latex] (y) edge (ry);
    \path[-latex] (u) edge (ry);

\end{tikzpicture}
}
\caption{The DAGs in $(a)$ through $(g)$ describe the most general missingness mechanism, MCAR, MAR, and four MNAR mechanisms, respectively, when missingness exists only in the outcome.}\label{fig: missingness in outcome}
\end{figure*}

\begin{table}[!t]
\centering
\begingroup
\footnotesize
\setlength{\tabcolsep}{2.4pt}
\renewcommand{\arraystretch}{1.16}
\begin{tabularx}{\textwidth}{@{}P{0.15\textwidth}P{0.25\textwidth}C{0.1\textwidth}C{0.1\textwidth}C{0.15\textwidth}C{0.15\textwidth}@{}}
\toprule
Mechanism & Missingness assumption & Positivity & Outcome & Noncompliance & Dependence \\
\midrule
1$ZD$
& $R^Y\independent (U,Y)\mid (Z,D)$
& \posmark
& \norestr
& \norestr
& \norestr \\

\addlinespace[0.25em]
1$UD$
& $R^Y\independent (Z,Y)\mid (U,D)$
& \posmark
& \norestr
& \norestr
& \norestr \\

\addlinespace[0.25em]
1$DY$
& $R^Y\independent (Z,U)\mid (D,Y)$
& \posmark
& Binary
& Two-sided
& \tabmech{$Y\notindependent Z\mid D=d$}{for $d=0,1$} \\

\addlinespace[0.25em]
1$ZY$
& $R^Y\independent (U,D)\mid (Z,Y)$
& \posmark
& Binary
& Two-sided
& \tabmech{$Y\notindependent D\mid Z=z$}{for $z=0,1$} \\

\addlinespace[0.25em]
1$UY$
& $R^Y\independent (Z,D)\mid (U,Y)$
& \posmark
& Binary
& \norestr
& \norestr \\
\bottomrule
\end{tabularx}
\endgroup
\caption{Summary of missing outcome mechanisms in Section \ref{sec:misoutcome}. Each row identifies the CACE under the listed conditions. A checkmark indicates that a positivity condition is required; the exact positivity condition is stated in Theorem \ref{the1}. The symbol $\ast$ means no restriction on that aspect.}
\label{tab:outcome-missing-summary}
\end{table}

\begin{assumption}[1\textit{ZD}]
$R^Y \independent  (U,Y) \mid (Z,D)$.
\end{assumption}

Assumption 1\textit{ZD} is the MAR mechanism, since the probability of observing $Y$ depends only on the fully observed IV $Z$ and treatment $D$. With pretreatment covariates $X$, it reads $R^Y\independent (U,Y)\mid (Z,D,X)$, and its plausibility is strengthened when $X$ contains rich baseline predictors of response.

\begin{assumption}[1\textit{UD}]
$R^Y \independent  (Z,Y) \mid (U, D)$.
\end{assumption}

Assumption 1\textit{UD} allows the missingness in $Y$ to depend on $U$ and $D$; it corresponds to latent ignorability together with the response exclusion restriction that $Z$ affects $R^Y$ only through $D$ \citep{frangakis1999addressing,Zhou2006ITT,O'Malley2005likelihood}.

\begin{assumption}[1\textit{DY}]
$R^{Y} \independent (Z,U) \mid (D,Y)$.
\end{assumption}

Assumption 1\textit{DY} is a self-censoring mechanism, since $R^Y$ may depend on the missing outcome value $Y$ itself \citep{li2023self}. It was introduced by \citet{Imai2009statistical}, who considered the setting of two-sided noncompliance and a binary outcome.

\begin{assumption}[1\textit{ZY}]
$R^{Y} \independent (U,D) \mid (Z,Y)$.
\end{assumption}

Assumption 1\textit{ZY} allows both the IV and the outcome value to affect the probability of observing $Y$. \citet{Small2009discussion} studied parametric identification under this mechanism, whereas we study its nonparametric identification.

\begin{assumption}[1\textit{UY}]
$R^Y \independent  (Z,D) \mid (U,Y)$.
\end{assumption}

Assumption 1\textit{UY} instead allows the latent compliance status and the outcome value to affect the probability of observing $Y$. As with Assumption 1\textit{ZY}, \citet{Small2009discussion} studied parametric identification, whereas we study its nonparametric identification.

The following theorem formalizes the identification conditions in Table \ref{tab:outcome-missing-summary}.

\begin{theorem}\label{the1} Suppose the IV assumptions hold. The CACE is nonparametrically identifiable in each of the following cases: 
\leavevmode\newline $(1ZD)$ under Assumption 1\textit{ZD}, if $\bP(R^Y=1\mid Z=z,D=d)>0$ for all $z$ and $d$;\\
$(1UD)$ under Assumption 1\textit{UD}, if $\bP(R^Y=1\mid U=c,D=d)>0$ for $d=0,1$;\\
$(1DY)$ under Assumption 1\textit{DY}, for a binary $Y$ with two-sided noncompliance, if $\bP(R^Y=1\mid D=d,Y=y)>0$ for all $d$ and $y$ and $Y \notindependent Z \mid D=d$ for $d=0,1$;\\
$(1ZY)$ under Assumption 1\textit{ZY}, for a binary $Y$ with two-sided noncompliance, if $\bP(R^Y=1\mid Z=z,Y=y)>0$ for all $z$ and $y$ and $Y \notindependent D \mid Z=z$ for $z=0,1$;\\
$(1UY)$ under Assumption 1\textit{UY}, for a binary $Y$, if $\bP(R^Y=1\mid U=c,Y=y)>0$ for $y=0,1$.
\end{theorem}



In Theorem \ref{the1}, part $(1ZD)$ is the MAR case; $(1UD)$ aligns with \citet{frangakis1999addressing}, \citet{Zhou2006ITT}, and \citet{O'Malley2005likelihood}; and $(1DY)$ corresponds to \citet{Imai2009statistical}, with the additional observation that two-sided noncompliance is necessary for identification. Parts $(1ZY)$ and $(1UY)$ are new nonparametric identification results under outcome MNAR.

All five parts of Theorem \ref{the1} include a positivity assumption, which ensures that observed data on $Y$ are available in all relevant strata. Identification of the CACE under MAR is standard, and identification under Assumption 1\textit{UD} has been extensively discussed in the literature; we refer readers to the appendix for details. Below, we focus on parts $(1DY)$ through $(1UY)$ and give the intuition behind the additional conditions required by these mechanisms involving $Y$.

Because one-sided noncompliance is a special case of two-sided noncompliance with no always-takers, one might expect identification under one-sided noncompliance whenever it holds under two-sided. Theorem \ref{the1} $(1DY)$ and $(1ZY)$ show otherwise: there the CACE is identifiable under two-sided but not one-sided noncompliance. We illustrate why using $(1DY)$. Under Assumption 1$DY$,
\begin{align}
\bP(D=d,Y=y\mid Z=z) &= \frac{\bP(D=d,Y=y,R^Y=1\mid Z=z)}{\bP(R^Y=1\mid D=d,Y=y)}.\nonumber
\end{align}
The conditional distribution $\bP(D, Y\mid Z)$ is identifiable, and so is the CACE if $\bP(R^Y\mid D,Y)$ is identifiable.
Define 
\begin{align*}
    \eta_{d}(y) &= \frac{\bP(R^Y=0\mid D=d,Y=y)}{\bP(R^Y=1\mid D=d,Y=y)}
\end{align*}
for all $d$ and $y$. For each $d=0,1$, we have the following system of linear equations with $\{\eta_{d}(y):y\in\mathcal{Y}\}$ as the unknowns:
\begin{align}
\bP(D=d,R^Y=0\mid Z=z) &= \sum_{y\in\mathcal{Y}} \bP(D=d,Y=y,R^Y=0\mid Z=z) \nonumber\\&=\sum_{y\in\mathcal{Y}} \bP(D=d,Y=y,R^Y=1\mid Z=z)\eta_{d}(y)\nonumber
\end{align}
for $z=0,1$. 
The solutions 
$\eta_{d}(y)$ are unique only if $Y$ is binary and $Y \notindependent Z \mid D=d$ for $d=0,1$. Under one-sided noncompliance, however, $D=1$ occurs only when $Z=1$, so for $d=1$ we are left with the single equation 
\begin{align}
\bP(D=1,R^Y=0\mid Z=1) &=\sum_{y\in\mathcal{Y}} \bP(D=1,Y=y,R^Y=1\mid Z=1)\eta_{1}(y).\nonumber
\end{align}
With one equation and two unknowns $\eta_{1}(0)$ and $\eta_{1}(1)$, $\eta_{1}(y)$ is not identifiable without further assumptions. In essence, when $D=1$ there is no variation in $Z$ relative to $Y$ to identify the CACE under one-sided noncompliance. The same reasoning applies to Theorem \ref{the1} $(1ZY)$, and Section \ref{subsec::counterexamples1} of the appendix gives explicit one-sided counterexamples for both Assumptions 1$DY$ and 1$ZY$.

As a side remark, if Assumptions 1$DY$ and 1$ZY$ are strengthened to Assumption 1$Y$ of \citet{Chen2009identifiability}, namely $R^Y\independent (Z, U, D)\mid Y$, the CACE becomes identifiable for a binary $Y$ under both one-sided and two-sided noncompliance.
Identification is then possible for a discrete $Y$ with up to three categories under one-sided noncompliance and up to four under two-sided noncompliance; see Section \ref{proof1Y} of the appendix.

Theorem \ref{the1} $(1UY)$ requires a binary $Y$; we sketch why. Under Assumption 1$UY$,
\begin{eqnarray*}
\bP(U=n,D=0,Y=y,R^Y=1\mid Z=0)=\bP(U=n,D=0,Y=y,R^Y=1\mid Z=1),
\end{eqnarray*}
which allows us to identify the information for compliers when $z=d=0$ by utilizing the information from never-takers. Specifically,
\begin{eqnarray*}
&&\bP(U=c,D=0,Y=y,R^Y=1\mid Z=0)\\&=& \bP(D=0,Y=y,R^Y=1\mid Z=0)-\bP(U=n,D=0,Y=y,R^Y=1\mid Z=0).\nonumber
\end{eqnarray*} 
Similarly, we identify $\bP(U=c,D=1,Y=y,R^Y=1\mid Z=1)$ using the information from always-takers. We can therefore identify the ratio 
$$\frac{\bP(Y = y \mid U = c, D = 0)}{\bP(Y = y \mid U = c, D = 1)}=\frac{\bP(U=c,D=0,Y=y,R^Y=1\mid Z=0)}{\bP(U=c,D=1,Y=y,R^Y=1\mid Z=1)}$$ for all $y$. In the special case where this ratio equals $1$ for all $y$, the CACE equals $0$ and is identified without recovering the individual probabilities $\bP(Y = y \mid U = c, D = d)$. In general, however, identifying the CACE requires those probabilities, which are recoverable only when $Y$ is binary.




\paragraph*{Summary}
Searching exhaustively over all outcome-missing mechanisms, Theorem \ref{the1} gives the most general ones that identify the CACE, collected in Table \ref{tab:outcome-missing-summary}, while the counterexamples in Section \ref{subsec::counterexamples1} of the appendix show that the rest are not identifiable; the classification is therefore complete. Beyond the CACE, the proof of Theorem \ref{the1} also settles identifiability of the full joint distribution $\bP(Z,U,D,Y)$, which is identifiable under $(1ZD)$, $(1DY)$, and $(1ZY)$, identifiable under $(1UD)$ with additional positivity conditions, and not generally identifiable under $(1UY)$.

\section{Missingness only in the treatment}
\label{sec: mistreatment}

This section considers missingness only in the treatment. Let $R^D$ be the response indicator for $D$, with $R^D=1$ if $D$ is observed and $R^D=0$ otherwise. The most general mechanism lets $R^D$ depend on all of $Z$, $U$, $D$, and $Y$, but in general the CACE is not nonparametrically identifiable under this mechanism (see Section \ref{subsec::counterexamples2} of the appendix for counterexamples). Figure \ref{fig: missingness in treatment} displays this most general mechanism along with MCAR, MAR, and five MNAR mechanisms that permit nonparametric identification of the CACE. We label each mechanism by 2 together with the variables $R^D$ may depend on; for example, Assumption 2$ZY$ describes the mechanism in which $R^D$ depends on $(Z,Y)$ but is conditionally independent of $(U,D)$.

When data are MCAR, i.e., $R^D\independent (Z, U, D, Y)$ and $\bP(Z,D,Y) = \bP(Z,D,Y\mid R^D=1)$, the complete-case analysis gives a consistent estimate of the CACE. Besides MCAR, the CACE is nonparametrically identifiable under the MAR mechanism and each of the five MNAR mechanisms shown in Figure \ref{fig: missingness in treatment} (d)-(h), under the conditions stated in Theorem \ref{the2} below. For nonparametric identification of the CACE, Assumptions $2ZY$ and $2UD$ require only positivity. Beyond positivity, Assumption $2UY$ requires a binary outcome, Assumptions $2ZD$ and $2DY$ require a conditional dependence assumption (for $2DY$, only under two-sided noncompliance), and Assumption $2ZU$ requires one-sided noncompliance. Table \ref{tab:treatment-missing-summary} summarizes these identifying conditions, and we present the formal statements in Theorem \ref{the2}. These five MNAR mechanisms are the most general that permit nonparametric identification: each lets $R^D$ depend on two variables, and allowing $R^D$ to depend on any further variable yields a mechanism under which the CACE is no longer identifiable. We provide counterexamples for the nonidentifiable mechanisms in Section \ref{subsec::counterexamples2} of the appendix, namely the three-variable mechanisms.

\begin{figure*}[ht]
\centering
\scalebox{0.8}{
\begin{tikzpicture}

    \node (z)  at (2,0) {$Z$};
    \node (d)  at (3.5,0) {$D$};
    \node (rd) at (3.5,1.5) {$R^D$};
    \node (y)  at (5,0) {$Y$};
    \node (u)  at (4.25,-0.8) {$U$};
    \node (c)  at (3.5,-1.5) {$(a)$ Most general missing model};

    \path[-latex] (z) edge (d);
    \path[-latex] (d) edge (y);
    \path[-latex] (u) edge (d);
    \path[-latex] (u) edge (y);
    \path[-latex] (z) edge (rd);
    \path[-latex] (d) edge (rd);
    \path[-latex] (u) edge (rd);
    \path[-latex] (y) edge (rd);
    
    \node (z)  at (7,0) {$Z$};
    \node (d)  at (8.5,0) {$D$};
    \node (rd) at (8.5,1.5) {$R^D$};
    \node (y)  at (10,0) {$Y$};
    \node (u)  at (9.25,-0.8) {$U$};
    \node (c)  at (8.5,-1.5) {$(b)$ MCAR};

    \path[-latex] (z) edge (d);
    \path[-latex] (d) edge (y);
    \path[-latex] (u) edge (d);
    \path[-latex] (u) edge (y);

    \node (z)  at (12,0) {$Z$};
    \node (d)  at (13.5,0) {$D$};
    \node (rd) at (13.5,1.5) {$R^D$};
    \node (y)  at (15,0) {$Y$};
    \node (u)  at (14.25,-0.8) {$U$};
    \node (c)  at (13.5,-1.5) {$(c)$ 2$ZY$ (MAR)};

    \path[-latex] (z) edge (d);
    \path[-latex] (d) edge (y);
    \path[-latex] (u) edge (d);
    \path[-latex] (u) edge (y);
    \path[-latex] (z) edge (rd);
    \path[-latex] (y) edge (rd);
    
    \node (z)  at (0,-3.8) {$Z$};
    \node (d)  at (1.5,-3.8) {$D$};
    \node (rd) at (1.5,-2.3) {$R^D$};
    \node (y)  at (3,-3.8) {$Y$};
    \node (u)  at (2.25,-4.6) {$U$};
    \node (c)  at (1.5,-5.3) {$(d)$ 2$UY$};

    \path[-latex] (z) edge (d);
    \path[-latex] (d) edge (y);
    \path[-latex] (u) edge (d);
    \path[-latex] (u) edge (y);
    \path[-latex] (y) edge (rd);
    \path[-latex] (u) edge (rd);

    \node (z)  at (3.5,-3.8) {$Z$};
    \node (d)  at (5,-3.8) {$D$};
    \node (rd) at (5,-2.3) {$R^D$};
    \node (y)  at (6.5,-3.8) {$Y$};
    \node (u)  at (5.75,-4.6) {$U$};
    \node (c)  at (5,-5.3) {$(e)$ 2$DY$};

    \path[-latex] (z) edge (d);
    \path[-latex] (d) edge (y);
    \path[-latex] (u) edge (d);
    \path[-latex] (u) edge (y);
    \path[-latex] (d) edge (rd);
    \path[-latex] (y) edge (rd);

    \node (z)  at (7,-3.8) {$Z$};
    \node (d)  at (8.5,-3.8) {$D$};
    \node (rd) at (8.5,-2.3) {$R^D$};
    \node (y)  at (10,-3.8) {$Y$};
    \node (u)  at (9.25,-4.6) {$U$};
    \node (c)  at (8.5,-5.3) {$(f)$ 2$UD$};

    \path[-latex] (z) edge (d);
    \path[-latex] (d) edge (y);
    \path[-latex] (u) edge (d);
    \path[-latex] (u) edge (y);
    \path[-latex] (d) edge (rd);
    \path[-latex] (u) edge (rd);

    \node (z)  at (10.5,-3.8) {$Z$};
    \node (d)  at (12,-3.8) {$D$};
    \node (rd) at (12,-2.3) {$R^D$};
    \node (y)  at (13.5,-3.8) {$Y$};
    \node (u)  at (12.75,-4.6) {$U$};
    \node (c)  at (12,-5.3) {$(g)$ 2$ZD$};

    \path[-latex] (z) edge (d);
    \path[-latex] (d) edge (y);
    \path[-latex] (u) edge (d);
    \path[-latex] (u) edge (y);
    \path[-latex] (z) edge (rd);
    \path[-latex] (d) edge (rd);

    \node (z)  at (14,-3.8) {$Z$};
    \node (d)  at (15.5,-3.8) {$D$};
    \node (rd) at (15.5,-2.3) {$R^D$};
    \node (y)  at (17,-3.8) {$Y$};
    \node (u)  at (16.25,-4.6) {$U$};
    \node (c)  at (15.5,-5.3) {$(h)$ 2$ZU$};
    
    \path[-latex] (z) edge (d);
    \path[-latex] (d) edge (y);
    \path[-latex] (u) edge (d);
    \path[-latex] (u) edge (y);
    \path[-latex] (z) edge (rd);
    \path[-latex] (u) edge (rd);
    
\end{tikzpicture}
}
\caption{The DAGs in $(a)$ through $(h)$ describe the most general missingness mechanism, MCAR, MAR, and five MNAR mechanisms, respectively, when missingness exists only in the treatment.}\label{fig: missingness in treatment}
\end{figure*}



\begin{table}[!t]
\centering
\begingroup
\footnotesize
\setlength{\tabcolsep}{2.0pt}
\renewcommand{\arraystretch}{1.16}
\begin{tabularx}{\textwidth}{@{}P{0.12\textwidth}P{0.24\textwidth}C{0.10\textwidth}C{0.10\textwidth}C{0.15\textwidth}C{0.15\textwidth}@{}}
\toprule
Mechanism & Missingness assumption & Positivity & Outcome & Noncompliance & Dependence \\
\midrule
2$ZY$
& $R^D\independent (U,D)\mid (Z,Y)$
& \posmark
& \norestr
& \norestr
& \norestr \\

\addlinespace[0.25em]
2$UY$
& $R^D\independent (Z,D)\mid (U,Y)$
& \posmark
& Binary
& \norestr
& \norestr \\

\addlinespace[0.25em]
2$DY$
& $R^D\independent (Z,U)\mid (D,Y)$
& \posmark
& \norestr
& \norestr
& \begin{tabular}[t]{@{}l@{}}
Two-sided:\\[-0.2ex]
$D\notindependent Z\mid Y=y$\\[-0.2ex]
{\scriptsize for all $y$}
\end{tabular} \\

\addlinespace[0.25em]
2$UD$
& $R^D\independent (Z,Y)\mid (U,D)$
& \posmark
& \norestr
& \norestr
& \norestr \\

\addlinespace[0.25em]
2$ZD$
& $R^D\independent (U,Y)\mid (Z,D)$
& \posmark
& \norestr
& \norestr
& \begin{tabular}[t]{@{}l@{}}
One-sided:\\[-0.2ex]
$Y\notindependent D\mid Z=1$\\[0.35em]
Two-sided:\\[-0.2ex]
$Y\notindependent D\mid Z=z$\\[-0.2ex]
{\scriptsize for $z=0,1$}
\end{tabular} \\

\addlinespace[0.25em]
2$ZU$
& $R^D\independent (D,Y)\mid (Z,U)$
& \posmark
& \norestr
& One-sided
& \norestr \\
\bottomrule
\end{tabularx}
\endgroup
\caption{Summary of missing treatment mechanisms in Section \ref{sec: mistreatment}. Each row identifies the CACE under the listed conditions. A checkmark indicates that a positivity condition is required; the exact positivity condition is stated in Theorem \ref{the2}. The symbol $\ast$ means no restriction on that aspect.}
\label{tab:treatment-missing-summary}
\end{table}

\begin{assumption}[2\textit{ZY}]
$R^D \independent (U,D) \mid (Z,Y)$.
\end{assumption}

Assumption 2\textit{ZY} is the MAR mechanism, since the probability of observing $D$ depends only on the fully observed IV $Z$ and outcome $Y$. In retrospective studies, $R^D$ may depend on $Y$ when $D$ is collected after $Y$ is measured. For example, researchers may ask about patients' smoking history retrospectively to investigate how smoking behavior affects a health outcome of interest. If patients in poorer health are less likely to participate in a survey than those in good health, then $R^D$ depends on $Y$.

\begin{assumption}[2\textit{UY}]
$R^D \independent (Z,D) \mid (U,Y)$.
\end{assumption}

Assumption 2\textit{UY} allows the treatment response indicator to depend on the latent compliance status $U$ and the fully observed outcome $Y$.

\begin{assumption}[2\textit{DY}]
$R^{D} \independent (Z,U) \mid (D,Y)$.
\end{assumption}

Assumption 2\textit{DY} allows the treatment response indicator to depend on the missing treatment value $D$ and the fully observed outcome $Y$.

\begin{assumption}[2\textit{UD}]
$R^D \independent (Z,Y) \mid (U,D)$.
\end{assumption}

Assumption 2\textit{UD} allows the treatment response indicator to depend on the latent compliance status $U$ and the missing treatment value $D$.

\begin{assumption}[2\textit{ZD}]
$R^{D} \independent (U,Y) \mid (Z,D)$.
\end{assumption}

Assumption 2\textit{ZD} allows the treatment response indicator to depend on the fully observed IV $Z$ and the missing treatment value $D$. It is analogous to Assumption 1 of \citet{zuo2024mediation} in the mediation analysis setting, where the authors focus on an MNAR mechanism in the mediator, allowing the fully observed treatment and the incompletely observed mediator to affect the likelihood of missingness. Specifically, in their Assumption 1, our $Z$ corresponds to the treatment, $D$ to the mediator, and $R^D$ to the missingness in the mediator.

\begin{assumption}[2\textit{ZU}]
$R^{D} \independent (D,Y) \mid (Z,U)$.
\end{assumption}

Assumption 2\textit{ZU} allows the treatment response indicator to depend on the fully observed IV $Z$ and the latent compliance status $U$. Because $Z$ and $U$ jointly determine $D$, this mechanism is more general than Assumptions 2\textit{UD} and 2\textit{ZD}.

The following theorem formalizes the identification conditions in Table \ref{tab:treatment-missing-summary}.

\begin{theorem}\label{the2} Suppose the IV assumptions hold. The CACE is nonparametrically identifiable in each of the following cases:
\leavevmode\newline $(2ZY)$ under Assumption 2\textit{ZY}, if $\bP(R^D=1\mid Z=z,Y=y)>0$ for all $z$ and $y$;\\
$(2UY)$ under Assumption 2\textit{UY}, for a binary $Y$, if $\bP(R^D=1\mid U=c,Y=y)>0$ for $y=0,1$;\\
$(2DY)$ under Assumption 2\textit{DY}, if $\bP(R^D=1\mid D=d,Y=y)>0$ for all $d$ and $y$, and either (i) one-sided noncompliance, or (ii) two-sided noncompliance with $D \notindependent Z \mid Y=y$ for all $y$;\\
$(2UD)$ under Assumption 2\textit{UD}, if $\bP(R^D=1\mid U=c,D=d)>0$ for $d=0,1$;\\
$(2ZD)$ under Assumption 2\textit{ZD}, if $\bP(R^D=1\mid Z=z,D=d)>0$ for all $z$ and $d$, and either (i) one-sided noncompliance with $Y \notindependent D \mid Z=1$, or (ii) two-sided noncompliance with $Y \notindependent D \mid Z=z$ for $z=0,1$;\\
$(2ZU)$ under Assumption 2\textit{ZU}, with one-sided noncompliance, if $\bP(R^D=1\mid Z=1,U=u)>0$ for $u=n,c$.
\end{theorem}

In Theorem \ref{the2}, part $(2ZY)$ is the MAR case and its identification is standard; parts $(2UY)$ through $(2ZU)$ are new nonparametric identification results under treatment MNAR. We now give the intuition behind the conditions these MNAR mechanisms require.

In Theorem \ref{the2} $(2UY)$, the binary-$Y$ requirement parallels that in Theorem \ref{the1} $(1UY)$. Under Assumption 2$UY$, we can identify the ratio $$\frac{\bP(Y=y\mid U=c, D=0)}{\bP(Y=y\mid U=c, D=1)}$$ for all $y$. When this ratio equals $1$, the CACE equals $0$; otherwise, identifying the CACE requires the probabilities $\bP(Y = y \mid U = c, D = d)$, which are recoverable only when $Y$ is binary. 

In Theorem \ref{the2} $(2DY)$, we require $D \notindependent Z \mid Y=y$ for all $y$ under two-sided noncompliance but not under one-sided noncompliance. To see why, note that under Assumption 2\textit{DY},
\begin{align}
\bP(D=d,Y=y\mid Z=z) &= \frac{\bP(D=d,Y=y,R^D=1\mid Z=z)}{\bP(R^D=1\mid D=d,Y=y)}.\nonumber
\end{align}
The conditional distribution $\bP(D,Y\mid Z)$ is identifiable, and so is the CACE if $\bP(R^D\mid D,Y)$ is identifiable.
Define 
\begin{align*}
    \zeta_{y}(d) &=  \frac{\bP(R^D=0\mid D=d,Y=y)}{\bP(R^D=1\mid D=d,Y=y)}
\end{align*}
for all $d$ and $y$. For each $y\in \mathcal{Y}$, we have the following system of linear equations with $\{\zeta_{y}(d):d=0,1\}$ as the unknowns:
\begin{align}
\bP(Y=y,R^D=0\mid Z=z) &= \sum_{d=0}^1 \bP(D=d,Y=y,R^D=0\mid Z=z) \nonumber\\&=\sum_{d=0}^1 \bP(D=d,Y=y,R^D=1\mid Z=z)\zeta_{y}(d)\nonumber
\end{align}
for $z=0,1$. Under two-sided noncompliance, the solutions $\zeta_{y}(d)$ are unique only if $D \notindependent Z \mid Y=y$ for all $y$. This condition is unnecessary under one-sided noncompliance: when $D(0)=0$, we identify $\zeta_{y}(0)$ as $$\zeta_{y}(0)=\frac{\bP(Y=y,R^D=0\mid Z=0)}{\bP(D=0,Y=y,R^D=1\mid Z=0)},$$ and then $\zeta_{y}(1)$ as $$\zeta_{y}(1)=\frac{\bP(Y=y,R^D=0\mid Z=1)-\bP(D=0,Y=y,R^D=1\mid Z=1)\zeta_{y}(0)}{\bP(D=1,Y=y,R^D=1\mid Z=1)}.$$

Theorem \ref{the2} $(2UD)$ requires no condition beyond positivity. The reason is that under Assumption 2\textit{UD},
\begin{eqnarray*}
\bP(U=u,D=d,Y=y,R^D=1\mid Z=0)=\bP(U=u,D=d,Y=y,R^D=1\mid Z=1)
\end{eqnarray*} for $(u, d) = (n, 0), (a, 1)$,
so we can recover complier information for $z=0,1$ by using information from never-takers with $z=1$ and always-takers with $z=0$, respectively. Specifically,
\begin{eqnarray*}
&&\bP(U=c,D=0,Y=y,R^D=1\mid Z=0)\\&=& \bP(D=0,Y=y,R^D=1\mid Z=0)-\bP(U=n,D=0,Y=y,R^D=1\mid Z=1).\nonumber
\end{eqnarray*} 
Therefore, we can identify $\bP(Y=y\mid U=c,D=0)$ by $$\bP(Y=y\mid U=c,D=0)=\frac{\bP(U=c,D=0,Y=y,R^D=1\mid Z=0)}{\bP(U=c,D=0,R^D=1 \mid Z=0)}.$$ 
Similarly, we can identify $\bP(Y=y\mid U=c,D=1)$ and therefore the CACE. 

In Theorem \ref{the2} $(2ZD)$, we require $Y \notindependent D \mid Z=z$ for $z=0,1$ under two-sided noncompliance, which reduces to $Y \notindependent D \mid Z=1$ under one-sided noncompliance. To see why, note that under Assumption 2\textit{ZD},
\begin{align}
\bP(D=d,Y=y\mid Z=z) &= \frac{\bP(D=d,Y=y,R^D=1\mid Z=z)}{\bP(R^D=1\mid Z=z,D=d)}.\nonumber
\end{align}
The conditional distribution $\bP(D,Y\mid Z)$ is identifiable, and so is the CACE if $\bP(R^D\mid Z,D)$ is identifiable.
Define 
\begin{align*}
    \zeta_{z}(d) &=  \frac{\bP(R^D=0\mid Z=z,D=d)}{\bP(R^D=1\mid Z=z,D=d)},
\end{align*}
we have the following system of linear equations with $\zeta_{z}(d)$ as the unknowns:
\begin{align}
\bP(Y=y,R^D=0\mid Z=z) =\sum_{d=0}^1 \bP(D=d,Y=y,R^D=1\mid Z=z)\zeta_{z}(d)\nonumber
\end{align}
for each $y\in \mathcal{Y}$. With two-sided noncompliance, the uniqueness of the solutions $\{\zeta_z(d): z=0,1, d=0,1\}$ requires $Y \notindependent D \mid Z=z$ for $z=0,1$. In the case of one-sided noncompliance, the uniqueness of the solution $\{\zeta_{1}(d),d=0,1\}$ requires that $Y \notindependent D \mid Z=1$. For $z=0$, we identify $\zeta_{0}(0)$ as
\begin{align}
\zeta_{0}(0)=\frac{\bP(Y=y,R^D=0\mid Z=0)}{\bP(D=0,Y=y,R^D=1\mid Z=0)}.\nonumber
\end{align}

In Theorem \ref{the2} $(2ZU)$, identification under one-sided noncompliance relies on two facts: $Y$ is conditionally independent of $R^D$ given $Z$ and $U$, and a unit's compliance status in the $Z=1$ group is known once its $D$ is observed. We first identify $\bP(Y = y \mid U = n, D = 0)$ and $\bP(Y = y \mid U = c, D = 1)$:
\begin{align*}
\bP(Y=y\mid U=n,D=0)&=\bP(Y=y\mid Z=1,U=n,D=0,R^D=1),\\
\bP(Y=y\mid U=c,D=1)&=\bP(Y=y\mid Z=1,U=c,D=1,R^D=1).
\end{align*} Since
\begin{align*}
\bP(Y=y\mid Z=1)&=\bP(Y=y\mid U=n,D=0)\bP(U=n,D=0\mid Z=1)\\&~~+\bP(Y=y\mid U=c,D=1)\{1-\bP(U=n,D=0\mid Z=1)\},
\end{align*}
we can subsequently identify 
$\bP(U=n,D=0\mid Z=1)$. Further, since
\begin{align*}
\bP(Y=y\mid Z=0)&=\bP(Y=y\mid U=n,D=0)\bP(U=n,D=0\mid Z=1)\\&~~+\bP(Y=y\mid U=c,D=0)\{1-\bP(U=n,D=0\mid Z=1)\},
\end{align*}
we can identify $\bP(Y=y\mid U=c, D=0)$, and hence the CACE. Under two-sided noncompliance, the CACE is not identifiable without further assumptions; Section \ref{subsec::counterexamples2} of the appendix gives a counterexample.

\paragraph*{Summary} An exhaustive search over all treatment-missing mechanisms shows that the cases in Theorem \ref{the2}, collected in Table \ref{tab:treatment-missing-summary}, are the most general ones that identify the CACE; the counterexamples in Section \ref{subsec::counterexamples2} of the appendix show that the more general mechanisms are not identifiable, so the classification is complete. Beyond the CACE, the proof of Theorem \ref{the2} also settles identifiability of the full joint distribution $\bP(Z,U,D,Y)$, which is identifiable under $(2ZY)$, $(2DY)$, $(2ZD)$, and $(2ZU)$, and under $(2UY)$ and $(2UD)$ with additional assumptions.

\section{Missingness in both the treatment and outcome}
\label{sec: mistreatmentoutcome}

This section addresses missingness in both the treatment and outcome. We focus on prospective studies in which $D$ is collected before $Y$, so the future variables $Y$ and $R^Y$ do not affect $R^D$. Each combined mechanism pairs one outcome-missing mechanism from Section \ref{sec:misoutcome} with one of Assumptions 2$UD$, 2$ZD$, and 2$ZU$ from Section \ref{sec: mistreatment}, the treatment-missing mechanisms compatible with the prospective design. The key distinction is whether the direct path $R^D\to R^Y$ may be included. Under 2$UD$ it may, because $R^D$ remains conditionally independent of $Z$ given $(U,D)$; under 2$ZD$ or 2$ZU$ it breaks identification, and we provide counterexamples in Section \ref{subsec::counterexamples3} of the appendix. For the mechanisms in which the direct path $R^D\to R^Y$ is not allowed, we further show that removing one of the other direct paths into $R^D$ or $R^Y$ restores identification, with details in Section \ref{sec: mistreatmentoutcome4} of the appendix.

We label the combined mechanisms with both 1 and 2: label 1 lists the variables $R^Y$ depends on (other than $R^D$), and label 2 lists the variables $R^D$ depends on. We write $\oplus$ when the direct path $R^D\to R^Y$ is allowed and $+$ when it is not, and we write $1\mathcal{M}$ for any of the five outcome-missing mechanisms 1$ZD$, 1$UD$, 1$DY$, 1$ZY$, and 1$UY$. For example, Assumption 1$ZD\oplus$2$UD$ allows $R^Y$ to depend on $(Z,D,R^D)$ and $R^D$ on $(U,D)$, whereas Assumption 1$ZD+$2$ZD$ excludes the path $R^D\to R^Y$ and allows both response indicators to depend on $(Z,D)$.

Figure \ref{fig: missingness in treatment and outcome 1} shows the DAGs for the combined mechanisms, and Table \ref{tab:both-missing-summary} summarizes the three classes of mechanisms studied in this section. The following subsections give the formal assumptions and theorems, grouped by the treatment-missingness component.

\begin{figure*}[ht]
\centering
\scalebox{0.8}{
\begin{tikzpicture}
    \node (z)  at (0,0) {$Z$};
    \node (d)  at (1.5,0) {$D$};
    \node (rd) at (1.5,1.5) {$R^D$};
    \node (ry) at (3,1.5) {$R^Y$};
    \node (y)  at (3,0) {$Y$};
    \node (u)  at (2.25,-0.8) {$U$};
    \node (c)  at (1.5,-1.5) {$(a)$ 1$ZD\oplus$2$UD$};
    
    \path[-latex] (z) edge (d);
    \path[-latex] (d) edge (y);
    \path[-latex] (u) edge (d);
    \path[-latex] (u) edge (y);
    \path[-latex] (d) edge (rd);
    \path[-latex] (u) edge (rd);
    \path[-latex] (d) edge (ry);
    \path[-latex] (z) edge (ry);
    \path[-latex] (rd) edge (ry);

    \node (z)  at (3.5,0) {$Z$};
    \node (d)  at (5,0) {$D$};
    \node (rd) at (5,1.5) {$R^D$};
    \node (ry) at (6.5,1.5) {$R^Y$};
    \node (y)  at (6.5,0) {$Y$};
    \node (u)  at (5.75,-0.8) {$U$};
    \node (c)  at (5,-1.5) {$(b)$ 1$UD\oplus$2$UD$};

    \path[-latex] (z) edge (d);
    \path[-latex] (d) edge (y);
    \path[-latex] (u) edge (d);
    \path[-latex] (u) edge (y);
    \path[-latex] (d) edge (rd);
    \path[-latex] (u) edge (rd);
    \path[-latex] (d) edge (ry);
    \path[-latex] (u) edge (ry);
    \path[-latex] (rd) edge (ry);
    
    \node (z)  at (7,0) {$Z$};
    \node (d)  at (8.5,0) {$D$};
    \node (rd) at (8.5,1.5) {$R^D$};
    \node (ry) at (10,1.5) {$R^Y$};
    \node (y)  at (10,0) {$Y$};
    \node (u)  at (9.25,-0.8) {$U$};
    \node (c)  at (8.5,-1.5) {$(c)$ 1$DY\oplus$2$UD$};

    \path[-latex] (z) edge (d);
    \path[-latex] (d) edge (y);
    \path[-latex] (u) edge (d);
    \path[-latex] (u) edge (y);
    \path[-latex] (d) edge (rd);
    \path[-latex] (u) edge (rd);
    \path[-latex] (d) edge (ry);
    \path[-latex] (y) edge (ry);
    \path[-latex] (rd) edge (ry);

    \node (z)  at (10.5,0) {$Z$};
    \node (d)  at (12,0) {$D$};
    \node (rd) at (12,1.5) {$R^D$};
    \node (ry) at (13.5,1.5) {$R^Y$};
    \node (y)  at (13.5,0) {$Y$};
    \node (u)  at (12.75,-0.8) {$U$};
    \node (c)  at (12,-1.5) {$(d)$ 1$ZY\oplus$2$UD$};

    \path[-latex] (z) edge (d);
    \path[-latex] (d) edge (y);
    \path[-latex] (u) edge (d);
    \path[-latex] (u) edge (y);
    \path[-latex] (d) edge (rd);
    \path[-latex] (u) edge (rd);
    \path[-latex] (z) edge (ry);
    \path[-latex] (y) edge (ry);
    \path[-latex] (rd) edge (ry);

    \node (z)  at (14,0) {$Z$};
    \node (d)  at (15.5,0) {$D$};
    \node (rd) at (15.5,1.5) {$R^D$};
    \node (ry) at (17,1.5) {$R^Y$};
    \node (y)  at (17,0) {$Y$};
    \node (u)  at (16.25,-0.8) {$U$};
    \node (c)  at (15.5,-1.5) {$(e)$ 1$UY\oplus$2$UD$};

    \path[-latex] (z) edge (d);
    \path[-latex] (d) edge (y);
    \path[-latex] (u) edge (d);
    \path[-latex] (u) edge (y);
    \path[-latex] (d) edge (rd);
    \path[-latex] (u) edge (rd);
    \path[-latex] (u) edge (ry);
    \path[-latex] (rd) edge (ry);
    \path[-latex] (y) edge (ry);

        \node (z)  at (0,-3.8) {$Z$};
    \node (d)  at (1.5,-3.8) {$D$};
    \node (rd) at (1.5,-2.3) {$R^D$};
    \node (ry) at (3,-2.3) {$R^Y$};
    \node (y)  at (3,-3.8) {$Y$};
    \node (u)  at (2.25,-4.6) {$U$};
    \node (c)  at (1.5,-5.3) {$(f)$ 1$ZD$+2$ZD$};

    \path[-latex] (z) edge (d);
    \path[-latex] (d) edge (y);
    \path[-latex] (u) edge (d);
    \path[-latex] (u) edge (y);
    \path[-latex] (z) edge (rd);
    \path[-latex] (d) edge (rd);
    \path[-latex] (d) edge (ry);
    \path[-latex] (z) edge (ry);

    \node (z)  at (3.5,-3.8) {$Z$};
    \node (d)  at (5,-3.8) {$D$};
    \node (rd) at (5,-2.3) {$R^D$};
    \node (ry) at (6.5,-2.3) {$R^Y$};
    \node (y)  at (6.5,-3.8) {$Y$};
    \node (u)  at (5.75,-4.6) {$U$};
    \node (c)  at (5,-5.3) {$(g)$ 1$UD$+2$ZD$};

    \path[-latex] (z) edge (d);
    \path[-latex] (d) edge (y);
    \path[-latex] (u) edge (d);
    \path[-latex] (u) edge (y);
    \path[-latex] (z) edge (rd);
    \path[-latex] (d) edge (rd);
    \path[-latex] (d) edge (ry);
    \path[-latex] (u) edge (ry);
    
    \node (z)  at (7,-3.8) {$Z$};
    \node (d)  at (8.5,-3.8) {$D$};
    \node (rd) at (8.5,-2.3) {$R^D$};
    \node (ry) at (10,-2.3) {$R^Y$};
    \node (y)  at (10,-3.8) {$Y$};
    \node (u)  at (9.25,-4.6) {$U$};
    \node (c)  at (8.5,-5.3) {$(h)$ 1$DY$+2$ZD$};

    \path[-latex] (z) edge (d);
    \path[-latex] (d) edge (y);
    \path[-latex] (u) edge (d);
    \path[-latex] (u) edge (y);
    \path[-latex] (z) edge (rd);
    \path[-latex] (d) edge (rd);
    \path[-latex] (d) edge (ry);
    \path[-latex] (y) edge (ry);

    \node (z)  at (10.5,-3.8) {$Z$};
    \node (d)  at (12,-3.8) {$D$};
    \node (rd) at (12,-2.3) {$R^D$};
    \node (ry) at (13.5,-2.3) {$R^Y$};
    \node (y)  at (13.5,-3.8) {$Y$};
    \node (u)  at (12.75,-4.6) {$U$};
    \node (c)  at (12,-5.3) {$(i)$ 1$ZY$+2$ZD$};

    \path[-latex] (z) edge (d);
    \path[-latex] (d) edge (y);
    \path[-latex] (u) edge (d);
    \path[-latex] (u) edge (y);
    \path[-latex] (z) edge (rd);
    \path[-latex] (d) edge (rd);
    \path[-latex] (z) edge (ry);
    \path[-latex] (y) edge (ry);

    \node (z)  at (14,-3.8) {$Z$};
    \node (d)  at (15.5,-3.8) {$D$};
    \node (rd) at (15.5,-2.3) {$R^D$};
    \node (ry) at (17,-2.3) {$R^Y$};
    \node (y)  at (17,-3.8) {$Y$};
    \node (u)  at (16.25,-4.6) {$U$};
    \node (c)  at (15.5,-5.3) {$(j)$ 1$UY$+2$ZD$};

    \path[-latex] (z) edge (d);
    \path[-latex] (d) edge (y);
    \path[-latex] (u) edge (d);
    \path[-latex] (u) edge (y);
    \path[-latex] (z) edge (rd);
    \path[-latex] (d) edge (rd);
    \path[-latex] (u) edge (ry);
    \path[-latex] (y) edge (ry);

        \node (z)  at (0,-7.6) {$Z$};
    \node (d)  at (1.5,-7.6) {$D$};
    \node (rd) at (1.5,-6.1) {$R^D$};
    \node (ry) at (3,-6.1) {$R^Y$};
    \node (y)  at (3,-7.6) {$Y$};
    \node (u)  at (2.25,-8.4) {$U$};
    \node (c)  at (1.5,-9.1) {$(k)$ 1$ZD$+2$ZU$};

    \path[-latex] (z) edge (d);
    \path[-latex] (d) edge (y);
    \path[-latex] (u) edge (d);
    \path[-latex] (u) edge (y);
    \path[-latex] (z) edge (rd);
    \path[-latex] (u) edge (rd);
    \path[-latex] (d) edge (ry);
    \path[-latex] (z) edge (ry);

    \node (z)  at (3.5,-7.6) {$Z$};
    \node (d)  at (5,-7.6) {$D$};
    \node (rd) at (5,-6.1) {$R^D$};
    \node (ry) at (6.5,-6.1) {$R^Y$};
    \node (y)  at (6.5,-7.6) {$Y$};
    \node (u)  at (5.75,-8.4) {$U$};
    \node (c)  at (5,-9.1) {$(l)$ 1$UD$+2$ZU$};

    \path[-latex] (z) edge (d);
    \path[-latex] (d) edge (y);
    \path[-latex] (u) edge (d);
    \path[-latex] (u) edge (y);
    \path[-latex] (z) edge (rd);
    \path[-latex] (u) edge (rd);
    \path[-latex] (d) edge (ry);
    \path[-latex] (u) edge (ry);
    
    \node (z)  at (7,-7.6) {$Z$};
    \node (d)  at (8.5,-7.6) {$D$};
    \node (rd) at (8.5,-6.1) {$R^D$};
    \node (ry) at (10,-6.1) {$R^Y$};
    \node (y)  at (10,-7.6) {$Y$};
    \node (u)  at (9.25,-8.4) {$U$};
    \node (c)  at (8.5,-9.1) {$(m)$ 1$DY$+2$ZU$};

    \path[-latex] (z) edge (d);
    \path[-latex] (d) edge (y);
    \path[-latex] (u) edge (d);
    \path[-latex] (u) edge (y);
    \path[-latex] (z) edge (rd);
    \path[-latex] (u) edge (rd);
    \path[-latex] (d) edge (ry);
    \path[-latex] (y) edge (ry);

    \node (z)  at (10.5,-7.6) {$Z$};
    \node (d)  at (12,-7.6) {$D$};
    \node (rd) at (12,-6.1) {$R^D$};
    \node (ry) at (13.5,-6.1) {$R^Y$};
    \node (y)  at (13.5,-7.6) {$Y$};
    \node (u)  at (12.75,-8.4) {$U$};
    \node (c)  at (12,-9.1) {$(n)$ 1$ZY$+2$ZU$};

    \path[-latex] (z) edge (d);
    \path[-latex] (d) edge (y);
    \path[-latex] (u) edge (d);
    \path[-latex] (u) edge (y);
    \path[-latex] (z) edge (rd);
    \path[-latex] (u) edge (rd);
    \path[-latex] (z) edge (ry);
    \path[-latex] (y) edge (ry);

    \node (z)  at (14,-7.6) {$Z$};
    \node (d)  at (15.5,-7.6) {$D$};
    \node (rd) at (15.5,-6.1) {$R^D$};
    \node (ry) at (17,-6.1) {$R^Y$};
    \node (y)  at (17,-7.6) {$Y$};
    \node (u)  at (16.25,-8.4) {$U$};
    \node (c)  at (15.5,-9.1) {$(o)$ 1$UY$+2$ZU$};

    \path[-latex] (z) edge (d);
    \path[-latex] (d) edge (y);
    \path[-latex] (u) edge (d);
    \path[-latex] (u) edge (y);
    \path[-latex] (z) edge (rd);
    \path[-latex] (u) edge (rd);
    \path[-latex] (u) edge (ry);
    \path[-latex] (y) edge (ry);

\end{tikzpicture}
}
\caption{The DAGs illustrate the combined missingness mechanisms obtained by superposing Assumptions 1$ZD$, 1$UD$, 1$DY$, 1$ZY$, and 1$UY$ with Assumptions 2$UD$, 2$ZD$, and 2$ZU$. Panels $(m)$ and $(n)$, corresponding to 1$DY$+2$ZU$ and 1$ZY$+2$ZU$, are not included in the identification results because the 1$DY$ and 1$ZY$ mechanisms require two-sided noncompliance, whereas 2$ZU$ requires one-sided.}\label{fig: missingness in treatment and outcome 1}
\end{figure*}

\begin{table}[ht]
\centering
\begingroup
\footnotesize
\setlength{\tabcolsep}{3.2pt}
\renewcommand{\arraystretch}{1.18}
\begin{tabularx}{\textwidth}{@{}P{0.15\textwidth}P{0.35\textwidth}C{0.10\textwidth}L@{}}
\toprule
Mechanism & Missingness assumption & $R^D\to R^Y$ & Conditions \\
\midrule
$1\mathcal{M}\oplus2UD$
&
\begin{tabular}[t]{@{}l@{}}
$R^D\independent (Z,Y)\mid (U,D)$;\\[-0.2ex]
$R^Y$ follows 1$\mathcal{M}$ conditional on $R^D$
\end{tabular}
&
\pathyes
&
1$\mathcal{M}$ condition + 2$UD$ condition. \\

\addlinespace[0.45em]
$1\mathcal{M}+2ZD$
&
\begin{tabular}[t]{@{}l@{}}
$R^D\independent (U,Y)\mid (Z,D)$;\\[-0.2ex]
$R^Y$ follows 1$\mathcal{M}$
\end{tabular}
&
\pathno
&
1$\mathcal{M}$ condition + 2$ZD$ condition. \\

\addlinespace[0.45em]
$1\mathcal{M}+2ZU$
&
\begin{tabular}[t]{@{}l@{}}
$R^D\independent (D,Y)\mid (Z,U)$;\\[-0.2ex]
$R^Y$ follows 1$\mathcal{M}$
\end{tabular}
&
\pathno
&
1$\mathcal{M}$ condition + 2$ZU$ condition. For 1$DY$ and 1$ZY$, the outcome and treatment mechanisms impose conflicting noncompliance conditions; no positive result is obtained. \\
\bottomrule
\end{tabularx}
\endgroup
\caption{Summary of combined missing mechanisms in Section \ref{sec: mistreatmentoutcome}. Here $1\mathcal{M}$ denotes any of 1$ZD$, 1$UD$, 1$DY$, 1$ZY$, and 1$UY$. A checkmark indicates that the direct path $R^D\to R^Y$ is allowed, and a $\times$ indicates that it is excluded. Theorems \ref{the3}--\ref{the5} state the exact conditions.}
\label{tab:both-missing-summary}
\end{table}

\subsection{The missing outcome models combined with Assumption 2\textbf{$UD$}}\label{subsec: mistreatmentoutcome1}

The first class pairs one of Assumptions 1$ZD$, 1$UD$, 1$DY$, 1$ZY$, and 1$UY$ with Assumption 2$UD$. It is the only class that allows the direct path $R^D\to R^Y$ without removing any other direct path, because the 2$UD$ mechanism preserves the cross-$Z$ comparisons used to isolate compliers.

\begin{assumption}[1\textit{ZD}$\oplus$2\textit{UD}]
$R^{D} \independent (Z,Y) \mid (U,D)$ and $R^Y \independent (U,Y)\mid (Z,D,R^D)$.
\end{assumption}

\begin{assumption}[1\textit{UD}$\oplus$2\textit{UD}]
$R^D \independent  (Z,Y) \mid (U,D)$ and $R^Y \independent (Z,Y)\mid (U,D,R^D)$.
\end{assumption}

\begin{assumption}[1\textit{DY}$\oplus$2\textit{UD}]
$R^D \independent (Z,Y) \mid (U,D)$ and $R^Y \independent (Z,U)\mid (D,Y,R^D)$.
\end{assumption}

\begin{assumption}[1\textit{ZY}$\oplus$2\textit{UD}]
$R^D \independent (Z,Y)\mid (U,D)$ and $R^Y \independent (U,D)\mid (Z,Y,R^D)$.
\end{assumption}

\begin{assumption}[1\textit{UY}$\oplus$2\textit{UD}]
$R^D \independent  (Z,Y) \mid (U,D)$ and $R^Y \independent (Z,D)\mid (U,Y,R^D)$.
\end{assumption}

The following theorem formalizes the identification conditions for this first class of mechanisms.

\begin{theorem}\label{the3} Suppose the IV assumptions hold. The CACE is nonparametrically identifiable in each of the following cases:
\leavevmode\newline $(1ZD$$\oplus$$2UD)$ under Assumption 1\textit{ZD}$\oplus$2\textit{UD}, if $\bP(R^D=1\mid U=c,D=d)>0$ for $d=0,1$, and $\bP(R^Y=1\mid Z=z,D=d,R^D=1)>0$ for all $z$ and $d$;\\
$(1UD$$\oplus$$2UD)$ under Assumption 1\textit{UD}$\oplus$2\textit{UD}, if $\bP(R^D = 1 \mid U = c, D = d) > 0$ and $\bP(R^Y = 1 \mid U = c, D = d, R^D = 1) > 0$ for $d = 0, 1$;\\
$(1DY$$\oplus$$2UD)$ under Assumption 1\textit{DY}$\oplus$2\textit{UD}, for a binary $Y$ with two-sided noncompliance, if $\bP(R^D=1\mid U=c,D=d)>0$ for $d=0,1$, $\bP(R^Y=1\mid D=d,Y=y,R^D=1)>0$ for all $d$ and $y$, and $Y \notindependent Z \mid (D=d,R^D=1)$ for $d=0,1$;\\
$(1ZY$$\oplus$$2UD)$ under Assumption 1\textit{ZY}$\oplus$2\textit{UD}, for a binary $Y$ with two-sided noncompliance, if $\bP(R^D=1\mid U=c,D=d)>0$ for $d=0,1$, $\bP(R^Y=1\mid Z=z,Y=y,R^D=1)>0$ for all $z$ and $y$, and $Y \notindependent D \mid (Z=z,R^D=1)$ for $z=0,1$;\\
$(1UY$$\oplus$$2UD)$ under Assumption 1\textit{UY}$\oplus$2\textit{UD}, for a binary $Y$, if $\bP(R^D=1\mid U=c,D=d)>0$ for $d=0,1$, and $\bP(R^Y=1\mid U=c,Y=y,R^D=1)>0$ for $y=0,1$.
\end{theorem}

The conditions in Theorem \ref{the3} inherit the corresponding outcome-missing conditions from Theorem \ref{the1}, conditioning on $R^D=1$ wherever $R^Y$ depends on $R^D$, and add the 2\textit{UD} positivity condition from Theorem \ref{the2}. Because $R^D\independent Z\mid(U,D)$, the direct path $R^D\to R^Y$ does not disrupt the cross-$Z$ comparisons used to isolate compliers. The case-specific recovery formulas mirror those of the corresponding outcome-missing mechanisms in Section \ref{sec:misoutcome}; see the proof for details.

\subsection{The missing outcome models combined with Assumption 2\textbf{$ZD$}}\label{subsec: mistreatmentoutcome2}

The second class pairs one of Assumptions 1$ZD$, 1$UD$, 1$DY$, 1$ZY$, and 1$UY$ with Assumption 2$ZD$. Here the positive results do not allow the direct path $R^D\to R^Y$; we give counterexamples in Section \ref{subsec::counterexamples3} of the appendix.

\begin{assumption}[1\textit{ZD}+2\textit{ZD}]
$R^D \independent (U,Y)\mid (Z,D)$ and $R^{Y}\independent (U,Y,R^D) \mid (Z,D)$.
\end{assumption}

\begin{assumption}[1\textit{UD}+2\textit{ZD}]
$R^D\independent  (U,Y) \mid (Z,D)$ and $R^Y \independent (Z,Y,R^D)\mid (U,D)$.
\end{assumption}

\begin{assumption}[1\textit{DY}+2\textit{ZD}]
$R^D \independent (U,Y)\mid (Z,D)$ and $R^Y \independent (Z,U,R^D)\mid (D,Y)$.
\end{assumption}

\begin{assumption}[1\textit{ZY}+2\textit{ZD}] 
$R^D \independent (U,Y)\mid (Z,D)$ and $R^Y \independent (U,D,R^D)\mid (Z,Y)$.
\end{assumption}

\begin{assumption}[1\textit{UY}+2\textit{ZD}]
$R^D\independent  (U,Y) \mid (Z,D)$ and $R^Y \independent (Z,D,R^D)\mid (U,Y)$.
\end{assumption}

The following theorem formalizes the identification conditions for this second class of mechanisms. Define a random vector $Y^\dagger=(Y \cdot R^Y, R^Y)$ such that 
$\bP\{Y^\dagger=(y,1)\}=\bP(Y=y,R^Y=1)$ for all $y\in \mathcal{Y}$ and
$\bP\{Y^\dagger=(0,0)\}=\bP(R^Y=0)$.

\begin{theorem}\label{the4} Suppose the IV assumptions hold. The CACE is nonparametrically identifiable in each of the following cases:
\leavevmode\newline $(1ZD$$+$$2ZD)$ under Assumption 1\textit{ZD}+2\textit{ZD}, if $\bP(R^D=1\mid Z=z,D=d) > 0$ and $\bP(R^Y=1\mid Z=z,D=d) > 0$ for all $z$ and $d$, and either (i) one-sided noncompliance with $Y^\dagger \notindependent D \mid Z=1$, or (ii) two-sided noncompliance with $Y^\dagger \notindependent D \mid Z=z$ for $z=0,1$;\\
$(1UD$$+$$2ZD)$ under Assumption 1\textit{UD}+2\textit{ZD}, if $\bP(R^D=1\mid Z=z,D=d)>0$ for all $z$ and $d$ and $\bP(R^Y=1\mid U=c,D=d)>0$ for $d=0,1$, and either (i) one-sided noncompliance with $Y^\dagger \notindependent D \mid Z=1$, or (ii) two-sided noncompliance with $Y^\dagger \notindependent D \mid Z=z$ for $z=0,1$;\\
$(1DY$$+$$2ZD)$ under Assumption 1\textit{DY}+2\textit{ZD}, for a binary $Y$ with two-sided noncompliance, if $\bP(R^D=1\mid Z=z,D=d) > 0$ for all $z$ and $d$, $\bP(R^Y=1\mid D=d,Y=y) > 0$ for all $d$ and $y$, $Y^\dagger \notindependent D \mid Z=z$ for $z=0,1$, and $Y \notindependent Z \mid D=d$ for $d=0,1$;\\
$(1ZY$$+$$2ZD)$ under Assumption 1\textit{ZY}+2\textit{ZD}, for a binary $Y$ with two-sided noncompliance, if $\bP(R^D=1\mid Z=z,D=d) > 0$ for all $z$ and $d$, $\bP(R^Y=1\mid Z=z,Y=y) > 0$ for all $z$ and $y$, and $Y \notindependent D \mid Z=z$ for $z=0,1$;\\
$(1UY$$+$$2ZD)$ under Assumption 1\textit{UY}+2\textit{ZD}, for a binary $Y$, if $\bP(R^D=1\mid Z=z,D=d)>0$ for all $z$ and $d$ and $\bP(R^Y=1\mid U=c,Y=y)>0$ for $y=0,1$, and either (i) one-sided noncompliance with $Y^\dagger \notindependent D \mid Z=1$, or (ii) two-sided noncompliance with $Y^\dagger \notindependent D \mid Z=z$ for $z=0,1$.
\end{theorem}

The conditions in Theorem \ref{the4} combine the corresponding outcome-missing conditions from Theorem \ref{the1} with the 2\textit{ZD} condition from Theorem \ref{the2}. Where stated, the dependence condition from Theorem \ref{the2} changes from $Y\notindependent D\mid Z=z$ to $Y^\dagger\notindependent D\mid Z=z$, so that dependence between $D$ and either the observed outcome value or the outcome response indicator can help recover $\bP(R^D=1\mid Z=z,D=d)$. Adding the direct path $R^D\to R^Y$ destroys the uniqueness of this recovery, and Section \ref{subsec::counterexamples3} of the appendix gives counterexamples.

\subsection{The missing outcome models combined with Assumption 2\textbf{$ZU$}}\label{subsec: mistreatmentoutcome3}

The third class pairs the outcome-missing mechanisms with Assumption 2$ZU$. Only Assumptions 1$ZD$+2$ZU$, 1$UD$+2$ZU$, and 1$UY$+2$ZU$ yield identification: Assumptions 1$DY$ and 1$ZY$ are excluded because they require two-sided noncompliance, whereas 2$ZU$ is identifiable only under one-sided noncompliance. As in Section \ref{subsec: mistreatmentoutcome2}, the direct path $R^D\to R^Y$ is not allowed.

\begin{assumption}[1\textit{ZD}+2\textit{ZU}]
$R^D \independent (D,Y) \mid (Z,U)$ and $R^Y \independent (U,Y,R^D) \mid (Z,D)$.
\end{assumption}

\begin{assumption}[1\textit{UD}+2\textit{ZU}]
$R^D \independent (D,Y) \mid (Z,U)$ and $R^Y \independent (Z,Y,R^D) \mid (U,D)$.
\end{assumption}

\begin{assumption}[1\textit{DY}+2\textit{ZU}]
$R^D \independent (D,Y) \mid (Z,U)$ and $R^Y \independent (Z,U,R^D) \mid (D,Y)$.
\end{assumption}

\begin{assumption}[1\textit{ZY}+2\textit{ZU}]
$R^D \independent (D,Y) \mid (Z,U)$ and $R^Y \independent (U,D,R^D) \mid (Z,Y)$.
\end{assumption}

\begin{assumption}[1\textit{UY}+2\textit{ZU}]
$R^D \independent (D,Y) \mid (Z,U)$ and $R^Y \independent (Z,D,R^D) \mid (U,Y)$.
\end{assumption}

The following theorem formalizes the identification conditions for this third class of mechanisms. 

\begin{theorem}\label{the5}
Suppose the IV assumptions hold. The CACE is nonparametrically identifiable in each of the following cases:
\leavevmode\newline $(1ZD$$+$$2ZU)$ under Assumption 1\textit{ZD}+2\textit{ZU}, with one-sided noncompliance, if $\bP(R^D=1\mid Z=1,U=u)>0$ for $u=n,c$, and $\bP(R^Y=1\mid Z=z,D=d)>0$ for all $z$ and $d$;\\
$(1UD$$+$$2ZU)$ under Assumption 1\textit{UD}+2\textit{ZU}, with one-sided noncompliance, if $\bP(R^D=1\mid Z=1,U=u)>0$ for $u=n,c$, and $\bP(R^Y=1\mid U=c,D=d)>0$ for $d=0,1$;\\
$(1UY$$+$$2ZU)$ under Assumption 1\textit{UY}+2\textit{ZU}, for a binary $Y$ with one-sided noncompliance, if $\bP(R^D=1\mid Z=1,U=u)>0$ for $u=n,c$, and $\bP(R^Y=1\mid U=c,Y=y)>0$ for $y=0,1$.
\end{theorem}

The conditions in Theorem \ref{the5} combine the 2\textit{ZU} condition from Theorem \ref{the2} with the outcome-missing mechanism conditions from Theorem \ref{the1} that do not require two-sided noncompliance. Thus 1\textit{ZD}, 1\textit{UD}, and 1\textit{UY} remain identifiable under one-sided noncompliance, whereas 1\textit{DY} and 1\textit{ZY} are not. We refer readers to the proofs for details.

\newcommand{\TrueCACE}{0.55}
\newcommand{\NonidentRangeLow}{0.26}
\newcommand{\NonidentRangeHigh}{0.41}

\section{Numerical study}\label{sec:numerical}
\begin{figure}[H]
\centering
\includegraphics[width=\textwidth]{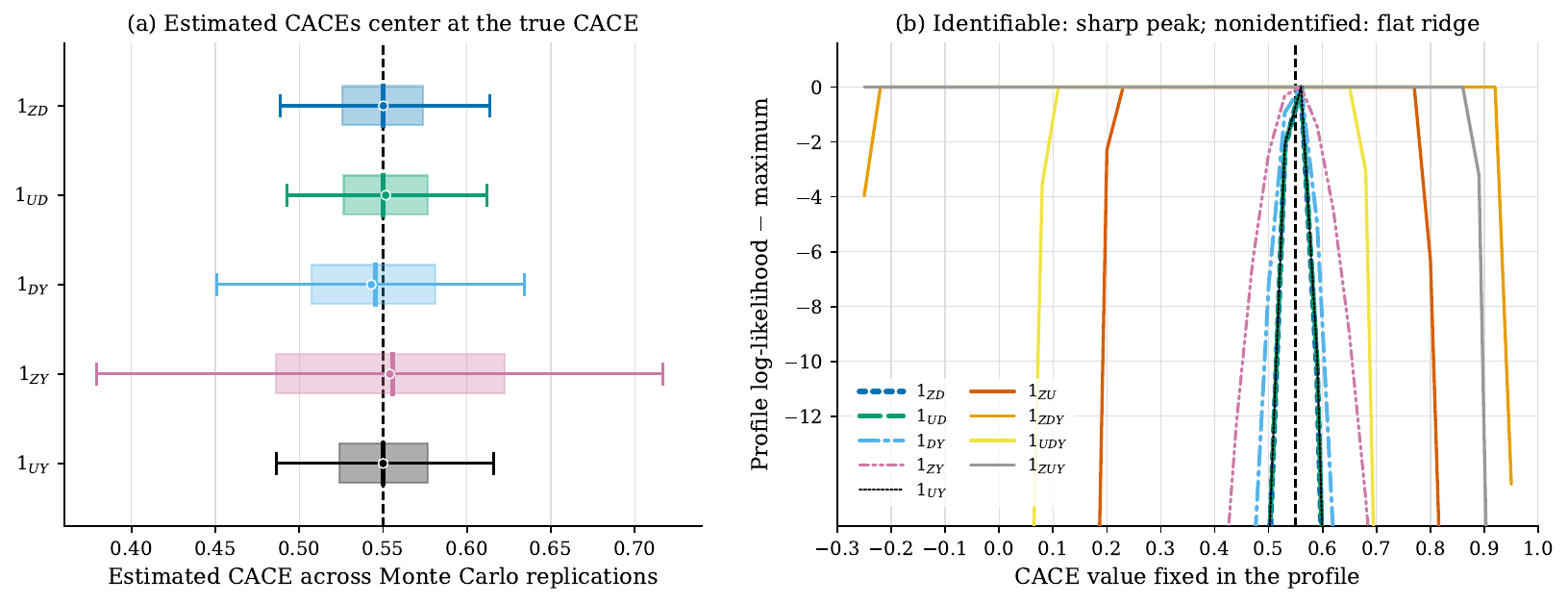}
\caption{Identifiability determines whether the CACE can be recovered. Mechanism labels follow Section~\ref{sec:misoutcome}: the prefix 1 marks outcome-missingness mechanisms and the subscript lists the variables $R^Y$ may depend on. \textbf{(a)} Sampling distributions of the estimator that assumes the true data-generating mechanism ($1000$ replications, sample size $n=5000$; bars: interquartile range and $5$th--$95$th percentiles, median and mean marked); all five concentrate on the true CACE (dashed line). \textbf{(b)} Profile log-likelihoods evaluated at the population level: identifiable mechanisms (dashed/dotted) peak at the truth, nonidentified ones (solid) stay flat.}
\label{fig:identification-headline}
\end{figure}

The identification results separate the mechanisms under which the CACE can be recovered from those under which it cannot. Taking missingness only in the outcome as our illustrative case, we show what that distinction means in practice: under an identifiable mechanism the correctly specified estimator (the estimator that assumes the true mechanism) recovers the CACE, whereas under a nonidentified mechanism the observed data are compatible with a wide range of CACE values. Figure \ref{fig:identification-headline} makes both sides of the contrast visible at a glance.

\paragraph{Setup.} The complete data follow the IV model of Section \ref{sec: notation} with two-sided noncompliance and a binary outcome:
\[
Z\sim \mathrm{Bernoulli}(0.5), \qquad
\big(\bP(U=a),\,\bP(U=c),\,\bP(U=n)\big)=(0.25,\,0.50,\,0.25),
\]
with $D=1$ for always-takers, $D=0$ for never-takers, and $D=Z$ for compliers. The outcome probabilities $\bP(Y=1\mid U=u,D=d)$ are $0.05$ for $(a,1)$, $0.65$ for $(n,0)$, $0.10$ for $(c,0)$, and $0.65$ for $(c,1)$, so the true CACE is $\tau=0.65-0.10=0.55$. We generate the outcome response indicator $R^Y$ under each of the five identifiable mechanisms 1\textit{ZD}, 1\textit{UD}, 1\textit{DY}, 1\textit{ZY}, and 1\textit{UY} from Figure \ref{fig: missingness in outcome}, and under four nonidentified mechanisms 1\textit{ZU}, 1\textit{ZDY}, 1\textit{UDY}, and 1\textit{ZUY}. Response probabilities are calibrated to an average response rate of about $65\%$ and bounded away from zero and one. For each mechanism, we fit the observed-data likelihood implied by the corresponding assumption, imposing no restrictions beyond the conditional independences the mechanism encodes: the model is parametrized by the compliance-class probabilities, the outcome probabilities $q_{ud}=\bP(Y=1\mid U=u,D=d)$, and the response probabilities $\bP(R^Y=1\mid\cdot)$ as a function of the variables that the mechanism allows. The CACE is estimated by maximum likelihood: $\hat\tau=\hat q_{c1}-\hat q_{c0}$. We use $1000$ Monte Carlo replications with sample size $n=5000$.

\paragraph{Identifiable mechanisms recover the CACE.} Panel (a) of Figure \ref{fig:identification-headline} shows the sampling distribution of the correctly specified estimator under each identifiable mechanism. All five are centered at the true CACE, confirming the identification results of Theorem \ref{the1}. In finite samples, the dispersion of the estimated CACE differs across mechanisms in the way the theory anticipates: the mechanisms requiring only positivity (1\textit{ZD}, 1\textit{UD}) and the complier-ratio mechanism (1\textit{UY}) are the most precise, whereas the self-censoring mechanisms 1\textit{DY} and 1\textit{ZY}, which rely on the conditional-dependence conditions, are noisier, with 1\textit{ZY} the most variable.

\paragraph{Nonidentified mechanisms leave the CACE undetermined.} 
Panel (b) plots a profile log-likelihood for the CACE, computed at the population level: in place of a sample log-likelihood, we use the expected log-likelihood under the observed-data distribution implied by the data-generating process. For each fixed $\tau_0$, we impose $q_{c1}-q_{c0}=\tau_0$ and maximize over all remaining parameters, yielding the profile log-likelihood $\ell_p(\tau_0)$; the panel shows the recentered curve $\ell_p(\tau_0)-\max_\tau\ell_p(\tau)$ against $\tau_0$, so its maximum is zero. Under the identifiable mechanisms, the true CACE is the unique maximizer, and the profile peaks sharply there. Under the nonidentified mechanisms, the profile is nearly flat over a broad interval: many CACE values are observationally equivalent, in the sense that they imply the same observed-data distribution and so cannot be distinguished by any amount of data. This is exactly the failure of identification that the counterexamples in Appendix \ref{subsec::counterexamples1} construct explicitly. Table \ref{tab:identification-diagnostics} quantifies this contrast through a complementary diagnostic: under every identifiable mechanism, the spread of CACE values across maximizers is numerically zero, whereas under the nonidentified mechanisms, it ranges from \NonidentRangeLow{} to \NonidentRangeHigh{}.

\begin{figure}[H] \centering \includegraphics[width=0.62\textwidth]{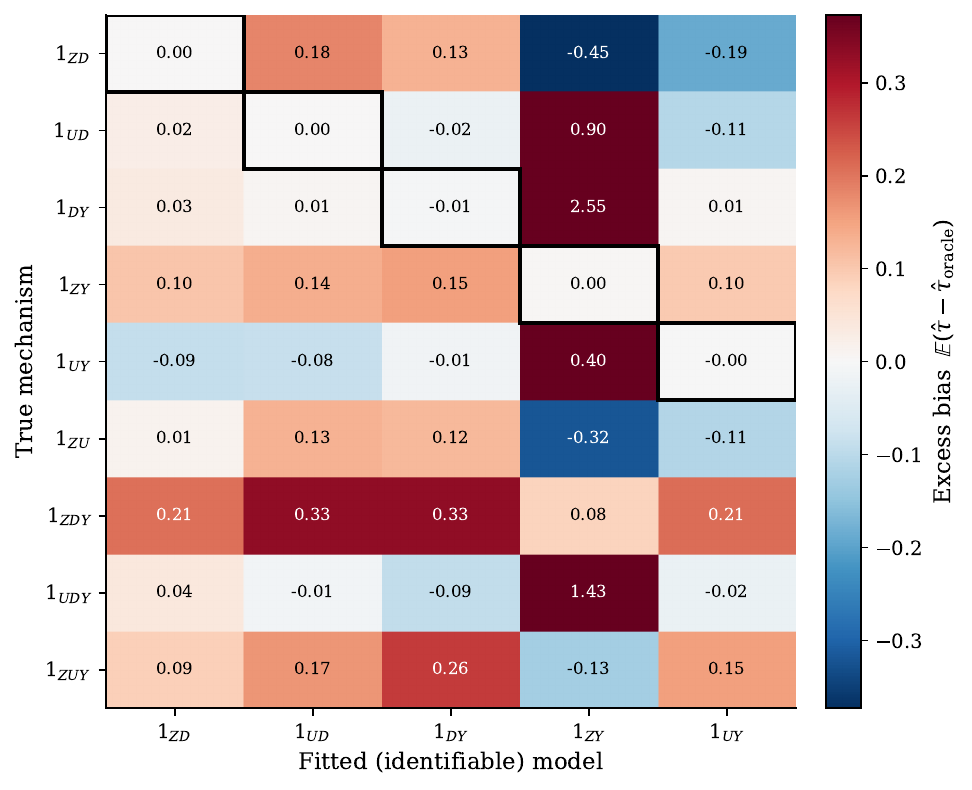} \caption{Excess bias relative to the no-missing-data oracle, $\mathbb{E}(\hat\tau-\hat\tau_{\mathrm{oracle}})$, by true mechanism (rows) and fitted identifiable model (columns). Outlined cells are correctly specified. } \label{fig:misspecification-heatmap} \end{figure}

\paragraph{Fitting the wrong mechanism biases the CACE.} Because the mechanism cannot be tested from the observed data, the analyst's chosen mechanism may differ from the true one. Figure \ref{fig:misspecification-heatmap} reports the excess bias of the CACE estimate relative to the no-missing-data oracle, for every pairing of a true outcome-missingness mechanism (rows) with a fitted identifiable mechanism (columns), averaged over $1000$ Monte Carlo replications at sample size $n=5000$. The outlined diagonal cells are correctly specified and are essentially unbiased, whereas the off-diagonal cells, which fit an identifiable mechanism other than the true one, can carry substantial bias. This quantifies the cost of misspecifying the missingness mechanism and motivates the strategy developed in Section \ref{sec: examples}: rather than commit to a single untestable mechanism, estimate the CACE under all mechanisms that are plausible for the application and report how much the estimate varies across them.  

\begin{table}[H]
\centering
\begin{tabular}{l c r r}
\toprule
Mechanism & Identifiable & CACE at maximum & CACE range \\
\midrule
$1_{ZD}$ & Yes & 0.55 & 0.00 \\
$1_{UD}$ & Yes & 0.55 & 0.00 \\
$1_{DY}$ & Yes & 0.55 & 0.00 \\
$1_{ZY}$ & Yes & 0.55 & 0.00 \\
$1_{UY}$ & Yes & 0.55 & 0.00 \\
\midrule
$1_{ZU}$ & No & 0.50 & 0.27 \\
$1_{ZDY}$ & No & 0.68 & 0.35 \\
$1_{UDY}$ & No & 0.39 & 0.26 \\
$1_{ZUY}$ & No & 0.29 & 0.41 \\
\bottomrule
\end{tabular}

\caption{Identifiability versus likelihood ambiguity, evaluated at the population level. CACE at maximum is the CACE at the top maximizer; CACE range is the spread of CACE values across all maximizers. Identifiable mechanisms: unique maximizer, so range $=0$ and CACE at maximum $=\TrueCACE$. Nonidentifiable mechanisms: maximizers tie, so CACE at maximum is numerical noise and the nonzero range shows how far the CACE is left undetermined.}
\label{tab:identification-diagnostics}
\end{table}

\FloatBarrier

\section{Empirical illustration: the National Job Corps Study}\label{sec:njcs}

We illustrate the theory using the National Job Corps Study \citep{schochet2001national,schochet2006national}, a randomized evaluation of the Job Corps training program. Random assignment to Job Corps is the instrument $Z$, actual enrollment status is the treatment $D$, weekly earnings in the fourth year after assignment is the outcome $Y$, and we adjust for the baseline covariates $X$ (gender, age, race, education, baseline earnings, child status, and arrest history) through parametric models, with details in the replication materials.\footnote{Replication code and data for both the simulation study and this empirical application are available at \url{https://github.com/sushi133/IV_MNAR}.} The study has one-sided noncompliance: those assigned to control had no access to the program, so $D=0$ whenever $Z=0$. Among the $5{,}084$ subjects assigned to Job Corps, $19.7\%$ are known not to have enrolled and enrollment is missing for a further $27.8\%$; the outcome is missing for $21.4\%$ of the full sample of $8{,}707$.

\begin{figure}[H]
\centering
\includegraphics[width=0.8\textwidth]{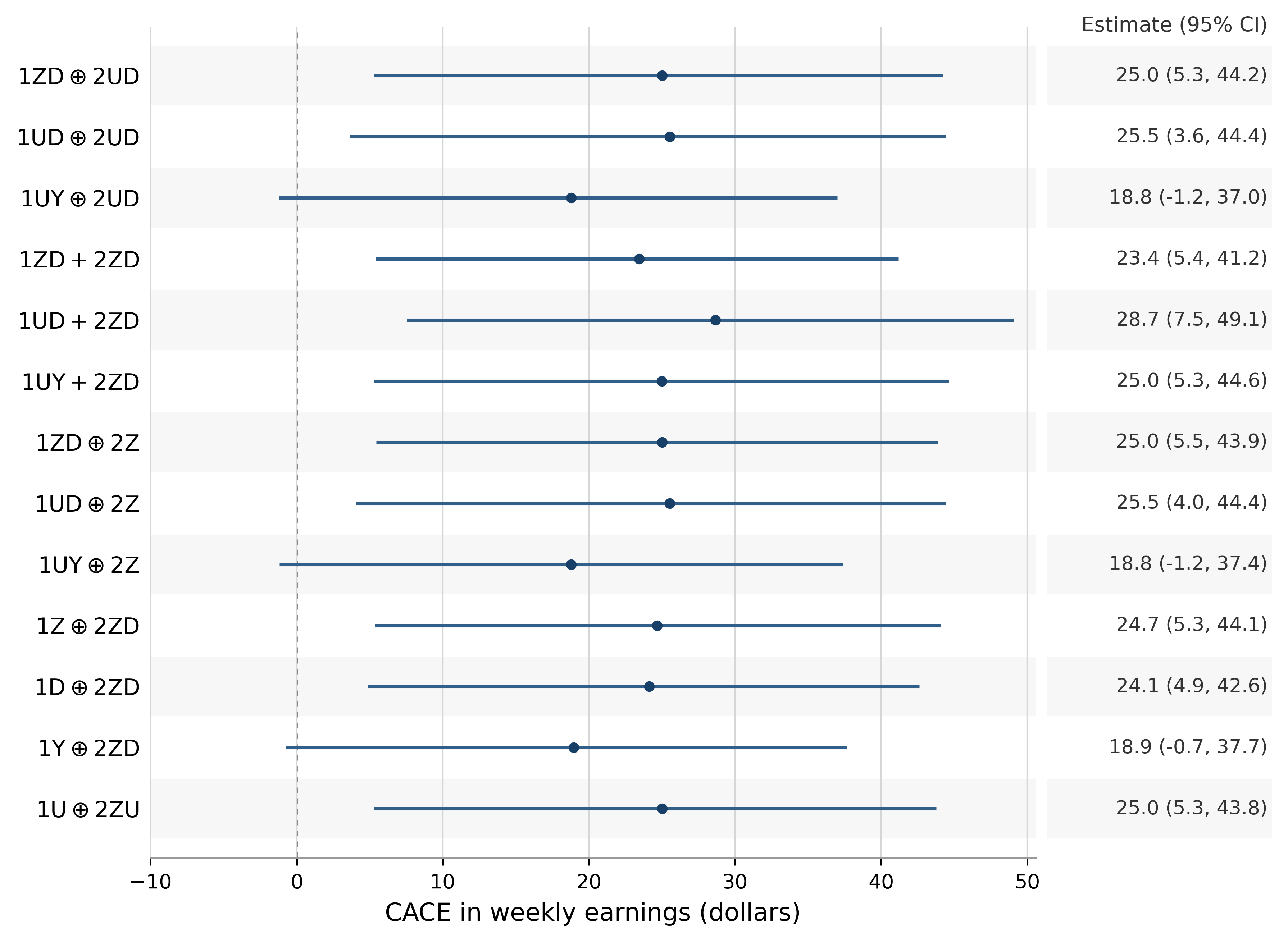}
\caption{Estimated CACE of Job Corps enrollment on weekly earnings, under each of the $13$ identified missingness mechanisms in Section~\ref{sec: mistreatmentoutcome} and Appendix~\ref{sec: mistreatmentoutcome4}. Points are point estimates and bars are $95\%$ percentile bootstrap confidence intervals.}
\label{fig:njcs-cace-forest}
\end{figure}

We estimate the CACE under every mechanism in Section~\ref{sec: mistreatmentoutcome} and Appendix~\ref{sec: mistreatmentoutcome4} that identifies the effect when both $D$ and $Y$ are missing, after excluding the mechanisms that require two-sided noncompliance and those $2\textit{ZU}$ mechanisms that duplicate a $2\textit{ZD}$ counterpart under one-sided noncompliance, since $D$ determines $U$ when $Z=1$ and $D$ is known by design when $Z=0$. This leaves $13$ distinct mechanisms. Several of them require a binary outcome; we assume that any outcome-dependent missingness acts only through the indicator of positive earnings $H=\mathbf{1}(Y>0)$, a meaningful dichotomization here since $14.9\%$ of subjects report zero earnings. Figure~\ref{fig:njcs-cace-forest} reports the resulting estimates and $95\%$ percentile bootstrap confidence intervals based on $500$ resamples.


The estimated effect is positive under all $13$ mechanisms (ranging from \$19 to \$29 per week), with $10$ of the $13$ confidence intervals excluding zero and the remaining 3 extending only marginally below it. Its sign is stable across the identified mechanisms while its statistical significance is mechanism-dependent.


\section{Discussion}\label{sec:disc}

For missingness in the outcome alone and in the treatment alone, we exhaustively searched all possible missingness mechanisms and characterized the most general ones permitting nonparametric identification of the CACE; the numerical study illustrates the practical value of this classification. We then extended the results to the case where both the treatment and outcome are missing, focusing on prospective studies in which the future variables $Y$ and $R^Y$ do not affect $R^D$.


Recall that all our identification results condition on pretreatment covariates. When the covariate dimension is high, the nuisance components of the identifying formulas can be estimated by flexible or machine-learning methods, provided suitable regularity conditions hold, including positivity, consistent nuisance estimation, and adequate control of overfitting. A full estimation theory for high-dimensional covariates is beyond the scope of this paper.

The instrument $Z$ may itself be missing. Our analysis never uses the marginal distribution of $Z$: each identification result relies on only one of the two conditional laws $\bP(Y,D\mid Z)$ or $\bP(Y\mid U=c,Z)$. If $R^Z\independent (U,D,Y)\mid Z$ for the results based on $\bP(Y,D\mid Z)$, or $R^Z\independent Y\mid (U,D,Z)$ for those based on $\bP(Y\mid U=c,Z)$, then the relevant law is unchanged on the units with $Z$ observed, so that it can be recovered from those units and the CACE remains identifiable.  When these conditions fail, however, restricting to the units with $Z$ observed distorts the relevant law, and identification would require modeling the missingness in $Z$ jointly with that in $D$ and $Y$. The same dichotomy applies to a missing pretreatment covariate $X$ when the target is the CACE within levels of $X$. A full treatment of missingness in $Z$ or $X$ is beyond the scope of this paper and is left to future work.

Several extensions remain useful for applied work. Many of our positive results require a binary outcome; discrete outcomes with more categories require finite-dimensional rank conditions, whereas continuous outcomes generally require stronger restrictions such as completeness or parametric structure. Another promising direction for future work is to incorporate auxiliary variables, such as follow-up reminders, survey incentives, or proxies for the response process, which may identify mechanisms beyond those considered here.


\acks{We thank the reviewers for helpful comments. Peng Ding was supported by the U.S. National Science Foundation (grant \# 1945136). Fan Yang is also affiliated with the Yanqi Lake Beijing Institute of Mathematical Sciences and Applications.}


\newpage

\appendix
\section{Alternative missingness mechanisms}\label{sec: mistreatmentoutcome4}

In the main text, Section \ref{sec: mistreatmentoutcome} treats the combined mechanisms in which the direct path $R^D\to R^Y$ is present only under Assumption 2\textit{UD}. This appendix shows that the path can also be accommodated under the other treatment-missing mechanisms, provided one of the remaining direct paths into $R^D$ or $R^Y$ is removed in exchange. Concretely, for Assumptions 1\textit{ZD}+2\textit{ZD}, 1\textit{UD}+2\textit{ZD}, 1\textit{UY}+2\textit{ZD}, 1\textit{DY}+2\textit{ZD}, and 1\textit{ZY}+2\textit{ZD} of Section \ref{subsec: mistreatmentoutcome2}, and for 1\textit{ZD}+2\textit{ZU}, 1\textit{UD}+2\textit{ZU}, and 1\textit{UY}+2\textit{ZU} of Section \ref{subsec: mistreatmentoutcome3}, this trade-off yields eighteen alternative mechanisms that retain identification. Five of these reduce to the more general mechanisms of Section \ref{subsec: mistreatmentoutcome1} and are omitted. A further reduction comes from noncompliance: because Assumptions 1\textit{U}$\oplus$2\textit{ZU} and 1\textit{U}$\oplus$2\textit{ZD} are both identifiable only under one-sided noncompliance, we keep the more general 1\textit{U}$\oplus$2\textit{ZU} and omit 1\textit{U}$\oplus$2\textit{ZD}; Section \ref{subsec::counterexamples3} gives a counterexample for 1\textit{U}$\oplus$2\textit{ZD} under two-sided noncompliance, confirming that nothing is lost. The twelve remaining mechanisms are shown in Figure \ref{fig: missingness in treatment and outcome 4} and summarized in Table \ref{tab:alt-missing-summary}. They fall into two groups according to what $R^D$ is allowed to depend on, treated in Sections \ref{subsec::alt-2Z} and \ref{subsec::alt-2ZDZU} below.

\begin{figure}[ht]
\centering
\scalebox{0.8}{
\begin{tikzpicture}

    \node (z)  at (0,-3.8) {$Z$};
    \node (d)  at (1.5,-3.8) {$D$};
    \node (rd) at (1.5,-2.3) {$R^D$};
    \node (ry) at (3,-2.3) {$R^Y$};
    \node (y)  at (3,-3.8) {$Y$};
    \node (u)  at (2.25,-4.6) {$U$};
    \node (c)  at (1.5,-5.3) {1$ZD\oplus$2$Z$};

    \path[-latex] (z) edge (d);
    \path[-latex] (d) edge (y);
    \path[-latex] (u) edge (d);
    \path[-latex] (u) edge (y);
    \path[-latex] (z) edge (rd);
    \path[-latex] (d) edge (ry);
    \path[-latex] (z) edge (ry);
    \path[-latex] (rd) edge (ry);

    \node (z)  at (3.5,-3.8) {$Z$};
    \node (d)  at (5,-3.8) {$D$};
    \node (rd) at (5,-2.3) {$R^D$};
    \node (ry) at (6.5,-2.3) {$R^Y$};
    \node (y)  at (6.5,-3.8) {$Y$};
    \node (u)  at (5.75,-4.6) {$U$};
    \node (c)  at (5,-5.3) {1$UD\oplus$2$Z$};

    \path[-latex] (z) edge (d);
    \path[-latex] (d) edge (y);
    \path[-latex] (u) edge (d);
    \path[-latex] (u) edge (y);
    \path[-latex] (z) edge (rd);
    \path[-latex] (d) edge (ry);
    \path[-latex] (u) edge (ry);
    \path[-latex] (rd) edge (ry);
    
    \node (z)  at (7,-3.8) {$Z$};
    \node (d)  at (8.5,-3.8) {$D$};
    \node (rd) at (8.5,-2.3) {$R^D$};
    \node (ry) at (10,-2.3) {$R^Y$};
    \node (y)  at (10,-3.8) {$Y$};
    \node (u)  at (9.25,-4.6) {$U$};
    \node (c)  at (8.5,-5.3) {1$UY\oplus$2$Z$};

    \path[-latex] (z) edge (d);
    \path[-latex] (d) edge (y);
    \path[-latex] (u) edge (d);
    \path[-latex] (u) edge (y);
    \path[-latex] (z) edge (rd);
    \path[-latex] (u) edge (ry);
    \path[-latex] (y) edge (ry);
    \path[-latex] (rd) edge (ry);

    \node (z)  at (10.5,-3.8) {$Z$};
    \node (d)  at (12,-3.8) {$D$};
    \node (rd) at (12,-2.3) {$R^D$};
    \node (ry) at (13.5,-2.3) {$R^Y$};
    \node (y)  at (13.5,-3.8) {$Y$};
    \node (u)  at (12.75,-4.6) {$U$};
    \node (c)  at (12,-5.3) {1$DY\oplus$2$Z$};

    \path[-latex] (z) edge (d);
    \path[-latex] (d) edge (y);
    \path[-latex] (u) edge (d);
    \path[-latex] (u) edge (y);
    \path[-latex] (z) edge (rd);
    \path[-latex] (d) edge (ry);
    \path[-latex] (y) edge (ry);
    \path[-latex] (rd) edge (ry);

    \node (z)  at (14,-3.8) {$Z$};
    \node (d)  at (15.5,-3.8) {$D$};
    \node (rd) at (15.5,-2.3) {$R^D$};
    \node (ry) at (17,-2.3) {$R^Y$};
    \node (y)  at (17,-3.8) {$Y$};
    \node (u)  at (16.25,-4.6) {$U$};
    \node (c)  at (15.5,-5.3) {1$ZY\oplus$2$Z$};

    \path[-latex] (z) edge (d);
    \path[-latex] (d) edge (y);
    \path[-latex] (u) edge (d);
    \path[-latex] (u) edge (y);
    \path[-latex] (z) edge (rd);
    \path[-latex] (z) edge (ry);
    \path[-latex] (y) edge (ry);
    \path[-latex] (rd) edge (ry);

    \node (z)  at (0,-7.6) {$Z$};
    \node (d)  at (1.5,-7.6) {$D$};
    \node (rd) at (1.5,-6.1) {$R^D$};
    \node (ry) at (3,-6.1) {$R^Y$};
    \node (y)  at (3,-7.6) {$Y$};
    \node (u)  at (2.25,-8.4) {$U$};
    \node (c)  at (1.5,-9.1) {1$Z\oplus$2$ZD$};

    \path[-latex] (z) edge (d);
    \path[-latex] (d) edge (y);
    \path[-latex] (u) edge (d);
    \path[-latex] (u) edge (y);
    \path[-latex] (z) edge (rd);
    \path[-latex] (d) edge (rd);
    \path[-latex] (z) edge (ry);
    \path[-latex] (rd) edge (ry);

    \node (z)  at (3.5,-7.6) {$Z$};
    \node (d)  at (5,-7.6) {$D$};
    \node (rd) at (5,-6.1) {$R^D$};
    \node (ry) at (6.5,-6.1) {$R^Y$};
    \node (y)  at (6.5,-7.6) {$Y$};
    \node (u)  at (5.75,-8.4) {$U$};
    \node (c)  at (5,-9.1) {1$D\oplus$2$ZD$};

    \path[-latex] (z) edge (d);
    \path[-latex] (d) edge (y);
    \path[-latex] (u) edge (d);
    \path[-latex] (u) edge (y);
    \path[-latex] (d) edge (rd);
    \path[-latex] (z) edge (rd);
    \path[-latex] (d) edge (ry);
    \path[-latex] (rd) edge (ry);

    \node (z)  at (7,-7.6) {$Z$};
    \node (d)  at (8.5,-7.6) {$D$};
    \node (rd) at (8.5,-6.1) {$R^D$};
    \node (ry) at (10,-6.1) {$R^Y$};
    \node (y)  at (10,-7.6) {$Y$};
    \node (u)  at (9.25,-8.4) {$U$};
    \node (c)  at (8.5,-9.1) {1$Y\oplus$2$ZD$};

    \path[-latex] (z) edge (d);
    \path[-latex] (d) edge (y);
    \path[-latex] (u) edge (d);
    \path[-latex] (u) edge (y);
    \path[-latex] (z) edge (rd);
    \path[-latex] (d) edge (rd);
    \path[-latex] (y) edge (ry);
    \path[-latex] (rd) edge (ry);

    \node (z)  at (0,-11.4) {$Z$};
    \node (d)  at (1.5,-11.4) {$D$};
    \node (rd) at (1.5,-9.9) {$R^D$};
    \node (ry) at (3,-9.9) {$R^Y$};
    \node (y)  at (3,-11.4) {$Y$};
    \node (u)  at (2.25,-12.2) {$U$};
    \node (c)  at (1.5,-12.9) {1$Z\oplus$2$ZU$};

    \path[-latex] (z) edge (d);
    \path[-latex] (d) edge (y);
    \path[-latex] (u) edge (d);
    \path[-latex] (u) edge (y);
    \path[-latex] (z) edge (rd);
    \path[-latex] (u) edge (rd);
    \path[-latex] (z) edge (ry);
    \path[-latex] (rd) edge (ry);

    \node (z)  at (3.5,-11.4) {$Z$};
    \node (d)  at (5,-11.4) {$D$};
    \node (rd) at (5,-9.9) {$R^D$};
    \node (ry) at (6.5,-9.9) {$R^Y$};
    \node (y)  at (6.5,-11.4) {$Y$};
    \node (u)  at (5.75,-12.2) {$U$};
    \node (c)  at (5,-12.9) {1$U\oplus$2$ZU$};

    \path[-latex] (z) edge (d);
    \path[-latex] (d) edge (y);
    \path[-latex] (u) edge (d);
    \path[-latex] (u) edge (y);
    \path[-latex] (z) edge (rd);
    \path[-latex] (u) edge (rd);
    \path[-latex] (u) edge (ry);
    \path[-latex] (rd) edge (ry);

    \node (z)  at (7,-11.4) {$Z$};
    \node (d)  at (8.5,-11.4) {$D$};
    \node (rd) at (8.5,-9.9) {$R^D$};
    \node (ry) at (10,-9.9) {$R^Y$};
    \node (y)  at (10,-11.4) {$Y$};
    \node (u)  at (9.25,-12.2) {$U$};
    \node (c)  at (8.5,-12.9) {1$D\oplus$2$ZU$};

    \path[-latex] (z) edge (d);
    \path[-latex] (d) edge (y);
    \path[-latex] (u) edge (d);
    \path[-latex] (u) edge (y);
    \path[-latex] (z) edge (rd);
    \path[-latex] (u) edge (rd);
    \path[-latex] (d) edge (ry);
    \path[-latex] (rd) edge (ry);

    \node (z)  at (10.5,-11.4) {$Z$};
    \node (d)  at (12,-11.4) {$D$};
    \node (rd) at (12,-9.9) {$R^D$};
    \node (ry) at (13.5,-9.9) {$R^Y$};
    \node (y)  at (13.5,-11.4) {$Y$};
    \node (u)  at (12.75,-12.2) {$U$};
    \node (c)  at (12,-12.9) {1$Y\oplus$2$ZU$};

    \path[-latex] (z) edge (d);
    \path[-latex] (d) edge (y);
    \path[-latex] (u) edge (d);
    \path[-latex] (u) edge (y);
    \path[-latex] (z) edge (rd);
    \path[-latex] (u) edge (rd);
    \path[-latex] (y) edge (ry);
    \path[-latex] (rd) edge (ry);

\end{tikzpicture}
}
\caption{The DAGs illustrate the alternative missingness mechanisms that allow for the direct path from $R^D$ to $R^Y$ for identification.}\label{fig: missingness in treatment and outcome 4}
\end{figure}

\begin{table}[!t]
\centering
\begingroup
\footnotesize
\setlength{\tabcolsep}{4pt}
\renewcommand{\arraystretch}{1.2}
\begin{tabularx}{\textwidth}{@{}P{0.12\textwidth}C{0.16\textwidth}C{0.16\textwidth}C{0.10\textwidth}C{0.095\textwidth}C{0.14\textwidth}C{0.11\textwidth}@{}}
\toprule
Mechanism & $R^D$ depends on & $R^Y$ depends on & Positivity & Outcome & Noncompliance & Dependence \\
\midrule
\multicolumn{7}{@{}l}{\textit{Group 1: $R^D$ depends on $Z$ only} (Theorem \ref{the6})}\\
\addlinespace[0.2em]
1$ZD\oplus$2$Z$ & $Z$ & $Z,D,R^D$ & \posmark & \norestr & \norestr & \norestr \\
1$UD\oplus$2$Z$ & $Z$ & $U,D,R^D$ & \posmark & \norestr & \norestr & \norestr \\
1$UY\oplus$2$Z$ & $Z$ & $U,Y,R^D$ & \posmark & Binary & \norestr & \norestr \\
1$DY\oplus$2$Z$ & $Z$ & $D,Y,R^D$ & \posmark & Binary & Two-sided & \posmark \\
1$ZY\oplus$2$Z$ & $Z$ & $Z,Y,R^D$ & \posmark & Binary & Two-sided & \posmark \\
\addlinespace[0.4em]
\multicolumn{7}{@{}l}{\textit{Group 2: $R^D$ depends on $(Z,D)$ or $(Z,U)$} (Theorem \ref{the7})}\\
\addlinespace[0.2em]
1$Z\oplus$2$ZD$ & $Z,D$ & $Z,R^D$ & \posmark & \norestr & \norestr & \posmark \\
1$D\oplus$2$ZD$ & $Z,D$ & $D,R^D$ & \posmark & \norestr & \norestr & \posmark \\
1$Y\oplus$2$ZD$ & $Z,D$ & $Y,R^D$ & \posmark & Binary & \norestr & \posmark \\
1$Z\oplus$2$ZU$ & $Z,U$ & $Z,R^D$ & \posmark & \norestr & One-sided & \norestr \\
1$U\oplus$2$ZU$ & $Z,U$ & $U,R^D$ & \posmark & \norestr & One-sided & \posmark \\
1$D\oplus$2$ZU$ & $Z,U$ & $D,R^D$ & \posmark & \norestr & One-sided & \posmark \\
1$Y\oplus$2$ZU$ & $Z,U$ & $Y,R^D$ & \posmark & Binary & One-sided & \posmark \\
\bottomrule
\end{tabularx}
\endgroup
\caption{Summary of the alternative combined mechanisms in Appendix \ref{sec: mistreatmentoutcome4} that allow the direct path $R^D\to R^Y$. Each row identifies the CACE under the listed conditions. A checkmark in the positivity or dependence column indicates that the corresponding condition is required; the exact conditions are stated in Theorems \ref{the6} and \ref{the7}. The symbol $\ast$ means no restriction on that aspect.}
\label{tab:alt-missing-summary}
\end{table}

\subsection{The outcome models combined with treatment missingness depending on \texorpdfstring{$Z$}{Z}}\label{subsec::alt-2Z}
In the first group, $R^D$ depends only on the instrument $Z$, and $R^Y$ follows one of the outcome-missing mechanisms of Section \ref{sec:misoutcome} while additionally depending on $R^D$.

\begin{assumption}[1\textit{ZD}$\oplus$2\textit{Z}]
$R^{D} \independent (U,D,Y) \mid Z$ and $R^{Y} \independent (U,Y) \mid (Z,D,R^D)$.
\end{assumption}

\begin{assumption}[1\textit{UD}$\oplus$2\textit{Z}]
$R^{D} \independent (U,D,Y) \mid Z$ and $R^{Y} \independent (Z,Y) \mid (U,D,R^D)$.
\end{assumption}

\begin{assumption}[1\textit{UY}$\oplus$2\textit{Z}]
$R^{D} \independent (U,D,Y) \mid Z$ and $R^{Y} \independent (Z,D) \mid (U,Y,R^D)$.
\end{assumption}

\begin{assumption}[1\textit{DY}$\oplus$2\textit{Z}]
$R^{D} \independent (U,D,Y) \mid Z$ and $R^{Y} \independent (Z,U) \mid (D,Y,R^D)$.
\end{assumption}

\begin{assumption}[1\textit{ZY}$\oplus$2\textit{Z}]
$R^{D} \independent (U,D,Y) \mid Z$ and $R^{Y} \independent (U,D) \mid (Z,Y,R^D)$.
\end{assumption}

Theorem \ref{the6} states the identification conditions for this group.
\begin{theorem}\label{the6} Suppose the IV assumptions hold. The CACE is nonparametrically identifiable in each of the following cases:
\leavevmode\newline 
$(1ZD$$\oplus$$2Z)$ under Assumption 1\textit{ZD}$\oplus$2\textit{Z}, if $\bP(R^D=1\mid Z=z)>0$ for $z=0,1$ and $\bP(R^Y=1\mid Z=z,D=d,R^D=1)>0$ for all $z$ and $d$;\\
$(1UD$$\oplus$$2Z)$ under Assumption 1\textit{UD}$\oplus$2\textit{Z}, if $\bP(R^D=1\mid Z=z)>0$ for $z=0,1$ and $\bP(R^Y=1\mid U=c,D=d,R^D=1)>0$ for $d=0,1$;\\
$(1UY$$\oplus$$2Z)$ under Assumption 1\textit{UY}$\oplus$2\textit{Z}, for a binary $Y$, if $\bP(R^D=1\mid Z=z)>0$ for $z=0,1$ and $\bP(R^Y=1\mid U=c,Y=y,R^D=1)>0$ for $y=0,1$;\\
$(1DY$$\oplus$$2Z)$ under Assumption 1\textit{DY}$\oplus$2\textit{Z}, for a binary $Y$ with two-sided noncompliance, if $\bP(R^D=1\mid Z=z) > 0$ for $z=0,1$, $\bP(R^Y=1\mid D=d,Y=y,R^D=1) > 0$ for all $d$ and $y$, and $Y \notindependent Z \mid (D=d,R^D=1)$ for $d=0,1$;\\
$(1ZY$$\oplus$$2Z)$ under Assumption 1\textit{ZY}$\oplus$2\textit{Z}, for a binary $Y$ with two-sided noncompliance, if $\bP(R^D=1\mid Z=z) > 0$ for $z=0,1$, $\bP(R^Y=1\mid Z=z,Y=y,R^D=1) > 0$ for all $z$ and $y$, and $Y \notindependent D \mid Z=z$ for $z=0,1$.
\end{theorem}

\subsection{The outcome models combined with treatment missingness depending on \texorpdfstring{$(Z,D)$ or $(Z,U)$}{(Z,D) or (Z,U)}}\label{subsec::alt-2ZDZU}
In the second group, $R^D$ depends on $(Z,D)$ or on $(Z,U)$, and $R^Y$ depends on a single variable in addition to $R^D$.

\begin{assumption}[1\textit{Z}$\oplus$2\textit{ZD}]
$R^{D} \independent (U,Y) \mid (Z,D)$ and $R^{Y} \independent (U,D,Y) \mid (Z,R^D)$.
\end{assumption}

\begin{assumption}[1\textit{D}$\oplus$2\textit{ZD}]
$R^D \independent (U,Y)\mid (Z,D)$ and $R^Y \independent (Z,U,Y)\mid (D,R^D)$.
\end{assumption}

\begin{assumption}[1\textit{Y}$\oplus$2\textit{ZD}]
$R^D \independent (U,Y)\mid (Z,D)$ and $R^Y \independent (Z,U,D)\mid (Y,R^D)$.
\end{assumption}

\begin{assumption}[1\textit{Z}$\oplus$2\textit{ZU}]
$R^D \independent (D,Y) \mid (Z,U)$ and $R^Y \independent (U,D,Y) \mid (Z,R^D)$.
\end{assumption}

\begin{assumption}[1\textit{U}$\oplus$2\textit{ZU}]
$R^D \independent (D,Y) \mid (Z,U)$ and $R^Y \independent (Z,D,Y) \mid (U,R^D)$.
\end{assumption}

\begin{assumption}[1\textit{D}$\oplus$2\textit{ZU}]
$R^D \independent (D,Y) \mid (Z,U)$ and $R^Y \independent (Z,U,Y) \mid (D,R^D)$.
\end{assumption}

\begin{assumption}[1\textit{Y}$\oplus$2\textit{ZU}]
$R^D \independent (D,Y) \mid (Z,U)$ and $R^Y \independent (Z,U,D) \mid (Y,R^D)$.
\end{assumption}

Theorem \ref{the7} states the identification conditions for this group.
\begin{theorem}\label{the7} Suppose the IV assumptions hold. The CACE is nonparametrically identifiable in each of the following cases:
\leavevmode\newline 
$(1Z$$\oplus$$2ZD)$ under Assumption 1\textit{Z}$\oplus$2\textit{ZD}, if $\bP(R^D=1\mid Z=z, D=d)>0$ for all $z$ and $d$ and $\bP(R^Y=1\mid Z=z, R^D=r^D)>0$ for all $z$ and $r^D$, and either (i) one-sided noncompliance with $Y \notindependent D \mid Z=1$, or (ii) two-sided noncompliance with $Y \notindependent D \mid Z=z$ for $z=0,1$;\\
$(1D$$\oplus$$2ZD)$ under Assumption 1\textit{D}$\oplus$2\textit{ZD}, if $\bP(R^D=1\mid Z=z,D=d) > 0$ for all $z$ and $d$, $\bP(R^Y=1\mid D=d,R^D=r^D) > 0$ for all $d$ and $r^D$, and $D \notindependent Z$, and either (i) one-sided noncompliance with $Y \notindependent D \mid Z=1$, or (ii) two-sided noncompliance with $Y \notindependent D \mid Z=z$ for $z=0,1$;\\
$(1Y$$\oplus$$2ZD)$ under Assumption 1\textit{Y}$\oplus$2\textit{ZD}, for a binary $Y$, if $\bP(R^D=1\mid Z=z,D=d) > 0$ for all $z$ and $d$, $\bP(R^Y=1\mid Y=y,R^D=r^D) > 0$ for all $y$ and $r^D$, $Y \notindependent (Z,D)\mid (R^D=1)$, and $Y \notindependent Z\mid (R^D=0)$, and either (i) one-sided noncompliance with $Y \notindependent D \mid Z=1$, or (ii) two-sided noncompliance with $Y \notindependent D \mid Z=z$ for $z=0,1$;\\
$(1Z$$\oplus$$2ZU)$ under Assumption 1\textit{Z}$\oplus$2\textit{ZU}, with one-sided noncompliance, if $\bP(R^D=1\mid Z=1,U=u)>0$ for $u=n,c$ and $\bP(R^Y=1\mid Z=z,R^D=r^D)>0$ for all $z$ and $r^D$;\\
$(1U$$\oplus$$2ZU)$ under Assumption 1\textit{U}$\oplus$2\textit{ZU}, with one-sided noncompliance, if $\bP(R^D=1\mid Z=1,U=n)>0$, $\bP(R^D=1\mid Z=z,U=c)>0$ for $z=0,1$, $\bP(R^Y=1\mid U=c,R^D=1)>0$, and $R^Y \notindependent U \mid (R^D=1)$;\\
$(1D$$\oplus$$2ZU)$ under Assumption 1\textit{D}$\oplus$2\textit{ZU}, with one-sided noncompliance, if $\bP(R^D=1\mid Z=1,U=u)>0$ for $u=n,c$, $\bP(R^Y=1\mid D=d,R^D=1)>0$ for $d=0,1$, $\bP(R^Y=1\mid D=0,R^D=0)>0$, and $Y \notindependent U \mid (Z=1)$;\\
$(1Y$$\oplus$$2ZU)$ under Assumption 1\textit{Y}$\oplus$2\textit{ZU}, for a binary $Y$ with one-sided noncompliance, if $\bP(R^D=1\mid Z=1,U=u)>0$ for $u=n,c$, $\bP(R^Y=1\mid Y=y,R^D=r^D)>0$ for all $y$ and $r^D$, $Y \notindependent (Z,D) \mid (R^D=1)$, and $Y \notindependent Z \mid (R^D=0)$.
\end{theorem}

\section{Proofs}\label{sec::proofs}
The main paper focuses on binary $Z$ and $D$. In this appendix, some arguments are stated first for more general supports using a completeness condition. Let $f(a\mid b)$ denote the conditional density or probability mass function of $A$ given $B=b$. We say that $f(a\mid b)$ is complete in $B$ if
\[
\int g(a)f(a\mid b)\textup{d}\nu(a)=0\quad \text{for all } b
\]
implies $g(A)=0$ almost surely for every square-integrable function $g$ \citep{newey2003instrumental}, where $\nu(\cdot)$ is the Lebesgue measure for a continuous variable and the counting measure for a discrete variable.

For the binary variables in the main theorems, this completeness requirement reduces to a finite-dimensional rank condition. To see this, suppose $A$ is binary and write $p_b=\bP(A=1\mid B=b)$. The display above then becomes
\[
(1-p_b)g(0)+p_b g(1)=0\quad \text{for all } b.
\]
If $B$ takes values $b_1,\ldots,b_m$, these equations can be written as
\[
\begin{pmatrix}
1-p_{b_1} & p_{b_1}\\
\vdots & \vdots\\
1-p_{b_m} & p_{b_m}
\end{pmatrix}
\begin{pmatrix}
g(0)\\
g(1)
\end{pmatrix}
=
\begin{pmatrix}
0\\
\vdots\\
0
\end{pmatrix}.
\]
The only solution is $g(0)=g(1)=0$ exactly when this matrix has rank two. Since the two columns are $(1-p_b)$ and $p_b$, this holds if and only if $p_b$ is not constant in $b$. Thus, for binary $B$, completeness is equivalent to
\[
\bP(A=1\mid B=0)\neq \bP(A=1\mid B=1),
\]
that is, $A\notindependent B$. When $B$ has more than two support points, the analogous condition is that $\bP(A=1\mid B=b)$ vary with $b$; and if $A$ has $K$ categories, completeness reduces to the requirement that the matrix with entries $\bP(A=a\mid B=b)$ have column rank $K$. The nonindependence conditions in the main text should therefore be read as the finite-support rank conditions arising in the binary cases, not as generic completeness conditions for arbitrary continuous variables.

\subsection{Proof of Theorem \ref{the1}}
\subsubsection{Assumption 1\textit{ZD}}\label{proof1ZD}
The identification of $\bP(D=d,Y=y\mid Z=z)$ follows from
\begin{eqnarray}
\bP(D=d,Y=y\mid Z=z)=\frac{\bP(D=d,Y=y,R^Y=1\mid Z=z)}{\bP(R^Y=1\mid Z=z,D=d)}.\nonumber
\end{eqnarray}

In addition, since
\begingroup
\allowdisplaybreaks
\begin{align*}
\bP(U=n,D=0,Y=y\mid Z=0)&=\bP(U=n,D=0,Y=y\mid Z=1),\\
\bP(U=a,D=1,Y=y\mid Z=1)&=\bP(U=a,D=1,Y=y\mid Z=0),\\
\bP(U=c,D=0,Y=y\mid Z=0) &= \bP(D=0,Y=y\mid Z=0)-\bP(U=n,D=0,Y=y\mid Z=0),\\
\bP(U=c,D=1,Y=y\mid Z=1) &= \bP(D=1,Y=y\mid Z=1)-\bP(U=a,D=1,Y=y\mid Z=1),
\end{align*}
we can identify $\bP(Z=z,U=u,D=d,Y=y)$. Therefore, identifying $\bP(D,Y\mid Z)$ implies identifying $\bP(Z,U,D,Y)$.
\endgroup
\subsubsection{Assumption 1\textit{UD}}\label{proof1DU}
We focus on the identification of $\bP(Y=y\mid U=c,D=d)$ to identify the CACE. Note that
\begin{eqnarray*}
&&\bP(Y=y\mid U=c,D=0)\\&=&\bP(Y=y\mid  Z=0,U=c,D=0,R^Y=1)\\&=&\frac{\bP(U=c,D=0,Y=y,R^Y=1\mid Z=0)}{\bP(U=c,D=0,R^Y=1\mid Z=0)}\\&=& \frac{\bP(D=0,Y=y,R^Y=1\mid Z=0)-\bP(U=n,D=0,Y=y,R^Y=1\mid Z=0)}{\bP(D=0,R^Y=1\mid Z=0)-\bP(U=n,D=0,R^Y=1\mid Z=0)}.\nonumber
\end{eqnarray*}
Since
\begin{eqnarray*}
&&\bP(U=n,D=0,Y=y,R^Y=1\mid Z=0)\\&=&\bP(Y=y,R^Y=1\mid Z=0,U=n,D=0)\bP(U=n,D=0\mid Z=0)\\&=&\bP(Y=y,R^Y=1\mid Z=1,U=n,D=0)\bP(U=n,D=0\mid Z=1)\\&=&\bP(U=n,D=0,Y=y,R^Y=1\mid Z=1),
\end{eqnarray*}
we can identify both $\bP(U=n,D=0,Y=y,R^Y=1\mid Z=0)$ and $\bP(U=n,D=0,R^Y=1\mid Z=0) = \int_{y\in \mathcal{Y}} \bP(U=n,D=0,Y=y,R^Y=1\mid Z=0)\textup{d}y$. Therefore, $\bP(Y=y \mid U=c, D=0)$ is identifiable. Similarly, we can identify $\bP(Y=y \mid U=c, D=1)$. Then, we identify $\mathrm{CACE} = \mathbb{E}(Y \mid U=c, D=1)-\mathbb{E}(Y \mid U=c, D=0)$.

In addition, we can identify $\bP(Z,U,D,Y)$ under additional positivity conditions. For $z$ that is consistent with the $(u, d)$ combination,
\begin{align*}
\bP(R^Y=1\mid U=u,D=d)&=\bP(R^Y=1\mid Z=z,U=u,D=d)\\&=\frac{\bP(U=u,D=d,R^Y=1\mid Z=z)}{\bP(U=u,D=d\mid Z=z)}.
\end{align*}
If $\bP(R^Y=1\mid U=u,D=d)>0$ for $(u,d)=(a,1),(n,0),(c,1),(c,0)$, the identification of $\bP(Z=z,U=u,D=d,Y=y)$ follows from 
\begin{align*}
\bP(Z=z,U=u,D=d,Y=y)&=\frac{\bP(Z=z,U=u,D=d,Y=y,R^Y=1)}{\bP(R^Y=1\mid U=u,D=d)}.
\end{align*}

\subsubsection{Assumption 1\textit{DY}}\label{proof1DY}
We focus on the identification of $\bP(D=d,Y=y\mid Z=z)$ to identify the CACE.
Define $\bP_{dy1\mid z} =\bP(D=d,Y=y,R^Y=1\mid Z=z)$,
    $\bP_{d+0\mid z} =\bP(D=d,R^Y=0\mid Z=z)$,
    $\eta_{d}(y) =\frac{\bP(R^Y=0\mid D=d,Y=y)}{\bP(R^Y=1\mid D=d,Y=y)}$.
Since
$$\bP_{dy1\mid z} = \bP(D=d,Y=y\mid Z=z)\bP(R^Y=1\mid D=d,Y=y),$$
we have
$$\bP_{d+0\mid z} = \int_{y\in \mathcal{Y}} \bP(D=d,Y=y,R^Y=0\mid Z=z)\textup{d}y =\int_{y\in \mathcal{Y}}\bP_{dy1\mid z}\eta_{d}(y)\textup{d}y$$
for each $z\in\mathcal{Z}$. The uniqueness of solutions $\eta_{d}(y)$ requires that $\bP(D=d,Y,R^Y=1 \mid Z)$ is complete in $Z$ for all $d$. For binary $Z$ and $D$, the uniqueness of solutions $\eta_{d}(y)$ requires that $Y$ is binary and $Y \notindependent Z \mid D=d$ for $d=0,1$ under two-sided noncompliance.

We can identify $\bP(R^Y=1\mid D=d,Y=y)$ once $\eta_{d}(y)$ is identified. Then, the identification of $\bP(D=d,Y=y\mid Z=z)$ follows from 
\begin{align}
\bP(D=d,Y=y\mid Z=z) &= \frac{\bP_{dy1\mid z}}{\bP(R^Y=1\mid D=d,Y=y)}.\nonumber
\end{align}

\subsubsection{Assumption 1\textit{ZY}}\label{proof1ZY}
We focus on the identification of $\bP(D=d, Y=y\mid Z=z)$ to identify the CACE. Define $\bP_{dy1\mid z} = \bP(D=d,Y=y,R^Y=1\mid Z=z)$, $
    \bP_{d+0\mid z} = \bP(D=d,R^Y=0\mid Z=z)$,
    $\eta_{z}(y) =  \frac{\bP(R^Y=0\mid Z=z,Y=y)}{\bP(R^Y=1\mid Z=z,Y=y)}$.
Since
$$\bP_{dy1\mid z} = \bP(D=d,Y=y\mid Z=z)\bP(R^Y=1\mid Z=z,Y=y),$$
we have
$$\bP_{d+0\mid z} = \int_{y\in \mathcal{Y}} \bP(D=d,Y=y, R^Y=0\mid Z=z)\textup{d}y =\int_{y\in \mathcal{Y}}\bP_{dy1\mid z}\eta_{z}(y)\textup{d}y
$$
for each $d\in\mathcal{D}$. The uniqueness of solutions $\eta_{z}(y)$ requires that $\bP(D,Y,R^Y=1\mid Z=z)$ is complete in $D$ for all $z$. For binary $Z$ and $D$, the uniqueness of solutions $\eta_{z}(y)$ requires that $Y$ is binary and $Y \notindependent  D\mid Z=z$ for $z=0,1$ under two-sided noncompliance.

We can identify $\bP(R^Y=1\mid Z=z,Y=y)$ once $\eta_{z}(y)$ is identified. Then, the identification of $\bP(D=d,Y=y\mid Z=z)$ follows from 
\begin{align}
\bP(D=d,Y=y\mid Z=z) &= \frac{\bP_{dy1\mid z}}{\bP(R^Y=1\mid Z=z,Y=y)}.\nonumber
\end{align}

\subsubsection{Assumption 1\textit{UY}}\label{proof1UY}
We focus on the identification of the CACE. Note that
\begin{eqnarray*}
&&\bP(U=c,D=0,Y=y,R^Y=1\mid Z=0)\\&=& \bP(D=0,Y=y,R^Y=1\mid Z=0)-\bP(U=n,D=0,Y=y,R^Y=1\mid Z=0).\nonumber
\end{eqnarray*} 
\allowdisplaybreaks
Since
\begin{eqnarray*}
&&\bP(U=n,D=0,Y=y,R^Y=1\mid Z=0)\\&=&\bP(Y=y,R^Y=1\mid Z=0,U=n,D=0)\bP(U=n,D=0\mid Z=0)\\&=&\bP(Y=y,R^Y=1\mid Z=1,U=n,D=0)\bP(U=n,D=0\mid Z=1)\\&=&\bP(U=n,D=0,Y=y,R^Y=1\mid Z=1),
\end{eqnarray*}
we can identify $\bP(U=c,D=0,Y=y,R^Y=1\mid Z=0)$. Similarly, we can identify $\bP(U=c,D=1,Y=y,R^Y=1\mid Z=1)$. Since
\begin{align*}
\frac{\bP(Y=1\mid U=c,D=0)}{\bP(Y=1\mid U=c,D=1)}&=\frac{\bP(U=c,D=0,Y=1,R^Y=1\mid Z=0)}{\bP(U=c,D=1,Y=1,R^Y=1\mid Z=1)},\\
\frac{1-\bP(Y=1\mid U=c,D=0)}{1-\bP(Y=1\mid U=c,D=1)}&=\frac{\bP(U=c,D=0,Y=0,R^Y=1\mid Z=0)}{\bP(U=c,D=1,Y=0,R^Y=1\mid Z=1)},
\end{align*}
we can identify the CACE.

However, $\bP(Z, U, D, Y)$ is not identifiable no matter with two-sided or one-sided noncompliance because there are more unknown parameters in $\bP(Z, U, D, Y)$ than the degree of freedom in the observable data frequencies. Consider two-sided noncompliance, the unknown parameters in $\bP(Z, U, D, Y)$ that describe the graphical model under Assumption 1\textit{UY} are: $\bP(Z=1)$; $\bP(U=u)$ for $u=a, n$; $\mathbb{P}(Y=1\mid U=u,D=d)$ for $(u,d)=(a,1),(n,0),(c,1),(c,0)$; and $\bP(R^Y=1\mid U=u,Y=y)$ for $u=a,n,c$ and $y=0,1$. In total, there are $13$ unknown parameters. In contrast, we have the following observable data frequencies: $\bP(Z=z,D=d,Y=y,R^Y=1)$ for $z=0,1$, $d=0,1$, and $y=0,1$; and $\bP(Z=z,D=d,R^Y=0)$ for $z=0,1$ and $d=0,1$. There are $12$ observable data frequencies. However, since they sum up to $1$, the degree of freedom is $11$. Since $11<13$, $\bP(Z, U, D, Y)$ is not identifiable. Similarly, with one-sided noncompliance, we obtain $9$ unknown parameters, whereas the degree of freedom in the observable data frequencies is $8$, $\bP(Z, U, D, Y)$ is not identifiable.

\subsubsection{Assumption 1\textit{Y}}\label{proof1Y}
Under Assumption 1$Y$, we have
\begin{align}
\bP(D=d,Y=y\mid Z=z) &= \frac{\bP(D=d,Y=y,R^Y=1\mid Z=z)}{\bP(R^Y=1\mid Y=y)}.\nonumber
\end{align}
The conditional distribution $\bP(D,Y\mid Z)$ is identifiable, and so is the CACE if $\bP(R^Y\mid Y)$ is identifiable. Define 
\begin{align*}
    \eta(y) &= \frac{\bP(R^Y=0\mid Y=y)}{\bP(R^Y=1\mid Y=y)}
\end{align*}
for all $y$, we have the following system of linear equations with $\{\eta(y):y\in\mathcal{Y}\}$ as the unknowns:
\begin{align}
\bP(D=d,R^Y=0\mid Z=z) &= \sum_{y\in\mathcal{Y}} \bP(D=d,Y=y,R^Y=0\mid Z=z) \nonumber\\&=\sum_{y\in\mathcal{Y}}\bP(D=d,Y=y,R^Y=1\mid Z=z)\eta(y)\nonumber
\end{align}
for all $z$ and $d$. For a binary $Y$, the uniqueness of the solutions $\eta(y)$ requires $Y \notindependent (Z, D)$. For a discrete $Y$, let $|\mathcal{Y}|$ denote the number of unique values that $Y$ can take. With two-sided noncompliance, we have $(z,d) = (0,0), (0,1), (1,0), (1,1)$. The uniqueness of the solution $\eta(y)$ requires that the rank of the $4 \times |\mathcal{Y}|$ matrix with $\bP(D=d,Y=y,R^Y=1\mid Z=z)$ as the entries equals $|\mathcal{Y}|$, which means that $|\mathcal{Y}| \leq 4$. With one-sided noncompliance, we have $(z,d) = (0,0), (1,0), (1,1)$. Here, the uniqueness of the solution $\eta(y)$ requires that the rank of the $3 \times |\mathcal{Y}|$ matrix with $\bP(D=d,Y=y,R^Y=1\mid Z=z)$ as the entries equals $|\mathcal{Y}|$, which means that $|\mathcal{Y}| \leq 3$. 

\subsection{Proof of Theorem \ref{the2}}
\subsubsection{Assumption 2\textit{ZY}}\label{proof2ZY}
The identification of $\bP(D=d,Y=y\mid Z=z)$ follows from
\begin{eqnarray}
\bP(D=d,Y=y\mid Z=z)=\frac{\bP(D=d,Y=y,R^D=1\mid Z=z)}{\bP(R^D=1\mid Z=z,Y=y)}.\nonumber
\end{eqnarray}

\subsubsection{Assumption 2\textit{UY}}\label{proof2UY}
We focus on the identification of the CACE. Note that
\begin{eqnarray*}
&&\bP(U=c,D=0,Y=y,R^D=1\mid Z=0)\\&=& \bP(D=0,Y=y,R^D=1\mid Z=0)-\bP(U=n,D=0,Y=y,R^D=1\mid Z=0).\nonumber
\end{eqnarray*}
Since
$$\bP(U=n,D=0,Y=y,R^D=1\mid Z=0)=\bP(U=n,D=0,Y=y,R^D=1\mid Z=1),
$$
we can identify $\bP(U=c,D=0,Y=y,R^D=1\mid Z=0)$. Similarly, we can identify $\bP(U=c,D=1,Y=y,R^D=1\mid Z=1)$. Since
\begin{eqnarray*}
\frac{\bP(Y=1\mid U=c,D=0)}{\bP(Y=1\mid U=c,D=1)}=\frac{\bP(U=c,D=0,Y=1,R^D=1\mid Z=0)}{\bP(U=c,D=1,Y=1,R^D=1\mid Z=1)},\\
\frac{1-\bP(Y=1\mid U=c,D=0)}{1-\bP(Y=1\mid U=c,D=1)}=\frac{\bP(U=c,D=0,Y=0,R^D=1\mid Z=0)}{\bP(U=c,D=1,Y=0,R^D=1\mid Z=1)},
\end{eqnarray*}
we can identify the CACE.

Different from Assumption 1\textit{UY}, under Assumption 2\textit{UY}, $\bP(Z,U,D,Y)$ is not identifiable with two-sided noncompliance but is identifiable with one-sided noncompliance under a full rank condition and additional positivity conditions. With two-sided noncompliance, following the discussion in \ref{proof1UY}, there are 13 unknown parameters in $\bP(Z,U,D,Y)$ whereas the degree of freedom in the observable data frequencies is 11; $\bP(Z,U,D,Y)$ is not identifiable. With one-sided noncompliance, the number of unknown parameters in $\bP(Z,U,D,Y)$ is $9$, which matches with the degree of freedom in the observable data frequencies, opening the possibility for identification. 

We now show the identifiability of $\bP(Z, U, D, Y)$ with one-sided noncompliance. Define
$\bP_{zudy1} = \bP(Z=z,U=u,D=d,Y=y,R^D=1)$, $ \bP_{zy+0} = \bP(Z=z,Y=y,R^D=0)$,
    $\zeta_{y}(u) =  \frac{\bP(R^D=0\mid U=u,Y=y)}{\bP(R^D=1\mid U=u,Y=y)}$. 
Since
$$\bP_{zudy1} = \bP(Z=z,U=u,D=d,Y=y)\bP(R^D=1\mid U=u,Y=y),$$
we have
\begin{align}
\bP_{1y+0} &= \bP_{1n0y1}\zeta_{y}(n)+\bP_{1c1y1}\zeta_{y}(c), \nonumber\\
\bP_{0y+0} &= \bP_{0n0y1}\zeta_{y}(n)+\bP_{0c0y1}\zeta_{y}(c), \nonumber
\end{align}
for $y=0,1$. The uniqueness of solutions $\zeta_{y}(u)$ requires that $\bP(Y=1\mid U=c,D=0) \neq \bP(Y=1\mid U=c,D=1)$. 

We can identify $\bP(R^D=1\mid U=u,Y=y)$ once $\zeta_{y}(u)$ is identified. Then, if $\bP(R^D=1\mid U=u,Y=y)>0$ for $u=n,c$ and $y=0,1$, the identification of $\bP(Z=z,U=u,D=d,Y=y)$ follows from 
\begin{align}
\bP(Z=z,U=u,D=d,Y=y)=\frac{\bP_{zudy1}}{\bP(R^D=1\mid U=u,Y=y)}. \nonumber
\end{align}

\subsubsection{Assumption 2\textit{DY}}\label{proof2DY}
We focus on the identification of $\bP(D=d,Y=y\mid Z=z)$ to identify the CACE. Define 
    $\bP_{dy1\mid z} = \bP(D=d,Y=y,R^D=1\mid Z=z)$,
    $\bP_{y+0\mid z} = \bP(Y=y,R^D=0\mid Z=z)$,
    $\zeta_{y}(d) =  \frac{\bP(R^D=0\mid D=d,Y=y)}{\bP(R^D=1\mid D=d,Y=y)}$.
Since
$$\bP_{dy1\mid z} = \bP(D=d,Y=y\mid Z=z)\bP(R^D=1\mid D=d,Y=y),$$
we have
$$\bP_{y+0\mid z} = \int_{d\in \mathcal{D}} \bP(D=d,Y=y,R^D=0\mid Z=z)\textup{d}d =\int_{d\in \mathcal{D}}\bP_{dy1\mid z}\zeta_{y}(d)\textup{d}d
$$
for each $z\in\mathcal{Z}$. The uniqueness of solutions $\zeta_{y}(d)$ requires that $\bP(D,Y=y,R^D=1\mid Z)$ is complete in $Z$ for all $y$. For binary $Z$ and $D$, we can identify $\zeta_{y}(d)$ directly under one-sided noncompliance, and the uniqueness of solutions $\zeta_{y}(d)$ requires $D \notindependent Z \mid Y = y$ for all $y$ under two-sided noncompliance.

We can identify $\bP(R^D=1\mid D=d,Y=y)$ once $\zeta_{y}(d)$ is identified. Then, the identification of $\bP(D=d,Y=y\mid Z=z)$ follows from 
\begin{align}
\bP(D=d,Y=y\mid Z=z) &= \frac{\bP_{dy1\mid z}}{\bP(R^D=1\mid D=d,Y=y)}.\nonumber
\end{align} 

\subsubsection{Assumption 2\textit{UD}}\label{proof2DU}
We focus on the identification of $\bP(Y=y\mid U=c,D=d)$ to identify the CACE. Note that
\begin{eqnarray*}
&&\bP(Y=y\mid U=c,D=0)\\&=&\frac{\bP(U=c,D=0,Y=y,R^D=1\mid Z=0)}{\bP(U=c,D=0,R^D=1 \mid Z=0)}\\&=&\frac{\bP(D=0,Y=y,R^D=1\mid Z=0)-\bP(U=n,D=0,Y=y,R^D=1\mid Z=0)}{\bP(D=0,R^D=1\mid Z=0)-\bP(U=n,D=0,R^D=1\mid Z=0)}.\nonumber
\end{eqnarray*}
Since
$$\bP(U=n,D=0,Y=y,R^D=1\mid Z=0)=\bP(U=n,D=0,Y=y,R^D=1\mid Z=1),
$$
we can identify $\bP(Y=y \mid U=c, D=0)$. Similarly, we can identify $\bP(Y=y \mid U=c, D=1)$, and therefore, the CACE.

In addition, we can identify $\bP(Z, U, D, Y)$ under a full rank condition and additional positivity conditions.
Define $\bP_{zudy1} = \bP(Z=z,U=u,D=d,Y=y,R^D=1)$,
    $\bP_{zy+0} = \bP(Z=z,Y=y,R^D=0)$,
    $\zeta(u,d) = \frac{\bP(R^D=0\mid U=u,D=d)}{\bP(R^D=1\mid U=u,D=d)}$.
Since
$$\bP_{zudy1} = \bP(Z=z,U=u,D=d,Y=y)\bP(R^D=1\mid U=u,D=d),$$
we have
\begin{align*}
\bP_{0y+0} &= \bP_{0n0y1}\zeta(n,0) + \bP_{0a1y1}\zeta(a,1) + \bP_{0c0y1}\zeta(c,0),\\
\bP_{1y+0} &= \bP_{1n0y1}\zeta(n,0) + \bP_{1a1y1}\zeta(a,1) + \bP_{1c1y1}\zeta(c,1),
\end{align*}
for each $y\in\mathcal{Y}$. The uniqueness of solutions $\zeta(u,d)$ requires that the above system of linear equations have full rank. 

We can identify $\bP(R^D=1\mid U=u,D=d)$ once $\zeta(u,d)$ is identified. Then, if $\bP(R^D=1\mid U=u,D=d)>0$ for $(u,d)=(a,1),(n,0),(c,1),(c,0)$, the identification of $\bP(Z=z,U=u,D=d,Y=y)$ follows from 
\begin{align}
\bP(Z=z,U=u,D=d,Y=y)=\frac{\bP_{zudy1}}{\bP(R^D=1\mid U=u,D=d)}. \nonumber
\end{align}

\subsubsection{Assumption 2\textit{ZD}}\label{proof2ZD}
We focus on the identification of $\bP(D=d,Y=y\mid Z=z)$ to identify the CACE. Define 
    $\bP_{dy1\mid z} = \bP(D=d,Y=y,R^D=1\mid Z=z)$,
    $\bP_{y+0\mid z} = \bP(Y=y,R^D=0\mid Z=z)$,
    $\zeta_{z}(d) =  \frac{\bP(R^D=0\mid Z=z,D=d)}{\bP(R^D=1\mid Z=z,D=d)}$.
Since
$$\bP_{dy1\mid z} = \bP(D=d,Y=y\mid Z=z)\bP(R^D=1\mid Z=z,D=d),$$
we have
$$\bP_{y+0\mid z} = \int_{d\in \mathcal{D}} \bP(D=d,Y=y,R^D=0\mid Z=z)\textup{d}d =\int_{d\in \mathcal{D}}\bP_{dy1\mid z}\zeta_{z}(d)\textup{d}d
$$
for each $y\in\mathcal{Y}$. The uniqueness of solutions $\zeta_{z}(d)$ requires that $\bP(D,Y,R^D=1\mid Z=z)$ is complete in $Y$ for all $z$. For binary $Z$ and $D$, the uniqueness of solutions $\zeta_{z}(d)$ requires $Y \notindependent D \mid Z = 1$ for one-sided noncompliance, and $Y \notindependent D \mid Z = z$ for $z=0,1$ under two-sided noncompliance.

We can identify $\bP(R^D=1\mid Z=z,D=d)$ once $\zeta_{z}(d)$ is identified. Then, the identification of $\bP(D=d,Y=y\mid Z=z)$ follows from 
\begin{align}
\bP(D=d,Y=y\mid Z=z) &= \frac{\bP_{dy1\mid z}}{\bP(R^D=1\mid Z=z,D=d)}.\nonumber
\end{align}

\subsubsection{Assumption 2\textit{ZU}}\label{proof2ZU}
The identification of $\bP(Y=y\mid U=n,D=0)$ and $\bP(Y=y\mid U=c,D=1)$ follows from 
\begin{align*}
\bP(Y=y\mid U=n,D=0)&=\bP(Y=y\mid Z=1,U=n,D=0,R^D=1),\\
\bP(Y=y\mid U=c,D=1)&=\bP(Y=y\mid Z=1,U=c,D=1,R^D=1).
\end{align*}
Since
\begin{align*}
\bP(Y=y\mid Z=1)&=\bP(U=n,D=0,Y=y\mid Z=1)+\bP(U=c,D=1,Y=y\mid Z=1)\\&=\bP(Y=y\mid U=n,D=0)\bP(U=n,D=0\mid Z=1)\\&~~+\bP(Y=y\mid U=c,D=1)\{1-\bP(U=n,D=0\mid Z=1)\},
\end{align*}
we can identify $\bP(U=n,D=0\mid Z=1)$. 
Since
\begin{align*}
\bP(Y=y\mid Z=0)&=\bP(U=n,D=0,Y=y\mid Z=0)+\bP(U=c,D=0,Y=y\mid Z=0)\\&=\bP(Y=y\mid U=n,D=0)\bP(U=n,D=0\mid Z=1)\\&~~+\bP(Y=y\mid U=c,D=0)\{1-\bP(U=n,D=0\mid Z=1)\},
\end{align*}
we can identify $\bP(Y=y\mid U=c, D=0)$. In addition, we can identify $\bP(Z=z, U = u, D = d, Y = y)$.

\subsection{Proof of Theorem \ref{the3}}
\subsubsection{Assumption 1\textit{ZD}$\oplus$2\textit{UD}}
Note that
\begin{eqnarray*}
&&\bP(Y=y\mid U=c,D=0)\\&=&\frac{\bP(U=c,D=0,Y=y,R^D=1,R^Y=1\mid Z=0)}{\bP(U=c,D=0,R^D=1,R^Y=1\mid Z=0)}\\&=&\frac{\bP(D=0,Y=y,R^D=1,R^Y=1\mid Z=0)-\bP(U=n,D=0,Y=y,R^D=1,R^Y=1\mid Z=0)}{\bP(D=0,R^D=1,R^Y=1\mid Z=0)-\bP(U=n,D=0,R^D=1,R^Y=1\mid Z=0)}.\nonumber
\end{eqnarray*}
Since
\begin{eqnarray*}
&&\bP(U=n,D=0,Y=y,R^D=1,R^Y=1\mid Z=0)\\&=&\bP(U=n,D=0,Y=y,R^D=1,R^Y=1\mid Z=1)\frac{\bP(R^Y=1\mid Z=0,D=0,R^D=1)}{\bP(R^Y=1\mid Z=1,D=0,R^D=1)},
\end{eqnarray*}
we can identify $\bP(Y=y\mid U=c,D=0)$, and similarly, $\bP(Y=y\mid U=c, D=1)$.

\subsubsection{Assumption 1\textit{UD}$\oplus$2\textit{UD}}
Note that
\begin{eqnarray*}
&&\bP(Y=y\mid U=c,D=0)\\&=&\frac{\bP(U=c,D=0,Y=y,R^D=1,R^Y=1\mid Z=0)}{\bP(U=c,D=0,R^D=1,R^Y=1\mid Z=0)}\\&=&\frac{\bP(D=0,Y=y,R^D=1,R^Y=1\mid Z=0)-\bP(U=n,D=0,Y=y,R^D=1,R^Y=1\mid Z=0)}{\bP(D=0,R^D=1,R^Y=1\mid Z=0)-\bP(U=n,D=0,R^D=1,R^Y=1\mid Z=0)}.\nonumber
\end{eqnarray*}
Since
\begin{eqnarray*}
\bP(U=n,D=0,Y=y,R^D=1,R^Y=1\mid Z=0)=\bP(U=n,D=0,Y=y,R^D=1,R^Y=1\mid Z=1),
\end{eqnarray*}
we can identify $\bP(Y=y\mid U=c,D=0)$, and similarly, $\bP(Y=y\mid U=c, D=1)$.

\subsubsection{Assumption 1\textit{DY}$\oplus$2\textit{UD}}
Note that
\begin{eqnarray*}
&&\bP(Y=y,R^Y=1\mid U=c,D=0,R^D=1)\\&=&\frac{\bP(U=c,D=0,Y=y,R^D=1,R^Y=1\mid Z=0)}{\bP(U=c,D=0,R^D=1\mid Z=0)}\\&=&\frac{\bP(D=0,Y=y,R^D=1,R^Y=1\mid Z=0)-\bP(U=n,D=0,Y=y,R^D=1,R^Y=1\mid Z=0)}{\bP(D=0,R^D=1\mid Z=0)-\bP(U=n,D=0,R^D=1\mid Z=0)}.\nonumber
\end{eqnarray*}
Since
\begin{align*}
\bP(U=n,D=0,Y=y,R^D=1,R^Y=1\mid Z=0)&=\bP(U=n,D=0,Y=y,R^D=1,R^Y=1\mid Z=1),\\
\bP(U=n,D=0,R^D=1\mid Z=0)&=\bP(U=n,D=0,R^D=1\mid Z=1),
\end{align*}
we can identify $\bP(Y=y,R^Y=1\mid U=c,D=0,R^D=1)$, and similarly, $\bP(Y=y,R^Y=1\mid U=c,D=1,R^D=1)$. The identification of $\bP(R^Y=1\mid D=d,Y=y,R^D=1)$ follows the same logic as in Assumption 1\textit{DY}$\oplus$2\textit{Z}, as detailed in \ref{proof1DY2Z}. The identification of $\bP(Y=y\mid U=c,D=d)$ follows from 
\begin{align*}
\bP(Y=y\mid U=c,D=d)&=\frac{\bP(Y=y,R^Y=1\mid U=c,D=d,R^D=1)}{\bP(R^Y=1\mid D=d,Y=y,R^D=1)}.
\end{align*}

\subsubsection{Assumption 1\textit{ZY}$\oplus$2\textit{UD}} 
Note that
\begin{eqnarray*}
&&\bP(Y=y,R^Y=1\mid Z=0,U=c,D=0,R^D=1)\\&=&\frac{\bP(U=c,D=0,Y=y,R^D=1,R^Y=1\mid Z=0)}{\bP(U=c,D=0,R^D=1\mid Z=0)}\\&=&\frac{\bP(D=0,Y=y,R^D=1,R^Y=1\mid Z=0)-\bP(U=n,D=0,Y=y,R^D=1,R^Y=1\mid Z=0)}{\bP(D=0,R^D=1\mid Z=0)-\bP(U=n,D=0,R^D=1\mid Z=0)}.\nonumber
\end{eqnarray*}
Since
\begin{eqnarray*}
&&\bP(U=n,D=0,Y=y,R^D=1,R^Y=1\mid Z=0)\\&=&\bP(U=n,D=0,Y=y,R^D=1,R^Y=1\mid Z=1)\frac{\bP(R^Y=1\mid Z=0,Y=y,R^D=1)}{\bP(R^Y=1\mid Z=1,Y=y,R^D=1)},\\
&&\bP(U=n,D=0,R^D=1\mid Z=0)=\bP(U=n,D=0,R^D=1\mid Z=1),
\end{eqnarray*}
we can identify $\bP(Y=y,R^Y=1\mid Z=0,U=c,D=0,R^D=1)$, and similarly, $\bP(Y=y,R^Y=1\mid Z=1,U=c,D=1,R^D=1)$. The identification of $\bP(R^Y=1\mid Z=z,Y=y,R^D=1)$ follows the same logic as in Assumption 1\textit{ZY}$\oplus$2\textit{Z}, as detailed in \ref{proof1ZY2Z}, with a slightly modified conditional dependence condition, $Y \notindependent D \mid (Z=z,R^D=1)$ for $z=0,1$, where $Z$ and $D$ are binary. The identification of $\bP(Y=y\mid U=c,D=d)$ follows from 
\begin{align*}
\bP(Y=y\mid U=c,D=d)&=\frac{\bP(Y=y,R^Y=1\mid Z=z,U=c,D=d,R^D=1)}{\bP(R^Y=1\mid Z=z,Y=y,R^D=1)}.
\end{align*}

\subsubsection{Assumption 1\textit{UY}$\oplus$2\textit{UD}}
Note that
\begin{eqnarray*}
&&\bP(U=c,D=0,Y=y,R^D=1,R^Y=1\mid Z=0)\\&=& \bP(D=0,Y=y,R^D=1,R^Y=1\mid Z=0)-\bP(U=n,D=0,Y=y,R^D=1,R^Y=1\mid Z=0),\\
&&\bP(U=c,D=0,R^D=1\mid Z=0)\\&=& \bP(D=0,R^D=1\mid Z=0)-\bP(U=n,D=0,R^D=1\mid Z=0).
\end{eqnarray*}
Since
\begin{align*}
\bP(U=n,D=0,Y=y,R^D=1,R^Y=1\mid Z=0)&=\bP(U=n,D=0,Y=y,R^D=1,R^Y=1\mid Z=1),\\
\bP(U=n,D=0,R^D=1\mid Z=0)&=\bP(U=n,D=0,R^D=1\mid Z=1),
\end{align*}
we can identify $\bP(U=c,D=0,Y=y,R^D=1,R^Y=1\mid Z=0)$ and $\bP(U=c,D=0,R^D=1\mid Z=0)$, and similarly, $\bP(U=c,D=1,Y=y,R^D=1,R^Y=1\mid Z=1)$ and $\bP(U=c,D=1,R^D=1\mid Z=1)$. Since
\begin{align*}
\frac{\bP(Y=1\mid U=c,D=0)}{\bP(Y=1\mid U=c,D=1)}&=\frac{\bP(U=c,D=0,Y=1,R^D=1,R^Y=1\mid Z=0)}{\bP(U=c,D=1,Y=1,R^D=1,R^Y=1\mid Z=1)}\\&~~\cdot\frac{\bP(U=c,D=1,R^D=1\mid Z=1)}{\bP(U=c,D=0,R^D=1\mid Z=0)},\\
\frac{1-\bP(Y=1\mid U=c,D=0)}{1-\bP(Y=1\mid U=c,D=1)}&=\frac{\bP(U=c,D=0,Y=0,R^D=1,R^Y=1\mid Z=0)}{\bP(U=c,D=1,Y=0,R^D=1,R^Y=1\mid Z=1)}\\&~~\cdot\frac{\bP(U=c,D=1,R^D=1\mid Z=1)}{\bP(U=c,D=0,R^D=1\mid Z=0)},
\end{align*}
we can identify the CACE.

\subsection{Proof of Theorem \ref{the4}}
\subsubsection{Assumption 1\textit{ZD}+2\textit{ZD}}\label{proof1ZD+2ZD}
Define 
    $\bP_{dy11\mid z} = \bP(D=d,Y=y,R^D=1,R^Y=1\mid Z=z)$,
    $\bP_{y+01\mid z} = \bP(Y=y,R^D=0,R^Y=1\mid Z=z)$,
    $\bP_{d1+0\mid z} = \bP(D=d,R^D=1,R^Y=0\mid Z=z)$,
    $\bP_{+0+0\mid z} = \bP(R^D=0,R^Y=0\mid Z=z)$,
    $\zeta_{z}(d) =  \frac{\bP(R^D=0\mid Z=z,D=d)}{\bP(R^D=1\mid Z=z,D=d)}$.
Since
\begin{align*}
\bP_{dy11\mid z}&=\bP(D=d,Y=y,R^Y=1\mid Z=z)\bP(R^D=1\mid Z=z,D=d),\\
\bP_{d1+0\mid z}&=\bP(D=d,R^Y=0\mid Z=z)\bP(R^D=1\mid Z=z,D=d),
\end{align*}
we have
$$\bP_{y+01\mid z} = \int_{d\in \mathcal{D}} \bP(D=d,Y=y,R^D=0,R^Y=1\mid Z=z)\textup{d}d=\int_{d\in \mathcal{D}}\bP_{dy11\mid z} \zeta_{z}(d)\textup{d}d$$
for each $y\in\mathcal{Y}$, and 
$$
\bP_{+0+0\mid z} = \int_{d\in \mathcal{D}} \bP(D=d,R^D=0,R^Y=0\mid Z=z)\textup{d}d =\int_{d\in \mathcal{D}}\bP_{d1+0\mid z} \zeta_{z}(d)\textup{d}d. $$
The uniqueness of solutions $\zeta_{z}(d)$ requires that $\bP(D,Y^\dagger,R^D=1\mid Z=z)$ is complete in $Y^\dagger$ for all $z$. For binary $Z$ and $D$, the uniqueness of solutions $\zeta_{z}(d)$ requires $Y^\dagger \notindependent D \mid Z = 1$ under one-sided noncompliance, and $Y^\dagger \notindependent D \mid Z = z$ for $z=0,1$ under two-sided noncompliance.

We can identify $\bP(R^D=1\mid Z=z,D=d)$ once $\zeta_{z}(d)$ is identified. The identification of $\bP(R^Y=1\mid Z=z,D=d)$ follows from
\begin{align}
\bP(R^Y=1\mid Z=z,D=d) = \bP(R^Y=1\mid Z=z,D=d,R^D=1)\nonumber.
\end{align} 
The identification of $\bP(D=d,Y=y\mid Z=z)$ follows from 
\begin{eqnarray}
\bP(D=d,Y=y\mid Z=z)= \frac{\bP_{dy11\mid z}}{\bP(R^D=1\mid Z=z,D=d)\bP(R^Y=1\mid Z=z,D=d)}.\nonumber
\end{eqnarray}

\subsubsection{Assumption 1\textit{UD}+2\textit{ZD}}
The identification of $\bP(R^D=1 \mid Z=z, D=d)$ follows the same logic as in Assumption 1\textit{ZD}+2\textit{ZD}, as detailed in \ref{proof1ZD+2ZD}. Note that
\begin{eqnarray*}
&&\bP(Y=y\mid U=c,D=0)\\&=&\frac{\bP(U=c,D=0,Y=y,R^D=1,R^Y=1\mid Z=0)}{\bP(U=c,D=0,R^D=1,R^Y=1\mid Z=0)}\\&=&\frac{\bP(D=0,Y=y,R^D=1,R^Y=1\mid Z=0)-\bP(U=n,D=0,Y=y,R^D=1,R^Y=1\mid Z=0)}{\bP(D=0,R^D=1,R^Y=1\mid Z=0)-\bP(U=n,D=0,R^D=1,R^Y=1\mid Z=0)}.\nonumber
\end{eqnarray*}
Since
\begin{eqnarray*}
&&\bP(U=n,D=0,Y=y,R^D=1,R^Y=1\mid Z=0)\\&=&\bP(U=n,D=0,Y=y,R^D=1,R^Y=1\mid Z=1)\frac{\bP(R^D=1\mid Z=0,D=0)}{\bP(R^D=1\mid Z=1,D=0)},
\end{eqnarray*}
we can identify $\bP(Y=y\mid U=c, D=0)$, and similarly, $\bP(Y=y\mid U=c, D=1)$. 

\subsubsection{Assumption 1\textit{DY}+2\textit{ZD}}
The identification of $\bP(R^D=1\mid Z=z,D=d)$ follows the same logic as in Assumption 1\textit{ZD}+2\textit{ZD}, as detailed in \ref{proof1ZD+2ZD}. Define 
    $\bP_{dy11\mid z} = \bP(D=d,Y=y,R^D=1,R^Y=1\mid Z=z)$,
    $\bP_{d1+0\mid z} = \bP(D=d,R^D=1,R^Y=0\mid Z=z)$,
    $\eta_{d}(y) =  \frac{\bP(R^Y=0\mid D=d,Y=y)}{\bP(R^Y=1\mid D=d,Y=y)}$.
Since
\begin{eqnarray*}
\bP_{dy11\mid z}=\bP(D=d,Y=y,R^D=1\mid Z=z)\bP(R^Y=1\mid D=d,Y=y),
\end{eqnarray*}
we have
$$\bP_{d1+0\mid z} = \int_{y\in \mathcal{Y}} \bP(D=d,Y=y,R^D=1,R^Y=0\mid Z=z)\textup{d}y =\int_{y\in \mathcal{Y}}\bP_{dy11\mid z}\eta_{d}(y)\textup{d}y
$$
for each $z\in\mathcal{Z}$. The uniqueness of solutions $\eta_{d}(y)$ requires that $\bP(D=d,Y,R^D=1,R^Y=1\mid Z)$ is complete in $Z$ for all $d$. For binary $Z$ and $D$, the uniqueness of solutions $\eta_{d}(y)$ requires that $Y$ is binary and $Y \notindependent Z \mid D = d$ for $d = 0,1$ under two-sided noncompliance.

We can identify $\bP(R^Y=1\mid D=d,Y=y)$ once $\eta_{d}(y)$ is identified. The identification of $\bP(D=d,Y=y\mid Z=z)$ follows from 
\begin{eqnarray}
\bP(D=d,Y=y\mid Z=z)= \frac{\bP_{dy11\mid z}}{\bP(R^D=1\mid Z=z,D=d)\bP(R^Y=1\mid D=d,Y=y)}.\nonumber
\end{eqnarray}

\subsubsection{Assumption 1\textit{ZY}+2\textit{ZD}}
Define 
    $\bP_{dy11\mid z} = \bP(D=d,Y=y,R^D=1,R^Y=1\mid Z=z)$,
    $\bP_{y+01\mid z} = \bP(Y=y,R^D=0,R^Y=1\mid Z=z)$,
    $\bP_{d1+0\mid z} = \bP(D=d,R^D=1,R^Y=0\mid Z=z)$,
    $\zeta_{z}(d) =  \frac{\bP(R^D=0\mid Z=z,D=d)}{\bP(R^D=1\mid Z=z,D=d)}$,
    $\eta_{z}(y) =  \frac{\bP(R^Y=0\mid Z=z,Y=y)}{\bP(R^Y=1\mid Z=z,Y=y)}$.
Since
\begin{eqnarray*}
\bP_{dy11\mid z}=\bP(D=d,Y=y\mid Z=z)\bP(R^Y=1\mid Z=z,Y=y)\bP(R^D=1\mid Z=z,D=d),
\end{eqnarray*}
we have
$$\bP_{y+01\mid z} = \int_{d\in \mathcal{D}} \bP(D=d,Y=y,R^D=0,R^Y=1\mid Z=z)\textup{d}d =\int_{d\in \mathcal{D}}\bP_{dy11\mid z}\zeta_{z}(d)\textup{d}d
$$
for each $y\in\mathcal{Y}$, and
$$
\bP_{d1+0\mid z} = \int_{y\in \mathcal{Y}} \bP(D=d,Y=y,R^D=1,R^Y=0\mid Z=z)\textup{d}y=\int_{y\in \mathcal{Y}}\bP_{dy11\mid z}\eta_{z}(y)\textup{d}y
$$
for each $d\in\mathcal{D}$. The uniqueness of solutions $\zeta_{z}(d)$ requires that $\bP(D,Y,R^D=1,R^Y=1\mid Z=z)$ is complete in $Y$ for all $z$, and the uniqueness of solutions $\eta_{z}(y)$ requires that $\bP(D,Y,R^D=1,R^Y=1\mid Z=z)$ is complete in $D$ for all $z$. For binary $Z$ and $D$, the uniqueness of solutions $\zeta_{z}(d)$ and $\eta_{z}(y)$ requires that $Y$ is binary and $Y \notindependent D \mid Z = z$ for $z = 0,1$ under two-sided noncompliance.

We can identify $\bP(R^D=1\mid Z=z,D=d)$ and $\bP(R^Y=1\mid Z=z,Y=y)$ once $\zeta_{z}(d)$ and $\eta_{z}(y)$ are identified. The identification of $\bP(D=d,Y=y\mid Z=z)$ follows from 
\begin{eqnarray}
\bP(D=d,Y=y\mid Z=z)=\frac{\bP_{dy11\mid z}}{\bP(R^D=1\mid Z=z,D=d)\bP(R^Y=1\mid Z=z,Y=y)}.\nonumber
\end{eqnarray}

\subsubsection{Assumption 1\textit{UY}+2\textit{ZD}}
The identification of $\bP(R^D=1\mid Z=z,D=d)$ follows the same logic as in Assumption 1\textit{ZD}+2\textit{ZD}, as detailed in \ref{proof1ZD+2ZD}. Note that
\begin{eqnarray*}
&&\bP(U=c,D=0,Y=y,R^D=1,R^Y=1\mid Z=0)\\&=& \bP(D=0,Y=y,R^D=1,R^Y=1\mid Z=0)\\&&-~\bP(U=n,D=0,Y=y,R^D=1,R^Y=1\mid Z=0).\nonumber
\end{eqnarray*}
Since
\begin{eqnarray*}
&&\bP(U=n,D=0,Y=y,R^D=1,R^Y=1\mid Z=0)\\&=&\bP(U=n,D=0,Y=y,R^D=1,R^Y=1\mid Z=1)\frac{\bP(R^D=1\mid Z=0,D=0)}{\bP(R^D=1\mid Z=1,D=0)},
\end{eqnarray*}
we can identify $\bP(U=c,D=0,Y=y,R^D=1,R^Y=1\mid Z=0)$, and similarly, $\bP(U=c,D=1,Y=y,R^D=1,R^Y=1\mid Z=1)$. Since
\begin{align*}
\frac{\bP(Y=1\mid U=c,D=0)}{\bP(Y=1\mid U=c,D=1)}&=\frac{\bP(U=c,D=0,Y=1,R^D=1,R^Y=1\mid Z=0)}{\bP(U=c,D=1,Y=1,R^D=1,R^Y=1\mid Z=1)}\\&~~\cdot\frac{\bP(R^D=1\mid Z=1,D=1)}{\bP(R^D=1\mid Z=0,D=0)},\\
\frac{1-\bP(Y=1\mid U=c,D=0)}{1-\bP(Y=1\mid U=c,D=1)}&=\frac{\bP(U=c,D=0,Y=0,R^D=1,R^Y=1\mid Z=0)}{\bP(U=c,D=1,Y=0,R^D=1,R^Y=1\mid Z=1)}\\&~~\cdot\frac{\bP(R^D=1\mid Z=1,D=1)}{\bP(R^D=1\mid Z=0,D=0)},
\end{align*}
we can identify the CACE.

\subsection{Proof of Theorem \ref{the5}}
\subsubsection{Assumption 1\textit{ZD}+2\textit{ZU}}
The identification of $\bP(Y=y\mid U=n,D=0)$ and $\bP(Y=y\mid U=c,D=1)$ follows from 
\begin{align*}
\bP(Y=y\mid U=n,D=0)&=\bP(Y=y\mid Z=1,U=n,D=0,R^D=1,R^Y=1),\\
\bP(Y=y\mid U=c,D=1)&=\bP(Y=y\mid Z=1,U=c,D=1,R^D=1,R^Y=1).
\end{align*} 
The identification of $\bP(R^Y=1\mid Z=z,D=d)$ follows from 
\begin{align*}
\bP(R^Y=1\mid Z=z,D=d)&=\bP(R^Y=1\mid Z=z,D=d,R^D=1).
\end{align*} 
Since
\begin{eqnarray*}
&&\bP(Y=y,R^Y=1\mid Z=1)\\&=&\bP(U=n,D=0,Y=y,R^Y=1\mid Z=1)+\bP(U=c,D=1,Y=y,R^Y=1\mid Z=1)\\&=&\bP(Y=y\mid U=n,D=0)\bP(U=n,D=0\mid Z=1)\bP(R^Y=1\mid Z=1,D=0)\\&&+\bP(Y=y\mid U=c,D=1)\{1-\bP(U=n,D=0\mid Z=1)\}\bP(R^Y=1\mid Z=1,D=1),
\end{eqnarray*}
we can identify $\bP(U=n,D=0\mid Z=1)$.
Since
\begin{eqnarray*}
&&\bP(Y=y,R^Y=1\mid Z=0)\\&=&\bP(U=n,D=0,Y=y,R^Y=1\mid Z=0)+\bP(U=c,D=0,Y=y,R^Y=1\mid Z=0)\\&=&\bP(Y=y\mid U=n,D=0)\bP(U=n,D=0\mid Z=1)\bP(R^Y=1\mid Z=0,D=0)\\&&+\bP(Y=y\mid U=c,D=0)\{1-\bP(U=n,D=0\mid Z=1)\}\bP(R^Y=1\mid Z=0,D=0),
\end{eqnarray*}
we can identify $\bP(Y=y\mid U=c, D=0)$. 

\subsubsection{Assumption 1\textit{UD}+2\textit{ZU}}\label{proof1DU+2ZU}
The identification of $\bP(Y=y,R^Y=1\mid U=n,D=0)$ and $\bP(Y=y,R^Y=1\mid U=c,D=1)$ follows from 
\begin{align*}
\bP(Y=y,R^Y=1\mid U=n,D=0)&=\bP(Y=y,R^Y=1\mid Z=1,U=n,D=0,R^D=1),\\
\bP(Y=y,R^Y=1\mid U=c,D=1)&=\bP(Y=y,R^Y=1\mid Z=1,U=c,D=1,R^D=1).
\end{align*} 
Since
\begin{eqnarray*}
&&\bP(Y=y,R^Y=1\mid Z=1)\\&=&\bP(U=n,D=0,Y=y,R^Y=1\mid Z=1)+\bP(U=c,D=1,Y=y,R^Y=1\mid Z=1)\\&=&\bP(Y=y,R^Y=1\mid U=n,D=0)\bP(U=n,D=0\mid Z=1)\\&&+\bP(Y=y,R^Y=1\mid U=c,D=1)\{1-\bP(U=n,D=0\mid Z=1)\},
\end{eqnarray*}
we can identify $\bP(U=n,D=0\mid Z=1)$. 
Since
\begin{eqnarray*}
&&\bP(Y=y,R^Y=1\mid Z=0)\\&=&\bP(U=n,D=0,Y=y,R^Y=1\mid Z=0)+\bP(U=c,D=0,Y=y,R^Y=1\mid Z=0)\\&=&\bP(Y=y,R^Y=1\mid U=n,D=0)\bP(U=n,D=0\mid Z=1)\\&&+\bP(Y=y,R^Y=1\mid U=c,D=0)\{1-\bP(U=n,D=0\mid Z=1)\},
\end{eqnarray*}
we can identify $\bP(Y=y,R^Y=1\mid U=c, D=0)$. The identification of $\bP(Y=y\mid U=c,D=d)$ follows from 
\begin{eqnarray*}
\bP(Y=y\mid U=c,D=d)=\frac{\bP(Y=y,R^Y=1\mid U=c, D=d)}{\bP(R^Y=1\mid U=c, D=d)}.
\end{eqnarray*}

\subsubsection{Assumption 1\textit{UY}+2\textit{ZU}}
The identification of $\bP(Y=y, R^Y=1 \mid U=c, D=d)$ follows the same logic as in Assumption 1\textit{UD}+2\textit{ZU}, as detailed in \ref{proof1DU+2ZU}. Since 
\begin{eqnarray*}
\frac{\bP(Y=1\mid U=c, D=0)}{\bP(Y=1\mid U=c, D=1)}=\frac{\bP(Y=1,R^Y=1\mid U=c,D=0)}{\bP(Y=1,R^Y=1\mid U=c, D=1)},\\
\frac{1-\bP(Y=1\mid U=c, D=0)}{1-\bP(Y=1\mid U=c, D=1)}=\frac{\bP(Y=0,R^Y=1\mid U=c,D=0)}{\bP(Y=0,R^Y=1\mid U=c,D=1)},
\end{eqnarray*}
we can identify the CACE.

\subsection{Proof of Theorem \ref{the6}}
\subsubsection{Assumption 1\textit{ZD}$\oplus$2\textit{Z}}
The identification of $\bP(D=d,Y=y\mid Z=z)$ follows from
\begin{eqnarray}
\bP(D=d,Y=y\mid Z=z)=\frac{\bP(D=d,Y=y,R^D=1,R^Y=1\mid Z=z)}{\bP(R^D=1\mid Z=z)\bP(R^Y=1\mid Z=z,D=d,R^D=1)}.\nonumber
\end{eqnarray}

\subsubsection{Assumption 1\textit{UD}$\oplus$2\textit{Z}}
Note that
\begin{eqnarray*}
&&\bP(Y=y\mid U=c,D=0)\\&=&\frac{\bP(U=c,D=0,Y=y,R^D=1,R^Y=1\mid Z=0)}{\bP(U=c,D=0,R^D=1,R^Y=1\mid Z=0)}\\&=&\frac{\bP(D=0,Y=y,R^D=1,R^Y=1\mid Z=0)-\bP(U=n,D=0,Y=y,R^D=1,R^Y=1\mid Z=0)}{\bP(D=0,R^D=1,R^Y=1\mid Z=0)-\bP(U=n,D=0,R^D=1,R^Y=1\mid Z=0)}.\nonumber
\end{eqnarray*}
Since
\begin{eqnarray*}
&&\bP(U=n,D=0,Y=y,R^D=1,R^Y=1\mid Z=0)\\&=&\bP(U=n,D=0,Y=y,R^D=1,R^Y=1\mid Z=1)\frac{\bP(R^D=1\mid Z=0)}{\bP(R^D=1\mid Z=1)},
\end{eqnarray*}
we can identify $\bP(Y=y\mid U=c, D=0)$, and similarly, $\bP(Y=y\mid U=c, D=1)$.

\subsubsection{Assumption 1\textit{UY}$\oplus$2\textit{Z}}
Note that
\begin{eqnarray*}
&&\bP(U=c,D=0,Y=y,R^D=1,R^Y=1\mid Z=0)\\&=& \bP(D=0,Y=y,R^D=1,R^Y=1\mid Z=0)-\bP(U=n,D=0,Y=y,R^D=1,R^Y=1\mid Z=0).\nonumber
\end{eqnarray*}
Since
\begin{eqnarray*}
&&\bP(U=n,D=0,Y=y,R^D=1,R^Y=1\mid Z=0)\\&=&\bP(U=n,D=0,Y=y,R^D=1,R^Y=1\mid Z=1)\frac{\bP(R^D=1\mid Z=0)}{\bP(R^D=1\mid Z=1)},
\end{eqnarray*}
we can identify $\bP(U=c,D=0,Y=y,R^D=1,R^Y=1\mid Z=0)$, and similarly, $\bP(U=c,D=1,Y=y,R^D=1,R^Y=1\mid Z=1)$. Since
\begin{align*}
\frac{\bP(Y=1\mid U=c,D=0)}{\bP(Y=1\mid U=c,D=1)}&=\frac{\bP(U=c,D=0,Y=1,R^D=1,R^Y=1\mid Z=0)}{\bP(U=c,D=1,Y=1,R^D=1,R^Y=1\mid Z=1)}\frac{\bP(R^D=1\mid Z=1)}{\bP(R^D=1\mid Z=0)},\\
\frac{1-\bP(Y=1\mid U=c,D=0)}{1-\bP(Y=1\mid U=c,D=1)}&=\frac{\bP(U=c,D=0,Y=0,R^D=1,R^Y=1\mid Z=0)}{\bP(U=c,D=1,Y=0,R^D=1,R^Y=1\mid Z=1)}\frac{\bP(R^D=1\mid Z=1)}{\bP(R^D=1\mid Z=0)},
\end{align*}
we can identify the CACE.

\subsubsection{Assumption 1\textit{DY}$\oplus$2\textit{Z}}\label{proof1DY2Z}
Define 
    $\bP_{dy11\mid z} = \bP(D=d,Y=y,R^D=1,R^Y=1\mid Z=z)$,
    $\bP_{d1+0\mid z} = \bP(D=d,R^D=1,R^Y=0\mid Z=z)$,
    $\eta_{d}(y) =  \frac{\bP(R^Y=0\mid D=d,Y=y,R^D=1)}{\bP(R^Y=1\mid D=d,Y=y,R^D=1)}$.
Since
\begin{eqnarray*}
\bP_{dy11\mid z}=\bP(D=d,Y=y,R^D=1\mid Z=z)\bP(R^Y=1\mid D=d,Y=y,R^D=1),
\end{eqnarray*}
we have
$$\bP_{d1+0\mid z} = \int_{y\in \mathcal{Y}} \bP(D=d,Y=y,R^D=1,R^Y=0\mid Z=z)\textup{d}y =\int_{y\in \mathcal{Y}}\bP_{dy11\mid z}\eta_{d}(y)\textup{d}y
$$
for each $z\in\mathcal{Z}$. The uniqueness of solutions $\eta_{d}(y)$ requires that $\bP(D=d,Y,R^D=1,R^Y=1\mid Z)$ is complete in $Z$ for all $d$. For binary $Z$ and $D$, the uniqueness of solutions $\eta_{d}(y)$ requires that $Y$ is binary and that $Y \notindependent Z \mid (D=d,R^D=1)$ for $d=0,1$ under two-sided noncompliance.

We can identify $\bP(R^Y=1\mid D=d,Y=y,R^D=1)$ once $\eta_{d}(y)$ is identified. The identification of $\bP(D=d,Y=y\mid Z=z)$ follows from 
\begin{eqnarray}
\bP(D=d,Y=y\mid Z=z)=\frac{\bP_{dy11\mid z}}{\bP(R^D=1\mid Z=z)\bP(R^Y=1\mid D=d,Y=y,R^D=1)}.\nonumber
\end{eqnarray}

\subsubsection{Assumption 1\textit{ZY}$\oplus$2\textit{Z}}\label{proof1ZY2Z}
Define 
    $\bP_{dy11\mid z} = \bP(D=d,Y=y,R^D=1,R^Y=1\mid Z=z)$,
    $\bP_{d1+0\mid z} = \bP(D=d,R^D=1,R^Y=0\mid Z=z)$,
    $\eta_{z}(y) =  \frac{\bP(R^Y=0\mid Z=z,Y=y,R^D=1)}{\bP(R^Y=1\mid Z=z,Y=y,R^D=1)}$.
Since
\begin{eqnarray*}
\bP_{dy11\mid z}=\bP(D=d,Y=y,R^D=1\mid Z=z)\bP(R^Y=1\mid Z=z,Y=y,R^D=1),
\end{eqnarray*}
we have
$$\bP_{d1+0\mid z} = \int_{y\in \mathcal{Y}} \bP(D=d,Y=y,R^D=1,R^Y=0\mid Z=z)\textup{d}y =\int_{y\in \mathcal{Y}}\bP_{dy11\mid z}\eta_{z}(y)\textup{d}y
$$
for each $d\in\mathcal{D}$. The uniqueness of solutions $\eta_{z}(y)$ requires that $\bP(D,Y,R^D=1,R^Y=1\mid Z=z)$ is complete in $D$ for all $z$. For binary $Z$ and $D$, the uniqueness of solutions $\eta_{z}(y)$ requires that $Y$ is binary and that $Y \notindependent D \mid Z=z$ for $z=0,1$ under two-sided noncompliance.

We can identify $\bP(R^Y=1\mid Z=z,Y=y,R^D=1)$ once $\eta_{z}(y)$ is identified. The identification of $\bP(D=d,Y=y\mid Z=z)$ follows from 
\begin{eqnarray}
\bP(D=d,Y=y\mid Z=z)=\frac{\bP_{dy11\mid z}}{\bP(R^D=1\mid Z=z)\bP(R^Y=1\mid Z=z,Y=y,R^D=1)}.\nonumber
\end{eqnarray}

\subsection{Proof of Theorem \ref{the7}}

\subsubsection{Assumption 1\textit{Z}$\oplus$2\textit{ZD}}
Define 
    $\bP_{dy11\mid z} = \bP(D=d,Y=y,R^D=1,R^Y=1\mid Z=z)$,
    $\bP_{y+01\mid z} = \bP(Y=y,R^D=0,R^Y=1\mid Z=z)$,
    $\zeta_{z}(d) = \frac{\bP(R^D=0\mid Z=z,D=d)}{\bP(R^D=1\mid Z=z,D=d)}$.
Since
\begin{eqnarray*}
\bP_{dy11\mid z}=\bP(D=d,Y=y\mid Z=z)\bP(R^D=1\mid Z=z,D=d)\bP(R^Y=1\mid Z=z,R^D=1),
\end{eqnarray*}
we have
\begin{align}
\bP_{y+01\mid z} &= \int_{d\in \mathcal{D}} \bP(D=d,Y=y,R^D=0,R^Y=1\mid Z=z)\textup{d}d \nonumber\\&=\int_{d\in \mathcal{D}}\bP_{dy11\mid z}\zeta_{z}(d)\frac{\bP(R^Y=1\mid Z=z,R^D=0)}{\bP(R^Y=1\mid Z=z,R^D=1)}\textup{d}d\nonumber
\end{align}
for each $y\in\mathcal{Y}$. The uniqueness of solutions $\zeta_{z}(d)$ requires that $\bP(D,Y,R^D=1,R^Y=1\mid Z=z)$ is complete in $Y$ for all $z$. For binary $Z$ and $D$, the uniqueness of solutions $\zeta_{z}(d)$ requires $Y \notindependent D \mid Z = 1$ under one-sided noncompliance, and $Y \notindependent D \mid Z = z$ for $z = 0,1$ under two-sided noncompliance.

We can identify $\bP(R^D=1\mid Z=z,D=d)$ once $\zeta_{z}(d)$ is identified. The identification of $\bP(D=d,Y=y\mid Z=z)$ follows from 
\begin{eqnarray}
\bP(D=d,Y=y\mid Z=z)=\frac{\bP_{dy11\mid z}}{\bP(R^D=1\mid Z=z,D=d)\bP(R^Y=1\mid Z=z,R^D=1)}.\nonumber
\end{eqnarray}

\subsubsection{Assumption 1\textit{D}$\oplus$2\textit{ZD}}
Define 
    $\bP_{dy11\mid z} = \bP(D=d,Y=y,R^D=1,R^Y=1\mid Z=z)$,
    $\bP_{y+01\mid z} = \bP(Y=y,R^D=0,R^Y=1\mid Z=z)$,
    $\zeta_{z}(d) = \frac{\bP(R^D=0\mid Z=z,D=d)}{\bP(R^D=1\mid Z=z,D=d)}$.
Since
\begin{eqnarray*}
\bP_{dy11\mid z}=\bP(D=d,Y=y\mid Z=z)\bP(R^D=1\mid Z=z,D=d)\bP(R^Y=1\mid D=d,R^D=1),
\end{eqnarray*}
we have
\begin{align}
\bP_{y+01\mid z} &= \int_{d\in \mathcal{D}} \bP(D=d,Y=y,R^D=0,R^Y=1\mid Z=z)\textup{d}d \nonumber\\&=\int_{d\in \mathcal{D}}\bP_{dy11\mid z}\zeta_{z}(d)\frac{\bP(R^Y=1\mid D=d,R^D=0)}{\bP(R^Y=1\mid D=d,R^D=1)}\textup{d}d\nonumber
\end{align} 
for each $y\in\mathcal{Y}$, and
\begin{eqnarray*}
\bP(R^D=0\mid Z=z)=\int_{d\in \mathcal{D}}\frac{\{\zeta_{z}(d)\bP(R^Y=1\mid D=d,R^D=0)\}\bP(D=d,R^D=1\mid Z=z)}{\bP(R^Y=1\mid D=d,R^D=0)}\textup{d}d
\end{eqnarray*}
for each $z\in\mathcal{Z}$.
The uniqueness of solutions $\{\zeta_{z}(d)\bP(R^Y=1\mid D=d,R^D=0)\}$ requires that $\bP(D,Y,R^D=1,R^Y=1\mid Z=z)$ is complete in $Y$ for all $z$, and the uniqueness of solutions $\bP(R^Y=1\mid D=d,R^D=0)$ requires that $\bP(D,R^D=1\mid Z)$ is complete in $Z$. For binary $Z$ and $D$, the uniqueness of solutions $\{\zeta_{z}(d)\bP(R^Y=1\mid D=d,R^D=0)\}$ requires $Y \notindependent D \mid Z = 1$ under one-sided noncompliance, and $Y \notindependent D \mid Z = z$ for $z = 0,1$ under two-sided noncompliance, and the uniqueness of solutions $\bP(R^Y=1\mid D = d, R^D = 0)$ requires $D \notindependent Z$.

We can identify $\bP(R^D=1\mid Z=z,D=d)$ once $\bP(R^Y=1\mid D=d,R^D=0)$ is identified. The identification of $\bP(D=d,Y=y\mid Z=z)$ follows from 
\begin{eqnarray}
\bP(D=d,Y=y\mid Z=z)=\frac{\bP_{dy11\mid z}}{\bP(R^D=1\mid Z=z,D=d)\bP(R^Y=1\mid D=d,R^D=1)}.\nonumber
\end{eqnarray}

\subsubsection{Assumption 1\textit{Y}$\oplus$2\textit{ZD}}\label{proof1Y2ZD}
Define 
    $\bP_{dy11\mid z} = \bP(D=d,Y=y,R^D=1,R^Y=1\mid Z=z)$,
    $\bP_{y+01\mid z} = \bP(Y=y,R^D=0,R^Y=1\mid Z=z)$,
    $\bP_{d1+0\mid z} = \bP(D=d,R^D=1,R^Y=0\mid Z=z)$,
    $\bP_{+0+0\mid z} = \bP(R^D=0,R^Y=0\mid Z=z)$,
    $\eta(y) =  \frac{\bP(R^Y=0\mid Y=y,R^D=1)}{\bP(R^Y=1\mid Y=y,R^D=1)}$,
    $\xi(y) =  \frac{\bP(R^Y=0\mid Y=y,R^D=0)}{\bP(R^Y=1\mid Y=y,R^D=0)}$,
    $\zeta_{z}(d) =  \frac{\bP(R^D=0\mid Z=z,D=d)}{\bP(R^D=1\mid Z=z,D=d)}$.
Since
\begin{align*}
\bP_{dy11\mid z}&=\bP(D=d,Y=y\mid Z=z)\bP(R^D=1\mid Z=z,D=d)\bP(R^Y=1\mid Y=y,R^D=1),\\
\bP_{y+01\mid z}&=\bP(Y=y,R^D=0\mid Z=z)\bP(R^Y=1\mid Y=y,R^D=0),
\end{align*}
we have
$$\bP_{d1+0\mid z} = \int_{y\in \mathcal{Y}} \bP(D=d,Y=y,R^D=1,R^Y=0\mid Z=z)\textup{d}y =\int_{y\in \mathcal{Y}}\bP_{dy11\mid z}\eta(y)\textup{d}y
$$
for each $(z,d)\in (\mathcal{Z},\mathcal{D})$,
$$
\bP_{+0+0\mid z} = \int_{y\in \mathcal{Y}} \bP(Y=y,R^D=0,R^Y=0\mid Z=z)\textup{d}y =\int_{y\in \mathcal{Y}}\bP_{y+01\mid z}\xi(y)\textup{d}y
$$
for each $z\in \mathcal{Z}$, and 
\begin{align}
\bP_{y+01\mid z} &= \int_{d\in \mathcal{D}} \bP(D=d,Y=y,R^D=0,R^Y=1\mid Z=z)\textup{d}d \nonumber\\&=\frac{\bP(R^Y=1\mid Y=y,R^D=0)}{\bP(R^Y=1\mid Y=y,R^D=1)}\int_{d\in \mathcal{D}}\bP_{dy11\mid z}\zeta_{z}(d)\textup{d}d\nonumber
\end{align}
for each $y\in\mathcal{Y}$. The uniqueness of solutions $\eta(y)$ requires that $\bP(D,Y,R^D=1,R^Y=1 \mid Z)$ is complete in $(Z,D)$, the uniqueness of solutions $\xi(y)$ requires that $\bP(Y,R^D=0,R^Y=1 \mid Z)$ is complete in $Z$, and the uniqueness of solutions $\zeta_{z}(d)$ requires that $\bP(D,Y,R^D=1,R^Y=1 \mid Z = z)$ is complete in $Y$ for all $z$. For binary $Z$ and $D$, the uniqueness of solutions $\eta(y)$ requires $Y \notindependent (Z,D) \mid (R^D = 1)$, the uniqueness of solutions $\xi(y)$ requires that $Y$ is binary and $Y \notindependent Z \mid (R^D = 0)$, and the uniqueness of solutions $\zeta_{z}(d)$ requires $Y \notindependent D \mid Z = 1$ under one-sided noncompliance, and $Y \notindependent D \mid Z = z$ for $z = 0,1$ under two-sided noncompliance.

We can identify $\bP(R^Y=1 \mid Y=y, R^D=1)$, $\bP(R^Y=1 \mid Y=y, R^D=0)$, and $\bP(R^D=1 \mid Z=z, D=d)$ once $\eta(y)$, $\xi(y)$, and $\zeta_{z}(d)$ are identified. The identification of $\bP(D=d,Y=y\mid Z=z)$ follows from 
\begin{eqnarray}
\bP(D=d,Y=y\mid Z=z)=\frac{\bP_{dy11\mid z}}{\bP(R^D=1\mid Z=z,D=d)\bP(R^Y=1\mid Y=y,R^D=1)}.\nonumber
\end{eqnarray}

\subsubsection{Assumption 1\textit{Z}$\oplus$2\textit{ZU}}
The identification of $\bP(Y=y\mid U=n,D=0)$ and $\bP(Y=y\mid U=c,D=1)$ follows from 
\begin{align*}
\bP(Y=y\mid U=n,D=0)&=\bP(Y=y\mid Z=1,U=n,D=0,R^D=1,R^Y=1),\\
\bP(Y=y\mid U=c,D=1)&=\bP(Y=y\mid Z=1,U=c,D=1,R^D=1,R^Y=1).
\end{align*} 
The identification of $\bP(Y=y\mid Z=z)$ follows from 
\begin{align*}
\bP(Y=y\mid Z=z)&=\frac{\bP(Y=y,R^D=0,R^Y=1\mid Z=z)}{\bP(R^Y=1\mid Z=z,R^D=0)}+\frac{\bP(Y=y,R^D=1,R^Y=1\mid Z=z)}{\bP(R^Y=1\mid Z=z,R^D=1)}.
\end{align*} 
The identification of $\bP(Y=y \mid U=c, D=0)$ follows the same logic as in Assumption 2\textit{ZU}, as detailed in \ref{proof2ZU}.

\subsubsection{Assumption 1\textit{U}$\oplus$2\textit{ZU}}
The identification of $\bP(Y=y\mid U=c,D=1)$ follows from 
\begin{align*}
\bP(Y=y\mid U=c,D=1)=\bP(Y=y\mid Z=1,U=c,D=1,R^D=1,R^Y=1).
\end{align*} 
Define 
    $\bP_{ud1r^Y\mid z} = \bP(U=u,D=d,R^D=1,R^Y=r^Y\mid Z=z)$,
    $\bP_{01r^Y\mid 0} = \bP(D=0,R^D=1,R^Y=r^Y\mid Z=0)$,
    $\zeta(u) = \frac{\bP(R^D=1\mid Z=0,U=u)}{\bP(R^D=1\mid Z=1,U=u)}$.
We have
\begin{align}
\bP_{01r^Y\mid 0}=\bP_{n01r^Y\mid 0}+\bP_{c01r^Y\mid 0}\nonumber=\bP_{n01r^Y\mid 1}\zeta(n)+\bP_{c11r^Y\mid 1}\zeta(c)\nonumber
\end{align}
for $r^Y=0,1$. The uniqueness of solutions $\zeta(u)$ requires that $R^Y \notindependent U \mid (R^D=1)$. Note that
\begin{eqnarray*}
&&\bP(U=c,D=0,Y=y,R^D=1,R^Y=1\mid Z=0)\\&=& \bP(D=0,Y=y,R^D=1,R^Y=1\mid Z=0)-\bP(U=n,D=0,Y=y,R^D=1,R^Y=1\mid Z=0).\nonumber
\end{eqnarray*}
Since
\begin{eqnarray*}
&&\bP(U=n,D=0,Y=y,R^D=1,R^Y=1\mid Z=0)\\&=&\bP(U=n,D=0,Y=y,R^D=1,R^Y=1\mid Z=1)\zeta(n),
\end{eqnarray*}
we can identify $\bP(U=c,D=0,Y=y,R^D=1,R^Y=1\mid Z=0)$. The identification of $\bP(Y=y\mid U=c,D=0)$ follows from 
\begin{eqnarray*}
\bP(Y=y\mid U=c,D=0)=\frac{\bP(U=c,D=0,Y=y,R^D=1,R^Y=1\mid Z=0)}{\bP(U=c,D=0,R^D=1,R^Y=1\mid Z=0)}.
\end{eqnarray*}

\subsubsection{Assumption 1\textit{D}$\oplus$2\textit{ZU}}
The identification of $\bP(Y=y\mid U=n,D=0)$ and $\bP(Y=y\mid U=c,D=1)$ follows from 
\begin{align*}
\bP(Y=y\mid U=n,D=0)&=\bP(Y=y\mid Z=1,U=n,D=0,R^D=1,R^Y=1),\\
\bP(Y=y\mid U=c,D=1)&=\bP(Y=y\mid Z=1,U=c,D=1,R^D=1,R^Y=1).
\end{align*} 
Define 
    $\bP_{udy11\mid 1} = \bP(U=u,D=d,Y=y,R^D=1,R^Y=1\mid Z=1)$,
    $\bP_{y+01\mid 1} = \bP(Y=y,R^D=0,R^Y=1\mid Z=1)$,
    $\eta(u) =  \frac{\bP(R^D=0\mid Z=1,U=u)}{\bP(R^D=1\mid Z=1,U=u)}$.
We have
\begin{align}
\bP_{y+01\mid 1} &= \bP_{n0y11\mid 1}\eta(n)\frac{\bP(R^Y=1\mid D=0,R^D=0)}{\bP(R^Y=1\mid D=0,R^D=1)}+\bP_{c1y11\mid 1}\eta(c)\frac{\bP(R^Y=1\mid D=1,R^D=0)}{\bP(R^Y=1\mid D=1,R^D=1)}\nonumber
\end{align}
for $y=0,1$. The uniqueness of solutions $\{\eta(u)\bP(R^Y=1\mid D=d,R^D=0)\}$ requires that $Y \notindependent U \mid (Z=1)$. Since
\begin{align}
\bP(R^Y=1\mid D=0,R^D=0)=\frac{\bP(D=0,R^D=0,R^Y=1\mid Z=0)}{\bP(D=0,R^D=0\mid Z=0)}\nonumber,
\end{align}
we can identify $\bP(R^D=1\mid Z=1,U=n)$. The identification of $\bP(U=n,D=0\mid Z=1)$ follows from 
\begin{align}
\bP(U=n,D=0\mid Z=1)=\frac{\bP(U=n,D=0,R^D=1\mid Z=1)}{\bP(R^D=1\mid Z=1,U=n)}\nonumber.
\end{align}
The identification of $\bP(Y=y\mid Z=0)$ follows from 
\begin{eqnarray*}
&&\bP(Y=y\mid Z=0)\\&=&\bP(D=0,Y=y,R^D=1\mid Z=0)+\bP(D=0,Y=y,R^D=0\mid Z=0)\nonumber\\&=&\frac{\bP(D=0,Y=y,R^D=1,R^Y=1\mid Z=0)}{\bP(R^Y=1\mid D=0,R^D=1)}+\frac{\bP(D=0,Y=y,R^D=0,R^Y=1\mid Z=0)}{\bP(R^Y=1\mid D=0,R^D=0)}\nonumber.
\end{eqnarray*}
Since
\begin{align*}
\bP(Y=y\mid Z=0)&=\bP(U=n,D=0,Y=y\mid Z=0)+\bP(U=c,D=0,Y=y\mid Z=0)\\&=\bP(Y=y\mid U=n,D=0)\bP(U=n,D=0\mid Z=1)\\&~~+\bP(Y=y\mid U=c,D=0)\{1-\bP(U=n,D=0\mid Z=1)\},
\end{align*}
we can identify $\bP(Y=y\mid U=c, D=0)$. 

\subsubsection{Assumption 1\textit{Y}$\oplus$2\textit{ZU}}
The identification of $\bP(R^Y=1 \mid Y=y, R^D=r^D)$ follows the same logic as in Assumption 1\textit{Y}$\oplus$2\textit{ZD}, as detailed in \ref{proof1Y2ZD}. The identification of $\bP(Y=y\mid U=n,D=0)$ and $\bP(Y=y\mid U=c,D=1)$ follows from 
\begin{align*}
\bP(Y=y\mid U=n,D=0)&=\frac{\bP(U=n,D=0,Y=y,R^D=1,R^Y=1\mid Z=1)}{\bP(U=n,D=0,R^D=1\mid Z=1)\bP(R^Y=1\mid Y=y,R^D=1)},\\
\bP(Y=y\mid U=c,D=1)&=\frac{\bP(U=c,D=1,Y=y,R^D=1,R^Y=1\mid Z=1)}{\bP(U=c,D=1,R^D=1\mid Z=1)\bP(R^Y=1\mid Y=y,R^D=1)}.
\end{align*} 
The identification of $\bP(Y=y\mid Z=z)$ follows from 
\begin{align*}
\bP(Y=y\mid Z=z)&=\frac{\bP(Y=y,R^D=0,R^Y=1\mid Z=z)}{\bP(R^Y=1\mid Y=y,R^D=0)}+\frac{\bP(Y=y,R^D=1,R^Y=1\mid Z=z)}{\bP(R^Y=1\mid Y=y,R^D=1)}.
\end{align*} 
The identification of $\bP(Y = y \mid U = c, D = 0)$ follows the same logic as in Assumption 2\textit{ZU}, as detailed in \ref{proof2ZU}. 

\section{Counterexamples}\label{sec::counterexamples}

\subsection{Counterexamples for the missing outcome models}\label{subsec::counterexamples1}
\begin{figure}[H]
\centering
\scalebox{0.8}{
\begin{tikzpicture}

    \node (z)  at (0,0) {$Z$};
    \node (d)  at (1.5,0) {$D$};
    \node (ry) at (3,1.5) {$R^Y$};
    \node (y)  at (3,0) {$Y$};
    \node (u)  at (2.25,-0.8) {$U$};
    \node (c)  at (1.5,-1.5) {1\textit{DY}};

    \path[-latex] (z) edge (d);
    \path[-latex] (d) edge (y);
    \path[-latex] (u) edge (d);
    \path[-latex] (u) edge (y);
    \path[-latex] (d) edge (ry);
    \path[-latex] (y) edge (ry);

    \node (z)  at (3.5,0) {$Z$};
    \node (d)  at (5,0) {$D$};
    \node (ry) at (6.5,1.5) {$R^Y$};
    \node (y)  at (6.5,0) {$Y$};
    \node (u)  at (5.75,-0.8) {$U$};
    \node (c)  at (5,-1.5) {1\textit{ZY}};

    \path[-latex] (z) edge (d);
    \path[-latex] (d) edge (y);
    \path[-latex] (u) edge (d);
    \path[-latex] (u) edge (y);
    \path[-latex] (z) edge (ry);
    \path[-latex] (y) edge (ry);

    \node (z)  at (7,0) {$Z$};
    \node (d)  at (8.5,0) {$D$};
    \node (ry) at (10,1.5) {$R^Y$};
    \node (y)  at (10,0) {$Y$};
    \node (u)  at (9.25,-0.8) {$U$};
    \node (c)  at (8.5,-1.5) {1\textit{ZU}};

    \path[-latex] (z) edge (d);
    \path[-latex] (d) edge (y);
    \path[-latex] (u) edge (d);
    \path[-latex] (u) edge (y);
    \path[-latex] (z) edge (ry);
    \path[-latex] (u) edge (ry);

    \node (z)  at (10.5,0) {$Z$};
    \node (d)  at (12,0) {$D$};
    \node (ry) at (13.5,1.5) {$R^Y$};
    \node (y)  at (13.5,0) {$Y$};
    \node (u)  at (12.75,-0.8) {$U$};
    \node (c)  at (12,-1.5) {1\textit{ZDY}};
    
    \path[-latex] (z) edge (d);
    \path[-latex] (d) edge (y);
    \path[-latex] (u) edge (d);
    \path[-latex] (u) edge (y);
    \path[-latex] (z) edge (ry);
    \path[-latex] (d) edge (ry);
    \path[-latex] (y) edge (ry);

    \node (z)  at (14,0) {$Z$};
    \node (d)  at (15.5,0) {$D$};
    \node (ry) at (17,1.5) {$R^Y$};
    \node (y)  at (17,0) {$Y$};
    \node (u)  at (16.25,-0.8) {$U$};
    \node (c)  at (15.5,-1.5) {1\textit{UDY}};

    \path[-latex] (z) edge (d);
    \path[-latex] (d) edge (y);
    \path[-latex] (u) edge (d);
    \path[-latex] (u) edge (y);
    \path[-latex] (d) edge (ry);
    \path[-latex] (u) edge (ry);
    \path[-latex] (y) edge (ry);

\end{tikzpicture}
}
\end{figure}

Under Assumptions 1$DY$ and 1$ZY$, identification holds with two-sided noncompliance, so we give counterexamples for one-sided noncompliance. Since Assumption 1$ZDY$ contains both 1$DY$ and 1$ZY$, identification fails even with two-sided noncompliance, for which we give a counterexample. Under Assumptions 1$ZU$ and 1$UDY$, identification fails under both one-sided and two-sided noncompliance; we give counterexamples for one-sided noncompliance, which is the special case of two-sided noncompliance with no always-takers. Define 
\begin{align*}
    \bP_{dy1\mid z} &= \bP(D=d,Y=y,R^Y=1\mid Z=z), \\
    \bP_{d+0\mid z} &= \bP(D=d,R^Y=0\mid Z=z).
\end{align*}
\subsubsection{Assumption 1\textit{DY}}\label{subsec::counterexamples1-dy}
For a binary $Y$ with one-sided noncompliance, we consider the following observable data probabilities:
\begin{eqnarray*}(\bP_{011\mid 0},\bP_{001\mid 0},\bP_{011\mid 1},\bP_{001\mid 1},\bP_{111\mid 1},\bP_{101\mid 1},\bP_{0+0\mid 0},\bP_{0+0\mid 1},\bP_{1+0\mid 1})=\left(\frac{1}{4},\frac{1}{6},\frac{1}{4},\frac{1}{12},\frac{1}{48},\frac{1}{32},\frac{7}{12},\frac{5}{12},\frac{19}{96}\right)\nonumber.
\end{eqnarray*}
Define the parameters:
\begingroup
\allowdisplaybreaks
\begin{align*}
&\Theta_{dy\mid z}=\bP(D=d,Y=y\mid Z=z)~\textup{for}~(z,d,y)=(0,0,1),(1,0,1),(1,0,0),(1,1,0),\\
&\Theta_{R^Y\mid dy}=\bP(R^Y=1\mid D=d,Y=y)~\textup{for}~d=0,1~\textup{and}~y=0,1.
\end{align*}
\endgroup
The following relationships between the observable data probabilities and the parameters hold,
\begingroup
\allowdisplaybreaks
\begin{align}
&\bP_{011\mid 0}=\Theta_{01\mid 0}\Theta_{R^Y\mid 01}, \\
&\bP_{001\mid 0}=(1-\Theta_{01\mid 0})\Theta_{R^Y\mid 00}, \\
&\bP_{011\mid 1}=\Theta_{01\mid 1}\Theta_{R^Y\mid 01}, \\
&\bP_{001\mid 1}=\Theta_{00\mid 1}\Theta_{R^Y\mid 00}, \\
&\bP_{111\mid 1}=(1-\Theta_{01\mid 1}-\Theta_{00\mid 1}-\Theta_{10\mid 1})\Theta_{R^Y\mid 11}, \\
&\bP_{101\mid 1}=\Theta_{10\mid 1}\Theta_{R^Y\mid 10}, \\
&\bP_{0+0\mid 0}=\Theta_{01\mid 0}(1-\Theta_{R^Y\mid 01})+(1-\Theta_{01\mid 0})(1-\Theta_{R^Y\mid 00}), \\
&\bP_{0+0\mid 1}=\Theta_{01\mid 1}(1-\Theta_{R^Y\mid 01})+\Theta_{00\mid 1}(1-\Theta_{R^Y\mid 00}), \\
&\bP_{1+0\mid 1}=\Theta_{10\mid 1}(1-\Theta_{R^Y\mid 10})+(1-\Theta_{01\mid 1}-\Theta_{00\mid 1}-\Theta_{10\mid 1})(1-\Theta_{R^Y\mid 11}).
\end{align}
\endgroup

Based on the observable data probabilities, we can have 
$\Theta_{01\mid 0}=\frac{1}{2}$, $\Theta_{00\mid 1}=\frac{1}{4}$, $\Theta_{01\mid 1}=\frac{1}{2}$, $\Theta_{10\mid 1}=\frac{1}{8}$, $\Theta_{R^Y\mid 00}=\frac{1}{3}$, $\Theta_{R^Y\mid 01}=\frac{1}{2}$, $\Theta_{R^Y\mid 10}=\frac{1}{4}$, $\Theta_{R^Y\mid 11}=\frac{1}{6}$, and the CACE $=\frac{1}{2}$. Alternatively, we can have $\Theta_{01\mid 0}=\frac{1}{2}$, $\Theta_{00\mid 1}=\frac{1}{4}$, $\Theta_{01\mid 1}=\frac{1}{2}$, $\Theta_{10\mid 1}=\frac{1}{12}$, $\Theta_{R^Y\mid 00}=\frac{1}{3}$, $\Theta_{R^Y\mid 01}=\frac{1}{2}$, $\Theta_{R^Y\mid 10}=\frac{3}{8}$, $\Theta_{R^Y\mid 11}=\frac{1}{8}$, and the CACE $=\frac{2}{3}$. Therefore, the CACE cannot be uniquely identified. 

\subsubsection{Assumption 1\textit{ZY}}\label{subsec::counterexamples1-zy}
For a binary $Y$ with one-sided noncompliance, we consider the following observable data probabilities:
\begin{eqnarray*}(\bP_{011\mid 0},\bP_{001\mid 0},\bP_{011\mid 1},\bP_{001\mid 1},\bP_{111\mid 1},\bP_{101\mid 1},\bP_{0+0\mid 0},\bP_{0+0\mid 1},\bP_{1+0\mid 1})=\left(\frac{1}{4},\frac{1}{6},\frac{1}{12},\frac{1}{16},\frac{1}{48},\frac{1}{32},\frac{7}{12},\frac{29}{48},\frac{19}{96}\right)\nonumber.
\end{eqnarray*}
Define the parameters:
\begingroup
\allowdisplaybreaks
\begin{align*}
&\Theta_{dy\mid z}=\bP(D=d,Y=y\mid Z=z)~\textup{for}~(z,d,y)=(0,0,1),(1,0,1),(1,0,0),(1,1,0),\\
&\Theta_{R^Y\mid zy}=\bP(R^Y=1\mid Z=z,Y=y)~\textup{for}~z=0,1~\textup{and}~y=0,1.
\end{align*}
\endgroup
The following relationships between the observable data probabilities and the parameters hold,
\begingroup
\allowdisplaybreaks
\begin{align}
&\bP_{011\mid 0}=\Theta_{01\mid 0}\Theta_{R^Y\mid 01}, \\
&\bP_{001\mid 0}=(1-\Theta_{01\mid 0})\Theta_{R^Y\mid 00}, \\
&\bP_{011\mid 1}=\Theta_{01\mid 1}\Theta_{R^Y\mid 11}, \\
&\bP_{001\mid 1}=\Theta_{00\mid 1}\Theta_{R^Y\mid 10}, \\
&\bP_{111\mid 1}=(1-\Theta_{01\mid 1}-\Theta_{00\mid 1}-\Theta_{10\mid 1})\Theta_{R^Y\mid 11}, \\
&\bP_{101\mid 1}=\Theta_{10\mid 1}\Theta_{R^Y\mid 10}, \\
&\bP_{0+0\mid 0}=\Theta_{01\mid 0}(1-\Theta_{R^Y\mid 01})+(1-\Theta_{01\mid 0})(1-\Theta_{R^Y\mid 00}), \\
&\bP_{0+0\mid 1}=\Theta_{01\mid 1}(1-\Theta_{R^Y\mid 11})+\Theta_{00\mid 1}(1-\Theta_{R^Y\mid 10}), \\
&\bP_{1+0\mid 1}=\Theta_{10\mid 1}(1-\Theta_{R^Y\mid 10})+(1-\Theta_{01\mid 1}-\Theta_{00\mid 1}-\Theta_{10\mid 1})(1-\Theta_{R^Y\mid 11}).
\end{align}
\endgroup

Based on the observable data probabilities, we can have $\Theta_{01\mid 0}=\frac{1}{2}$, $\Theta_{00\mid 1}=\frac{1}{4}$, $\Theta_{01\mid 1}=\frac{1}{2}$, $\Theta_{10\mid 1}=\frac{1}{8}$, $\Theta_{R^Y\mid 00}=\frac{1}{3}$, $\Theta_{R^Y\mid 01}=\frac{1}{2}$, $\Theta_{R^Y\mid 10}=\frac{1}{4}$, $\Theta_{R^Y\mid 11}=\frac{1}{6}$, and the CACE $=\frac{1}{2}$. Alternatively, we can have $\Theta_{01\mid 0}=\frac{2}{3}$, $\Theta_{00\mid 1}=\frac{1}{4}$, $\Theta_{01\mid 1}=\frac{1}{2}$, $\Theta_{10\mid 1}=\frac{1}{8}$, $\Theta_{R^Y\mid 00}=\frac{1}{2}$, $\Theta_{R^Y\mid 01}=\frac{3}{8}$, $\Theta_{R^Y\mid 10}=\frac{1}{4}$, $\Theta_{R^Y\mid 11}=\frac{1}{6}$, and the CACE $=-\frac{1}{6}$. Therefore, the CACE cannot be uniquely identified. 

\subsubsection{Assumption 1\textit{ZU}}\label{subsec::counterexamples1-zu}
For a binary $Y$ with one-sided noncompliance, we consider the following observable data probabilities:
\begin{eqnarray*}(\bP_{011\mid 0},\bP_{001\mid 0},\bP_{011\mid 1},\bP_{001\mid 1},\bP_{111\mid 1},\bP_{101\mid 1},\bP_{0+0\mid 0},\bP_{0+0\mid 1},\bP_{1+0\mid 1})=\left(\frac{1}{8},\frac{3}{16},\frac{1}{16},\frac{1}{16},\frac{3}{64},\frac{9}{64},\frac{11}{16},\frac{1}{8},\frac{9}{16}\right)\nonumber.
\end{eqnarray*}
Define the parameters:
\begingroup
\allowdisplaybreaks
\begin{align*}
&\Theta_{n}=\bP(U=n),\\
&\Theta_{Y\mid ud}=\bP(Y=1\mid U=u,D=d)~\textup{for}~(u,d)=(n,0),(c,1),(c,0),\\
&\Theta_{R^Y\mid zu}=\bP(R^Y=1\mid Z=z,U=u)~\textup{for}~z=0,1~\textup{and}~u=n,c.
\end{align*}
\endgroup
The following relationships between the observable data probabilities and the parameters hold,
\begingroup
\allowdisplaybreaks
\begin{align}
&\bP_{011\mid 0}=\Theta_{Y\mid n0}\Theta_{R^Y\mid 0n}\Theta_{n}+\Theta_{Y\mid c0}\Theta_{R^Y\mid 0c}(1-\Theta_{n}), \\
&\bP_{001\mid 0}=(1-\Theta_{Y\mid n0})\Theta_{R^Y\mid 0n}\Theta_{n}+(1-\Theta_{Y\mid c0})\Theta_{R^Y\mid 0c}(1-\Theta_{n}),\\
&\bP_{011\mid 1}=\Theta_{Y\mid n0}\Theta_{R^Y\mid 1n}\Theta_{n}, \\
&\bP_{001\mid 1}=(1-\Theta_{Y\mid n0})\Theta_{R^Y\mid 1n}\Theta_{n}, \\
&\bP_{111\mid 1}=\Theta_{Y\mid c1}\Theta_{R^Y\mid 1c}(1-\Theta_{n}), \\
&\bP_{101\mid 1}=(1-\Theta_{Y\mid c1})\Theta_{R^Y\mid 1c}(1-\Theta_{n}),\\
&\bP_{0+0\mid 0}=(1-\Theta_{R^Y\mid 0n})\Theta_{n}+(1-\Theta_{R^Y\mid 0c})(1-\Theta_{n}),\\
&\bP_{0+0\mid 1}=(1-\Theta_{R^Y\mid 1n})\Theta_{n},\\
&\bP_{1+0\mid 1}=(1-\Theta_{R^Y\mid 1c})(1-\Theta_{n}).
\end{align}
\endgroup
Based on the observable data probabilities, we can have $\Theta_{n}=\frac{1}{4}$, $\Theta_{Y\mid n0}=\frac{1}{2}$, $\Theta_{Y\mid c0}=\frac{1}{3}$, $\Theta_{Y\mid c1}=\frac{1}{4}$, $\Theta_{R^Y\mid 0n}=\frac{1}{2}$, $\Theta_{R^Y\mid 1n}=\frac{1}{2}$, $\Theta_{R^Y\mid 0c}=\frac{1}{4}$, $\Theta_{R^Y\mid 1c}=\frac{1}{4}$, and the CACE $=-\frac{1}{12}$. Alternatively, we can have $\Theta_{n}=\frac{1}{4}$, $\Theta_{Y\mid n0}=\frac{1}{2}$, $\Theta_{Y\mid c0}=\frac{3}{8}$, $\Theta_{Y\mid c1}=\frac{1}{4}$, $\Theta_{R^Y\mid 0n}=\frac{1}{4}$, $\Theta_{R^Y\mid 1n}=\frac{1}{2}$, $\Theta_{R^Y\mid 0c}=\frac{1}{3}$, $\Theta_{R^Y\mid 1c}=\frac{1}{4}$, and the CACE $=-\frac{1}{8}$. Therefore, the CACE cannot be uniquely identified. 

\subsubsection{Assumption 1\textit{ZDY}}\label{subsec::counterexamples1-zdy}
For a binary $Y$ with two-sided noncompliance, we consider the following observable data probabilities:
\begin{eqnarray*}
&&(\bP_{011\mid 0},\bP_{001\mid 0},\bP_{111\mid 0},\bP_{101\mid 0},\bP_{011\mid 1},\bP_{001\mid 1},\bP_{111\mid 1},\bP_{101\mid 1},\bP_{0+0\mid 0},\bP_{1+0\mid 0},\bP_{0+0\mid 1},\bP_{1+0\mid 1})\\&=&\left(\frac{1}{8},\frac{1}{12},\frac{1}{16},\frac{1}{32},\frac{1}{4},\frac{1}{48},\frac{1}{16},\frac{1}{16},\frac{13}{24},\frac{5}{32},\frac{17}{48},\frac{1}{4}\right)\nonumber.
\end{eqnarray*}
Define the parameters:
\begingroup
\allowdisplaybreaks
\begin{align*}
&\Theta_{dy\mid z}=\bP(D=d,Y=y\mid Z=z)~\textup{for}~(z,d,y)=(0,0,0),(0,1,1),(0,1,0),(1,0,1),(1,0,0),(1,1,0),\\
&\Theta_{R^Y\mid zdy}=\bP(R^Y=1\mid Z=z,D=d,Y=y)~\textup{for}~z=0,1,~d=0,1,~\textup{and}~y=0,1.
\end{align*}
\endgroup
The following relationships between the observable data probabilities and the parameters hold,
\begingroup
\allowdisplaybreaks
\begin{align}
&\bP_{011\mid 0}=(1-\Theta_{00\mid 0}-\Theta_{11\mid 0}-\Theta_{10\mid 0})\Theta_{R^Y\mid 001}, \\
&\bP_{001\mid 0}=\Theta_{00\mid 0}\Theta_{R^Y\mid 000}, \\
&\bP_{111\mid 0}=\Theta_{11\mid 0}\Theta_{R^Y\mid 011}, \\
&\bP_{101\mid 0}=\Theta_{10\mid 0}\Theta_{R^Y\mid 010}, \\
&\bP_{011\mid 1}=\Theta_{01\mid 1}\Theta_{R^Y\mid 101}, \\
&\bP_{001\mid 1}=\Theta_{00\mid 1}\Theta_{R^Y\mid 100}, \\
&\bP_{111\mid 1}=(1-\Theta_{01\mid 1}-\Theta_{00\mid 1}-\Theta_{10\mid 1})\Theta_{R^Y\mid 111}, \\
&\bP_{101\mid 1}=\Theta_{10\mid 1}\Theta_{R^Y\mid 110}, \\
&\bP_{0+0\mid 0}=(1-\Theta_{00\mid 0}-\Theta_{11\mid 0}-\Theta_{10\mid 0})(1-\Theta_{R^Y\mid 001})+\Theta_{00\mid 0}(1-\Theta_{R^Y\mid 000}), \\
&\bP_{1+0\mid 0}=\Theta_{11\mid 0}(1-\Theta_{R^Y\mid 011})+\Theta_{10\mid 0}(1-\Theta_{R^Y\mid 010}), \\
&\bP_{0+0\mid 1}=\Theta_{01\mid 1}(1-\Theta_{R^Y\mid 101})+\Theta_{00\mid 1}(1-\Theta_{R^Y\mid 100}), \\
&\bP_{1+0\mid 1}=(1-\Theta_{01\mid 1}-\Theta_{00\mid 1}-\Theta_{10\mid 1})(1-\Theta_{R^Y\mid 111})+\Theta_{10\mid 1}(1-\Theta_{R^Y\mid 110}).
\end{align}
\endgroup

Based on the observable data probabilities, we can have $\Theta_{10\mid 0}=\frac{1}{8}$, $\Theta_{11\mid 0}=\frac{1}{8}$, $\Theta_{00\mid 0}=\frac{1}{4}$, $\Theta_{00\mid 1}=\frac{1}{8}$, $\Theta_{01\mid 1}=\frac{1}{2}$, $\Theta_{10\mid 1}=\frac{1}{8}$, $\Theta_{R^Y\mid 000}=\frac{1}{3}$, $\Theta_{R^Y\mid 001}=\frac{1}{4}$, $\Theta_{R^Y\mid 010}=\frac{1}{4}$, $\Theta_{R^Y\mid 011}=\frac{1}{2}$, $\Theta_{R^Y\mid 100}=\frac{1}{6}$, $\Theta_{R^Y\mid 101}=\frac{1}{2}$, $\Theta_{R^Y\mid 110}=\frac{1}{2}$, $\Theta_{R^Y\mid 111}=\frac{1}{4}$, and the CACE $=1$. Alternatively, we can have $\Theta_{10\mid 0}=\frac{1}{12}$, $\Theta_{11\mid 0}=\frac{1}{6}$, $\Theta_{00\mid 0}=\frac{1}{2}$, $\Theta_{00\mid 1}=\frac{1}{4}$, $\Theta_{01\mid 1}=\frac{3}{8}$, $\Theta_{10\mid 1}=\frac{1}{6}$, $\Theta_{R^Y\mid 000}=\frac{1}{6}$, $\Theta_{R^Y\mid 001}=\frac{1}{2}$, $\Theta_{R^Y\mid 010}=\frac{3}{8}$, $\Theta_{R^Y\mid 011}=\frac{3}{8}$, $\Theta_{R^Y\mid 100}=\frac{1}{12}$, $\Theta_{R^Y\mid 101}=\frac{2}{3}$, $\Theta_{R^Y\mid 110}=\frac{3}{8}$, $\Theta_{R^Y\mid 111}=\frac{3}{10}$, and the CACE $=\frac{4}{3}$. Therefore, the CACE cannot be uniquely identified. 

\subsubsection{Assumption 1\textit{UDY}}\label{subsec::counterexamples1-udy}
For a binary $Y$ with one-sided noncompliance, we consider the following observable data probabilities:
\begin{eqnarray*}(\bP_{011\mid 0},\bP_{001\mid 0},\bP_{011\mid 1},\bP_{001\mid 1},\bP_{111\mid 1},\bP_{101\mid 1},\bP_{0+0\mid 0},\bP_{0+0\mid 1},\bP_{1+0\mid 1})=\left(\frac{3}{16},\frac{3}{16},\frac{1}{8},\frac{1}{16},\frac{1}{18},\frac{1}{12},\frac{5}{8},\frac{5}{16},\frac{13}{36}\right)\nonumber.
\end{eqnarray*}
Define the parameters:
\begingroup
\allowdisplaybreaks
\begin{align*}
&\Theta_{n}=\bP(U=n),\\
&\Theta_{Y\mid ud}=\bP(Y=1\mid U=u,D=d)~\textup{for}~(u,d)=(n,0),(c,1),(c,0),\\
&\Theta_{R^Y\mid udy}=\bP(R^Y=1\mid U=u,D=d,Y=y)~\textup{for}~(u,d)=(n,0),(c,1),(c,0)~\textup{and}~y=0,1.
\end{align*}
\endgroup
The following relationships between the observable data probabilities and the parameters hold,
\begingroup
\allowdisplaybreaks
\begin{align}
&\bP_{011\mid 0}=\Theta_{Y\mid n0}\Theta_{R^Y\mid n01}\Theta_{n}+\Theta_{Y\mid c0}\Theta_{R^Y\mid c01}(1-\Theta_{n}), \\
&\bP_{001\mid 0}=(1-\Theta_{Y\mid n0})\Theta_{R^Y\mid n00}\Theta_{n}+(1-\Theta_{Y\mid c0})\Theta_{R^Y\mid c00}(1-\Theta_{n}), \\
&\bP_{011\mid 1}=\Theta_{Y\mid n0}\Theta_{R^Y\mid n01}\Theta_{n}, \\
&\bP_{001\mid 1}=(1-\Theta_{Y\mid n0})\Theta_{R^Y\mid n00}\Theta_{n}, \\
&\bP_{111\mid 1}=\Theta_{Y\mid c1}\Theta_{R^Y\mid c11}(1-\Theta_{n}), \\
&\bP_{101\mid 1}=(1-\Theta_{Y\mid c1})\Theta_{R^Y\mid c10}(1-\Theta_{n}), \\
&\bP_{0+0\mid 0}=\Theta_{Y\mid n0}(1-\Theta_{R^Y\mid n01})\Theta_{n}+\Theta_{Y\mid c0}(1-\Theta_{R^Y\mid c01})(1-\Theta_{n})\nonumber\\&\hspace{1.5cm}+(1-\Theta_{Y\mid n0})(1-\Theta_{R^Y\mid n00})\Theta_{n}+(1-\Theta_{Y\mid c0})(1-\Theta_{R^Y\mid c00})(1-\Theta_{n}),\\
&\bP_{0+0\mid 1}=\Theta_{Y\mid n0}(1-\Theta_{R^Y\mid n01})\Theta_{n}+(1-\Theta_{Y\mid n0})(1-\Theta_{R^Y\mid n00})\Theta_{n},\\
&\bP_{1+0\mid 1}=\Theta_{Y\mid c1}(1-\Theta_{R^Y\mid c11})(1-\Theta_{n})+(1-\Theta_{Y\mid c1})(1-\Theta_{R^Y\mid c10})(1-\Theta_{n}).
\end{align}
\endgroup
Based on the observable data probabilities, we can have $\Theta_{n}=\frac{1}{2}$, $\Theta_{Y\mid n0}=\frac{1}{2}$, $\Theta_{Y\mid c0}=\frac{1}{4}$, $\Theta_{Y\mid c1}=\frac{1}{3}$, $\Theta_{R^Y\mid n00}=\frac{1}{4}$, $\Theta_{R^Y\mid n01}=\frac{1}{2}$, $\Theta_{R^Y\mid c00}=\frac{1}{3}$, $\Theta_{R^Y\mid c01}=\frac{1}{2}$, $\Theta_{R^Y\mid c10}=\frac{1}{4}$, $\Theta_{R^Y\mid c11}=\frac{1}{3}$, and the CACE $=\frac{1}{12}$. Alternatively, we can have $\Theta_{n}=\frac{1}{2}$, $\Theta_{Y\mid n0}=\frac{3}{4}$, $\Theta_{Y\mid c0}=\frac{1}{2}$, $\Theta_{Y\mid c1}=\frac{4}{9}$, $\Theta_{R^Y\mid n00}=\frac{1}{2}$, $\Theta_{R^Y\mid n01}=\frac{1}{3}$, $\Theta_{R^Y\mid c00}=\frac{1}{2}$, $\Theta_{R^Y\mid c01}=\frac{1}{4}$, $\Theta_{R^Y\mid c10}=\frac{3}{10}$, $\Theta_{R^Y\mid c11}=\frac{1}{4}$, and the CACE $=-\frac{1}{18}$. Therefore, the CACE cannot be uniquely identified. 

\subsection{Counterexamples for the missing treatment models}\label{subsec::counterexamples2}
\begin{figure}[H]
\centering
\scalebox{0.8}{
\begin{tikzpicture}

    \node (z)  at (0,0) {$Z$};
    \node (d)  at (1.5,0) {$D$};
    \node (rd) at (1.5,1.5) {$R^D$};
    \node (y)  at (3,0) {$Y$};
    \node (u)  at (2.25,-0.8) {$U$};
    \node (c)  at (1.5,-1.5) {2\textit{ZU}};

    \path[-latex] (z) edge (d);
    \path[-latex] (d) edge (y);
    \path[-latex] (u) edge (d);
    \path[-latex] (u) edge (y);
    \path[-latex] (z) edge (rd);
    \path[-latex] (u) edge (rd);

    \node (z)  at (4,0) {$Z$};
    \node (d)  at (5.5,0) {$D$};
    \node (rd) at (5.5,1.5) {$R^D$};
    \node (y)  at (7,0) {$Y$};
    \node (u)  at (6.25,-0.8) {$U$};
    \node (c)  at (5.5,-1.5) {2\textit{ZDY}};

    \path[-latex] (z) edge (d);
    \path[-latex] (d) edge (y);
    \path[-latex] (u) edge (d);
    \path[-latex] (u) edge (y);
    \path[-latex] (z) edge (rd);
    \path[-latex] (d) edge (rd);
    \path[-latex] (y) edge (rd);

    \node (z)  at (8,0) {$Z$};
    \node (d)  at (9.5,0) {$D$};
    \node (rd) at (9.5,1.5) {$R^D$};
    \node (y)  at (11,0) {$Y$};
    \node (u)  at (10.25,-0.8) {$U$};
    \node (c)  at (9.5,-1.5) {2\textit{UDY}};

    \path[-latex] (z) edge (d);
    \path[-latex] (d) edge (y);
    \path[-latex] (u) edge (d);
    \path[-latex] (u) edge (y);
    \path[-latex] (d) edge (rd);
    \path[-latex] (u) edge (rd);
    \path[-latex] (y) edge (rd);
    
\end{tikzpicture}
}
\end{figure}

Under Assumption 2$ZU$, identification holds with one-sided noncompliance, so we give a counterexample for two-sided noncompliance. Under Assumptions 2$ZDY$ and 2$UDY$, identification fails under both one-sided and two-sided noncompliance; we give counterexamples for one-sided noncompliance, which is the special case of two-sided noncompliance with no always-takers. Define 
\begin{align*}
    \bP_{dy1\mid z} &= \bP(D=d,Y=y,R^D=1\mid Z=z),\\
    \bP_{y+0\mid z} &= \bP(Y=y,R^D=0\mid Z=z).
\end{align*}
\subsubsection{Assumption 2\textit{ZU}}\label{subsec::counterexamples2-zu}
For a binary $Y$ with two-sided noncompliance, we consider the following observable data probabilities:
\begin{eqnarray*}
&&(\bP_{011\mid 0},\bP_{001\mid 0},\bP_{011\mid 1},\bP_{001\mid 1},\bP_{111\mid 0},\bP_{101\mid 0},\bP_{111\mid 1},\bP_{101\mid 1},\bP_{1+0\mid 0},\bP_{0+0\mid 0},\bP_{1+0\mid 1},\bP_{0+0\mid 1})\\&=&\left(\frac{5}{48},\frac{5}{48},\frac{1}{12},\frac{1}{12},\frac{1}{16},\frac{1}{16},\frac{7}{96},\frac{11}{96},\frac{1}{3},\frac{1}{3},\frac{29}{96},\frac{11}{32}\right)\nonumber.
\end{eqnarray*}
Define the parameters:
\begingroup
\allowdisplaybreaks
\begin{align*}
&\Theta_{u}=\bP(U=u)~\textup{for}~u=a,n,\\
&\Theta_{Y\mid ud}=\bP(Y=1\mid U=u,D=d)~\textup{for}~(u,d)=(a,1),(n,0),(c,1),(c,0),\\
&\Theta_{R^D\mid zu}=\bP(R^D=1\mid Z=z,U=u)~\textup{for}~z=0,1 ~\textup{and}~u=a,n,c.
\end{align*}
\endgroup
The following relationships between the observable data probabilities and the parameters hold,
\begingroup
\allowdisplaybreaks
\begin{align}
&\bP_{011\mid 0}=\Theta_{Y\mid n0}\Theta_{n}\Theta_{R^D\mid 0n}+\Theta_{Y\mid c0}(1-\Theta_{n}-\Theta_{a})\Theta_{R^D\mid 0c}, \\
&\bP_{001\mid 0}=(1-\Theta_{Y\mid n0})\Theta_{n}\Theta_{R^D\mid 0n}+(1-\Theta_{Y\mid c0})(1-\Theta_{n}-\Theta_{a})\Theta_{R^D\mid 0c}, \\
&\bP_{011\mid 1}=\Theta_{Y\mid n0}\Theta_{n}\Theta_{R^D\mid 1n}, \\
&\bP_{001\mid 1}=(1-\Theta_{Y\mid n0})\Theta_{n}\Theta_{R^D\mid 1n}, \\
&\bP_{111\mid 0}=\Theta_{Y\mid a1}\Theta_{a}\Theta_{R^D\mid 0a}, \\
&\bP_{101\mid 0}=(1-\Theta_{Y\mid a1})\Theta_{a}\Theta_{R^D\mid 0a}, \\
&\bP_{111\mid 1}=\Theta_{Y\mid a1}\Theta_{a}\Theta_{R^D\mid 1a}+\Theta_{Y\mid c1}(1-\Theta_{n}-\Theta_{a})\Theta_{R^D\mid 1c}, \\
&\bP_{101\mid 1}=(1-\Theta_{Y\mid a1})\Theta_{a}\Theta_{R^D\mid 1a}+(1-\Theta_{Y\mid c1})(1-\Theta_{n}-\Theta_{a})\Theta_{R^D\mid 1c}, \\
&\bP_{1+0\mid 0}=\Theta_{Y\mid n0}\Theta_{n}(1-\Theta_{R^D\mid 0n})+\Theta_{Y\mid c0}(1-\Theta_{n}-\Theta_{a})(1-\Theta_{R^D\mid 0c})\nonumber\\&\hspace{1.5cm}+\Theta_{Y\mid a1}\Theta_{a}(1-\Theta_{R^D\mid 0a}),\\
&\bP_{0+0\mid 0}=(1-\Theta_{Y\mid n0})\Theta_{n}(1-\Theta_{R^D\mid 0n})+(1-\Theta_{Y\mid c0})(1-\Theta_{n}-\Theta_{a})(1-\Theta_{R^D\mid 0c})\nonumber\\&\hspace{1.5cm}+(1-\Theta_{Y\mid a1})\Theta_{a}(1-\Theta_{R^D\mid 0a}),\\
&\bP_{1+0\mid 1}=\Theta_{Y\mid n0}\Theta_{n}(1-\Theta_{R^D\mid 1n})+\Theta_{Y\mid a1}\Theta_{a}(1-\Theta_{R^D\mid 1a})\nonumber\\&\hspace{1.5cm}+\Theta_{Y\mid c1}(1-\Theta_{n}-\Theta_{a})(1-\Theta_{R^D\mid 1c}),\\
&\bP_{0+0\mid 1}=(1-\Theta_{Y\mid n0})\Theta_{n}(1-\Theta_{R^D\mid 1n})+(1-\Theta_{Y\mid a1})\Theta_{a}(1-\Theta_{R^D\mid 1a})\nonumber\\&\hspace{1.5cm}+(1-\Theta_{Y\mid c1})(1-\Theta_{n}-\Theta_{a})(1-\Theta_{R^D\mid 1c}).
\end{align}
\endgroup
Based on the observable data probabilities, we can have  $\Theta_{n}=\frac{1}{2}$, $\Theta_{a}=\frac{1}{4}$, $\Theta_{Y\mid n0}=\frac{1}{2}$, $\Theta_{Y\mid c0}=\frac{1}{2}$, $\Theta_{Y\mid a1}=\frac{1}{2}$, $\Theta_{Y\mid c1}=\frac{1}{3}$, $\Theta_{R^D\mid 0n}=\frac{1}{4}$, $\Theta_{R^D\mid 1n}=\frac{1}{3}$, $\Theta_{R^D\mid 0a}=\frac{1}{2}$, $\Theta_{R^D\mid 1a}=\frac{1}{4}$, $\Theta_{R^D\mid 0c}=\frac{1}{3}$, $\Theta_{R^D\mid 1c}=\frac{1}{2}$, and the CACE $=-\frac{1}{6}$. Alternatively, we can have $\Theta_{n}=\frac{1}{3}$, $\Theta_{a}=\frac{1}{3}$, $\Theta_{Y\mid n0}=\frac{1}{2}$, $\Theta_{Y\mid c0}=\frac{1}{2}$, $\Theta_{Y\mid a1}=\frac{1}{2}$, $\Theta_{Y\mid c1}=\frac{3}{8}$, $\Theta_{R^D\mid 0n}=\frac{1}{4}$, $\Theta_{R^D\mid 1n}=\frac{1}{2}$, $\Theta_{R^D\mid 0a}=\frac{3}{8}$, $\Theta_{R^D\mid 1a}=\frac{1}{16}$, $\Theta_{R^D\mid 0c}=\frac{3}{8}$, $\Theta_{R^D\mid 1c}=\frac{1}{2}$, and the CACE $=-\frac{1}{8}$. Therefore, the CACE cannot be uniquely identified. 

\subsubsection{Assumption 2\textit{ZDY}}\label{subsec::counterexamples2-zdy}
For a binary $Y$ with one-sided noncompliance, we consider the following observable data probabilities:
\begin{eqnarray*}
&&(\bP_{011\mid 0},\bP_{001\mid 0},\bP_{011\mid 1},\bP_{001\mid 1},\bP_{111\mid 1},\bP_{101\mid 1},\bP_{1+0\mid 0},\bP_{0+0\mid 0},\bP_{1+0\mid 1},\bP_{0+0\mid 1})\\&=&\left(\frac{1}{8},\frac{1}{6},\frac{1}{4},\frac{1}{24},\frac{1}{64},\frac{1}{16},\frac{3}{8},\frac{1}{3},\frac{23}{64},\frac{13}{48}\right)\nonumber.
\end{eqnarray*}
Define the parameters:
\begingroup
\allowdisplaybreaks
\begin{align*}
&\Theta_{dy\mid z}=\bP(D=d,Y=y\mid Z=z)~\textup{for}~(z,d,y)=(0,0,1),(1,0,1),(1,0,0),(1,1,0),\\
&\Theta_{R^D\mid zdy}=\bP(R^D=1\mid Z=z,D=d,Y=y)~\textup{for}~(z,d)=(0,0),(1,0),(1,1)~\textup{and}~y=0,1.
\end{align*}
\endgroup
The following relationships between the observable data probabilities and the parameters hold,
\begingroup
\allowdisplaybreaks
\begin{align}
&\bP_{011\mid 0}=\Theta_{01\mid 0}\Theta_{R^D\mid 001}, \\
&\bP_{001\mid 0}=(1-\Theta_{01\mid 0})\Theta_{R^D\mid 000}, \\
&\bP_{011\mid 1}=\Theta_{01\mid 1}\Theta_{R^D\mid 101}, \\
&\bP_{001\mid 1}=\Theta_{00\mid 1}\Theta_{R^D\mid 100}, \\
&\bP_{111\mid 1}=(1-\Theta_{01\mid 1}-\Theta_{00\mid 1}-\Theta_{10\mid 1})\Theta_{R^D\mid 111}, \\
&\bP_{101\mid 1}=\Theta_{10\mid 1}\Theta_{R^D\mid 110}, \\
&\bP_{1+0\mid 0}=\Theta_{01\mid 0}(1-\Theta_{R^D\mid 001}), \\
&\bP_{0+0\mid 0}=(1-\Theta_{01\mid 0})(1-\Theta_{R^D\mid 000}),\\
&\bP_{1+0\mid 1}=\Theta_{01\mid 1}(1-\Theta_{R^D\mid 101})+(1-\Theta_{01\mid 1}-\Theta_{00\mid 1}-\Theta_{10\mid 1})(1-\Theta_{R^D\mid 111}),\\
&\bP_{0+0\mid 1}=\Theta_{00\mid 1}(1-\Theta_{R^D\mid 100})+\Theta_{10\mid 1}(1-\Theta_{R^D\mid 110}).
\end{align}
\endgroup

Based on the observable data probabilities, we can have $\Theta_{01\mid 0}=\frac{1}{2}$, $\Theta_{00\mid 1}=\frac{1}{4}$, $\Theta_{01\mid 1}=\frac{1}{2}$, $\Theta_{10\mid 1}=\frac{1}{8}$, $\Theta_{R^D\mid 000}=\frac{1}{3}$, $\Theta_{R^D\mid 001}=\frac{1}{4}$, $\Theta_{R^D\mid 100}=\frac{1}{6}$, $\Theta_{R^D\mid 101}=\frac{1}{2}$, $\Theta_{R^D\mid 110}=\frac{1}{2}$, $\Theta_{R^D\mid 111}=\frac{1}{8}$, and the CACE $=\frac{1}{2}$. Alternatively, we can have $\Theta_{01\mid 0}=\frac{1}{2}$, $\Theta_{00\mid 1}=\frac{1}{8}$, $\Theta_{01\mid 1}=\frac{3}{8}$, $\Theta_{10\mid 1}=\frac{1}{4}$, $\Theta_{R^D\mid 000}=\frac{1}{3}$, $\Theta_{R^D\mid 001}=\frac{1}{4}$, $\Theta_{R^D\mid 100}=\frac{1}{3}$, $\Theta_{R^D\mid 101}=\frac{2}{3}$, $\Theta_{R^D\mid 110}=\frac{1}{4}$, $\Theta_{R^D\mid 111}=\frac{1}{16}$, and the CACE $=\frac{1}{4}$. Therefore, the CACE cannot be uniquely identified. 

\subsubsection{Assumption 2\textit{UDY}}\label{subsec::counterexamples2-udy}
For a binary $Y$ with one-sided noncompliance, we consider the following observable data probabilities:
\begin{eqnarray*}
&&(\bP_{011\mid 0},\bP_{001\mid 0},\bP_{011\mid 1},\bP_{001\mid 1},\bP_{111\mid 1},\bP_{101\mid 1},\bP_{1+0\mid 0},\bP_{0+0\mid 0},\bP_{1+0\mid 1},\bP_{0+0\mid 1})\\&=&\left(\frac{3}{16},\frac{3}{16},\frac{1}{8},\frac{1}{16},\frac{1}{18},\frac{1}{12},\frac{3}{16},\frac{7}{16},\frac{17}{72},\frac{7}{16}\right)\nonumber.
\end{eqnarray*}
Define the parameters:
\begingroup
\allowdisplaybreaks
\begin{align*}
&\Theta_{n}=\bP(U=n),\\
&\Theta_{Y\mid ud}=\bP(Y=1\mid U=u,D=d)~\textup{for}~(u,d)=(n,0),(c,1),(c,0),\\
&\Theta_{R^D\mid udy}=\bP(R^D=1\mid U=u,D=d,Y=y)~\textup{for}~(u,d)=(n,0),(c,1),(c,0)~\textup{and}~y=0,1.
\end{align*}
\endgroup
The following relationships between the observable data probabilities and the parameters hold,
\begingroup
\allowdisplaybreaks
\begin{align}
&\bP_{011\mid 0}=\Theta_{Y\mid n0}\Theta_{n}\Theta_{R^D\mid n01}+\Theta_{Y\mid c0}(1-\Theta_{n})\Theta_{R^D\mid c01}, \\
&\bP_{001\mid 0}=(1-\Theta_{Y\mid n0})\Theta_{n}\Theta_{R^D\mid n00}+(1-\Theta_{Y\mid c0})(1-\Theta_{n})\Theta_{R^D\mid c00}, \\
&\bP_{011\mid 1}=\Theta_{Y\mid n0}\Theta_{n}\Theta_{R^D\mid n01}, \\
&\bP_{001\mid 1}=(1-\Theta_{Y\mid n0})\Theta_{n}\Theta_{R^D\mid n00}, \\
&\bP_{111\mid 1}=\Theta_{Y\mid c1}(1-\Theta_{n})\Theta_{R^D\mid c11}, \\
&\bP_{101\mid 1}=(1-\Theta_{Y\mid c1})(1-\Theta_{n})\Theta_{R^D\mid c10}, \\
&\bP_{1+0\mid 0}=\Theta_{Y\mid n0}\Theta_{n}(1-\Theta_{R^D\mid n01})+\Theta_{Y\mid c0}(1-\Theta_{n})(1-\Theta_{R^D\mid c01}), \\
&\bP_{0+0\mid 0}=(1-\Theta_{Y\mid n0})\Theta_{n}(1-\Theta_{R^D\mid n00})+(1-\Theta_{Y\mid c0})(1-\Theta_{n})(1-\Theta_{R^D\mid c00}), \\
&\bP_{1+0\mid 1}=\Theta_{Y\mid n0}\Theta_{n}(1-\Theta_{R^D\mid n01})+\Theta_{Y\mid c1}(1-\Theta_{n})(1-\Theta_{R^D\mid c11}), \\
&\bP_{0+0\mid 1}=(1-\Theta_{Y\mid n0})\Theta_{n}(1-\Theta_{R^D\mid n00})+(1-\Theta_{Y\mid c1})(1-\Theta_{n})(1-\Theta_{R^D\mid c10}). 
\end{align}
\endgroup
Based on the observable data probabilities, we can have $\Theta_{n}=\frac{1}{2}$, $\Theta_{Y\mid n0}=\frac{1}{2}$, $\Theta_{Y\mid c0}=\frac{1}{4}$, $\Theta_{Y\mid c1}=\frac{1}{3}$, $\Theta_{R^D\mid n00}=\frac{1}{4}$, $\Theta_{R^D\mid n01}=\frac{1}{2}$, $\Theta_{R^D\mid c00}=\frac{1}{3}$, $\Theta_{R^D\mid c01}=\frac{1}{2}$, $\Theta_{R^D\mid c10}=\frac{1}{4}$, $\Theta_{R^D\mid c11}=\frac{1}{3}$, and the CACE $=\frac{1}{12}$. Alternatively, we can have $\Theta_{n}=\frac{1}{3}$, $\Theta_{Y\mid n0}=\frac{1}{2}$, $\Theta_{Y\mid c0}=\frac{5}{16}$, $\Theta_{Y\mid c1}=\frac{3}{8}$, $\Theta_{R^D\mid n00}=\frac{3}{8}$, $\Theta_{R^D\mid n01}=\frac{3}{4}$, $\Theta_{R^D\mid c00}=\frac{3}{11}$, $\Theta_{R^D\mid c01}=\frac{3}{10}$, $\Theta_{R^D\mid c10}=\frac{1}{5}$, $\Theta_{R^D\mid c11}=\frac{2}{9}$, and the CACE $=\frac{1}{16}$. Therefore, the CACE cannot be uniquely identified. 

\subsection{Counterexamples for the missing treatment and outcome models}\label{subsec::counterexamples3}
\begin{figure}[H]
\centering
\scalebox{0.8}{
\begin{tikzpicture}

    \node (z)  at (0,-3.8) {$Z$};
    \node (d)  at (1.5,-3.8) {$D$};
    \node (rd) at (1.5,-2.3) {$R^D$};
    \node (ry) at (3,-2.3) {$R^Y$};
    \node (y)  at (3,-3.8) {$Y$};
    \node (u)  at (2.25,-4.6) {$U$};
    \node (c)  at (1.5,-5.3) {1$ZD\oplus$2$ZD$};

    \path[-latex] (z) edge (d);
    \path[-latex] (d) edge (y);
    \path[-latex] (u) edge (d);
    \path[-latex] (u) edge (y);
    \path[-latex] (z) edge (rd);
    \path[-latex] (d) edge (rd);
    \path[-latex] (d) edge (ry);
    \path[-latex] (z) edge (ry);
    \path[-latex] (rd) edge (ry);

    \node (z)  at (3.5,-3.8) {$Z$};
    \node (d)  at (5,-3.8) {$D$};
    \node (rd) at (5,-2.3) {$R^D$};
    \node (ry) at (6.5,-2.3) {$R^Y$};
    \node (y)  at (6.5,-3.8) {$Y$};
    \node (u)  at (5.75,-4.6) {$U$};
    \node (c)  at (5,-5.3) {1$UD\oplus$2$ZD$};

    \path[-latex] (z) edge (d);
    \path[-latex] (d) edge (y);
    \path[-latex] (u) edge (d);
    \path[-latex] (u) edge (y);
    \path[-latex] (z) edge (rd);
    \path[-latex] (d) edge (rd);
    \path[-latex] (d) edge (ry);
    \path[-latex] (u) edge (ry);
    \path[-latex] (rd) edge (ry);
    
    \node (z)  at (7,-3.8) {$Z$};
    \node (d)  at (8.5,-3.8) {$D$};
    \node (rd) at (8.5,-2.3) {$R^D$};
    \node (ry) at (10,-2.3) {$R^Y$};
    \node (y)  at (10,-3.8) {$Y$};
    \node (u)  at (9.25,-4.6) {$U$};
    \node (c)  at (8.5,-5.3) {1$DY\oplus$2$ZD$};

    \path[-latex] (z) edge (d);
    \path[-latex] (d) edge (y);
    \path[-latex] (u) edge (d);
    \path[-latex] (u) edge (y);
    \path[-latex] (z) edge (rd);
    \path[-latex] (d) edge (rd);
    \path[-latex] (d) edge (ry);
    \path[-latex] (y) edge (ry);
    \path[-latex] (rd) edge (ry);

    \node (z)  at (10.5,-3.8) {$Z$};
    \node (d)  at (12,-3.8) {$D$};
    \node (rd) at (12,-2.3) {$R^D$};
    \node (ry) at (13.5,-2.3) {$R^Y$};
    \node (y)  at (13.5,-3.8) {$Y$};
    \node (u)  at (12.75,-4.6) {$U$};
    \node (c)  at (12,-5.3) {1$ZY\oplus$2$ZD$};

    \path[-latex] (z) edge (d);
    \path[-latex] (d) edge (y);
    \path[-latex] (u) edge (d);
    \path[-latex] (u) edge (y);
    \path[-latex] (z) edge (rd);
    \path[-latex] (d) edge (rd);
    \path[-latex] (z) edge (ry);
    \path[-latex] (y) edge (ry);
    \path[-latex] (rd) edge (ry);

    \node (z)  at (14,-3.8) {$Z$};
    \node (d)  at (15.5,-3.8) {$D$};
    \node (rd) at (15.5,-2.3) {$R^D$};
    \node (ry) at (17,-2.3) {$R^Y$};
    \node (y)  at (17,-3.8) {$Y$};
    \node (u)  at (16.25,-4.6) {$U$};
    \node (c)  at (15.5,-5.3) {1$UY\oplus$2$ZD$};

    \path[-latex] (z) edge (d);
    \path[-latex] (d) edge (y);
    \path[-latex] (u) edge (d);
    \path[-latex] (u) edge (y);
    \path[-latex] (z) edge (rd);
    \path[-latex] (d) edge (rd);
    \path[-latex] (u) edge (ry);
    \path[-latex] (y) edge (ry);
    \path[-latex] (rd) edge (ry);

    \node (z)  at (0,-7.6) {$Z$};
    \node (d)  at (1.5,-7.6) {$D$};
    \node (rd) at (1.5,-6.1) {$R^D$};
    \node (ry) at (3,-6.1) {$R^Y$};
    \node (y)  at (3,-7.6) {$Y$};
    \node (u)  at (2.25,-8.4) {$U$};
    \node (c)  at (1.5,-9.1) {1\textit{U}$\oplus$2\textit{ZD}};

    \path[-latex] (z) edge (d);
    \path[-latex] (d) edge (y);
    \path[-latex] (u) edge (d);
    \path[-latex] (u) edge (y);
    \path[-latex] (z) edge (rd);
    \path[-latex] (d) edge (rd);
    \path[-latex] (u) edge (ry);
    \path[-latex] (rd) edge (ry);

\end{tikzpicture}
}
\end{figure}

Under Assumptions 1$ZD\oplus$2$ZD$, 1$UD\oplus$2$ZD$, and 1$UY\oplus$2$ZD$, identification fails under both one-sided and two-sided noncompliance; we give counterexamples for one-sided noncompliance, which is the special case of two-sided noncompliance with no always-takers. Assumptions 1$DY\oplus$2$ZD$ and 1$ZY\oplus$2$ZD$ contain Assumptions 1$DY$ and 1$ZY$, respectively, so identification fails under one-sided noncompliance, and we give counterexamples with two-sided noncompliance. Under Assumption 1$U\oplus$2$ZD$, identification holds with one-sided noncompliance, so we give a counterexample with two-sided noncompliance. Define 
\begin{align*}
    \bP_{dy11\mid z} &= \bP(D=d,Y=y,R^D=1,R^Y=1\mid Z=z),\\
    \bP_{y+01\mid z} &= \bP(Y=y,R^D=0,R^Y=1\mid Z=z),\\
    \bP_{d1+0\mid z} &= \bP(D=d,R^D=1,R^Y=0\mid Z=z),\\
    \bP_{+0+0\mid z} &= \bP(R^D=0,R^Y=0\mid Z=z).
\end{align*}
\subsubsection{Assumption 1\textit{ZD}$\oplus$2\textit{ZD}}\label{subsec::counterexamples3-1zd+2zd}
For a binary $Y$ with one-sided noncompliance, we consider the following observable data probabilities:
\begin{eqnarray*}
&&(\bP_{0111\mid 0},\bP_{0011\mid 0},\bP_{0111\mid 1},\bP_{0011\mid 1},\bP_{1111\mid 1},\bP_{1011\mid 1},\\&&\bP_{1+01\mid 0},\bP_{0+01\mid 0},\bP_{1+01\mid 1},\bP_{0+01\mid 1},\bP_{01+0\mid 0},\bP_{01+0\mid 1},\bP_{11+0\mid 1},\bP_{+0+0\mid 0},\bP_{+0+0\mid 1})\\&=&\left(\frac{1}{16},\frac{1}{16},\frac{1}{32},\frac{1}{16},\frac{1}{96},\frac{1}{96},\frac{1}{8},\frac{1}{8},\frac{13}{192},\frac{17}{192},\frac{1}{8},\frac{9}{32},\frac{1}{24},\frac{1}{2},\frac{13}{32}\right)\nonumber.
\end{eqnarray*}
Define the parameters:
\begingroup
\allowdisplaybreaks
\begin{align*}
&\Theta_{dy\mid z}=\bP(D=d,Y=y\mid Z=z)~\textup{for}~(z,d,y)=(0,0,1),(1,0,1),(1,0,0),(1,1,0),\\
&\Theta_{R^D\mid zd}=\bP(R^D=1\mid Z=z,D=d)~\textup{for}~(z,d)=(0,0),(1,0),(1,1),\\
&\Theta_{R^Y\mid zdr^D}=\bP(R^Y=1\mid Z=z,D=d,R^D=r^D)~\textup{for}~(z,d)=(0,0),(1,0),(1,1)~\textup{and}~r^D=0,1.
\end{align*}
\endgroup
The following relationships between the observable data probabilities and the parameters hold,
\begingroup
\allowdisplaybreaks
\begin{align}
&\bP_{0111\mid 0}=\Theta_{01\mid 0}\Theta_{R^D\mid 00}\Theta_{R^Y\mid 001},\\
&\bP_{0011\mid 0}=(1-\Theta_{01\mid 0})\Theta_{R^D\mid 00}\Theta_{R^Y\mid 001}, \\
&\bP_{0111\mid 1}=\Theta_{01\mid 1}\Theta_{R^D\mid 10}\Theta_{R^Y\mid 101},\\
&\bP_{0011\mid 1}=\Theta_{00\mid 1}\Theta_{R^D\mid 10}\Theta_{R^Y\mid 101},\\
&\bP_{1111\mid 1}=(1-\Theta_{01\mid 1}-\Theta_{00\mid 1}-\Theta_{10\mid 1})\Theta_{R^D\mid 11}\Theta_{R^Y\mid 111}, \\
&\bP_{1011\mid 1}=\Theta_{10\mid 1}\Theta_{R^D\mid 11}\Theta_{R^Y\mid 111},\\
&\bP_{1+01\mid 0}=\Theta_{01\mid 0}(1-\Theta_{R^D\mid 00})\Theta_{R^Y\mid 000},\\
&\bP_{0+01\mid 0}=(1-\Theta_{01\mid 0})(1-\Theta_{R^D\mid 00})\Theta_{R^Y\mid 000},\\
&\bP_{1+01\mid 1}=\Theta_{01\mid 1}(1-\Theta_{R^D\mid 10})\Theta_{R^Y\mid 100}\nonumber\\&\hspace{1.5cm}+(1-\Theta_{01\mid 1}-\Theta_{00\mid 1}-\Theta_{10\mid 1})(1-\Theta_{R^D\mid 11})\Theta_{R^Y\mid 110},\\
&\bP_{0+01\mid 1}=\Theta_{00\mid 1}(1-\Theta_{R^D\mid 10})\Theta_{R^Y\mid 100}+\Theta_{10\mid 1}(1-\Theta_{R^D\mid 11})\Theta_{R^Y\mid 110},\\
&\bP_{01+0\mid 0}=(1-\Theta_{01\mid 0})\Theta_{R^D\mid 00}(1-\Theta_{R^Y\mid 001})+\Theta_{01\mid 0}\Theta_{R^D\mid 00}(1-\Theta_{R^Y\mid 001}),\\
&\bP_{01+0\mid 1}=\Theta_{00\mid 1}\Theta_{R^D\mid 10}(1-\Theta_{R^Y\mid 101})+\Theta_{01\mid 1}\Theta_{R^D\mid 10}(1-\Theta_{R^Y\mid 101}),\\
&\bP_{11+0\mid 1}=\Theta_{10\mid 1}\Theta_{R^D\mid 11}(1-\Theta_{R^Y\mid 111})\nonumber\\&\hspace{1.5cm}+(1-\Theta_{01\mid 1}-\Theta_{00\mid 1}-\Theta_{10\mid 1})\Theta_{R^D\mid 11}(1-\Theta_{R^Y\mid 111}),\\
&\bP_{+0+0\mid 0}=(1-\Theta_{01\mid 0})(1-\Theta_{R^D\mid 00})(1-\Theta_{R^Y\mid 000})+\Theta_{01\mid 0}(1-\Theta_{R^D\mid 00})(1-\Theta_{R^Y\mid 000}),\\
&\bP_{+0+0\mid 1}=\Theta_{00\mid 1}(1-\Theta_{R^D\mid 10})(1-\Theta_{R^Y\mid 100})+\Theta_{01\mid 1}(1-\Theta_{R^D\mid 10})(1-\Theta_{R^Y\mid 100})\nonumber\\&\hspace{1.5cm}+\Theta_{10\mid 1}(1-\Theta_{R^D\mid 11})(1-\Theta_{R^Y\mid 110})\nonumber\\&\hspace{1.5cm}+(1-\Theta_{01\mid 1}-\Theta_{00\mid 1}-\Theta_{10\mid 1})(1-\Theta_{R^D\mid 11})(1-\Theta_{R^Y\mid 110}).
\end{align}
\endgroup
Based on the observable data probabilities, we can have $\Theta_{01\mid 0}=\frac{1}{2}$, $\Theta_{00\mid 1}=\frac{1}{2}$, $\Theta_{01\mid 1}=\frac{1}{4}$, $\Theta_{10\mid 1}=\frac{1}{8}$, $\Theta_{R^D\mid 00}=\frac{1}{4}$, $\Theta_{R^D\mid 10}=\frac{1}{2}$, $\Theta_{R^D\mid 11}=\frac{1}{4}$, $\Theta_{R^Y\mid 001}=\frac{1}{2}$, $\Theta_{R^Y\mid 101}=\frac{1}{4}$, $\Theta_{R^Y\mid 111}=\frac{1}{3}$, $\Theta_{R^Y\mid 000}=\frac{1}{3}$, $\Theta_{R^Y\mid 100}=\frac{1}{6}$, $\Theta_{R^Y\mid 110}=\frac{1}{2}$, and the CACE $=-\frac{1}{2}$. Alternatively, we can have $\Theta_{01\mid 0}=\frac{1}{2}$, $\Theta_{00\mid 1}=\frac{1}{3}$, $\Theta_{01\mid 1}=\frac{1}{6}$, $\Theta_{10\mid 1}=\frac{1}{4}$, $\Theta_{R^D\mid 00}=\frac{1}{4}$, $\Theta_{R^D\mid 10}=\frac{3}{4}$, $\Theta_{R^D\mid 11}=\frac{1}{8}$, $\Theta_{R^Y\mid 001}=\frac{1}{2}$, $\Theta_{R^Y\mid 101}=\frac{1}{4}$, $\Theta_{R^Y\mid 111}=\frac{1}{3}$, $\Theta_{R^Y\mid 000}=\frac{1}{3}$, $\Theta_{R^Y\mid 100}=\frac{1}{2}$, $\Theta_{R^Y\mid 110}=\frac{3}{14}$, and the CACE $=-\frac{1}{6}$. Therefore, the CACE cannot be uniquely identified. 

\subsubsection{Assumption 1\textit{UD}$\oplus$2\textit{ZD}}\label{subsec::counterexamples3-1ud+2zd}
For a binary $Y$ with one-sided noncompliance, we consider the following observable data probabilities:
\begin{eqnarray*}
&&(\bP_{0111\mid 0},\bP_{0011\mid 0},\bP_{0111\mid 1},\bP_{0011\mid 1},\bP_{1111\mid 1},\bP_{1011\mid 1},\\&&\bP_{1+01\mid 0},\bP_{0+01\mid 0},\bP_{1+01\mid 1},\bP_{0+01\mid 1},\bP_{01+0\mid 0},\bP_{01+0\mid 1},\bP_{11+0\mid 1},\bP_{+0+0\mid 0},\bP_{+0+0\mid 1})\\&=&\left(\frac{11}{192},\frac{19}{192},\frac{1}{128},\frac{1}{128},\frac{1}{16},\frac{1}{16},\frac{5}{192},\frac{1}{24},\frac{5}{64},\frac{5}{64},\frac{11}{32},\frac{3}{64},\frac{1}{8},\frac{83}{192},\frac{17}{32}\right)\nonumber.
\end{eqnarray*}
Define the parameters:
\begingroup
\allowdisplaybreaks
\begin{align*}
&\Theta_{n}=\bP(U=n),\\
&\Theta_{Y\mid ud}=\bP(Y=1\mid U=u,D=d)~\textup{for}~(u,d)=(n,0),(c,1),(c,0),\\
&\Theta_{R^D\mid zd}=\bP(R^D=1\mid Z=z,D=d)~\textup{for}~(z,d)=(0,0),(1,0),(1,1),\\
&\Theta_{R^Y\mid udr^D}=\bP(R^Y=1\mid U=u,D=d,R^D=r^D)~\textup{for}~(u,d)=(n,0),(c,1),(c,0)~\textup{and}~r^D=0,1.
\end{align*}
\endgroup
The following relationships between the observable data probabilities and the parameters hold,
\begingroup
\allowdisplaybreaks
\begin{align}
&\bP_{0111\mid 0}=\Theta_{Y\mid n0}\Theta_{n}\Theta_{R^D\mid 00}\Theta_{R^Y\mid n01}+\Theta_{Y\mid c0}(1-\Theta_{n})\Theta_{R^D\mid 00}\Theta_{R^Y\mid c01}, \\
&\bP_{0011\mid 0}=(1-\Theta_{Y\mid n0})\Theta_{n}\Theta_{R^D\mid 00}\Theta_{R^Y\mid n01}+(1-\Theta_{Y\mid c0})(1-\Theta_{n})\Theta_{R^D\mid 00}\Theta_{R^Y\mid c01}, \\
&\bP_{0111\mid 1}=\Theta_{Y\mid n0}\Theta_{n}\Theta_{R^D\mid 10}\Theta_{R^Y\mid n01}, \\
&\bP_{0011\mid 1}=(1-\Theta_{Y\mid n0})\Theta_{n}\Theta_{R^D\mid 10}\Theta_{R^Y\mid n01}, \\
&\bP_{1111\mid 1}=\Theta_{Y\mid c1}(1-\Theta_{n})\Theta_{R^D\mid 11}\Theta_{R^Y\mid c11}, \\
&\bP_{1011\mid 1}=(1-\Theta_{Y\mid c1})(1-\Theta_{n})\Theta_{R^D\mid 11}\Theta_{R^Y\mid c11},\\
&\bP_{1+01\mid 0}=\Theta_{Y\mid n0}\Theta_{n}(1-\Theta_{R^D\mid 00})\Theta_{R^Y\mid n00}+\Theta_{Y\mid c0}(1-\Theta_{n})(1-\Theta_{R^D\mid 00})\Theta_{R^Y\mid c00}, \\
&\bP_{0+01\mid 0}=(1-\Theta_{Y\mid n0})\Theta_{n}(1-\Theta_{R^D\mid 00})\Theta_{R^Y\mid n00}\nonumber\\&\hspace{1.5cm}+(1-\Theta_{Y\mid c0})(1-\Theta_{n})(1-\Theta_{R^D\mid 00})\Theta_{R^Y\mid c00}, \\
&\bP_{1+01\mid 1}=\Theta_{Y\mid n0}\Theta_{n}(1-\Theta_{R^D\mid 10})\Theta_{R^Y\mid n00}+\Theta_{Y\mid c1}(1-\Theta_{n})(1-\Theta_{R^D\mid 11})\Theta_{R^Y\mid c10}, \\
&\bP_{0+01\mid 1}=(1-\Theta_{Y\mid n0})\Theta_{n}(1-\Theta_{R^D\mid 10})\Theta_{R^Y\mid n00}\nonumber\\&\hspace{1.5cm}+(1-\Theta_{Y\mid c1})(1-\Theta_{n})(1-\Theta_{R^D\mid 11})\Theta_{R^Y\mid c10},\\
&\bP_{01+0\mid 0}=\Theta_{n}\Theta_{R^D\mid 00}(1-\Theta_{R^Y\mid n01})+(1-\Theta_{n})\Theta_{R^D\mid 00}(1-\Theta_{R^Y\mid c01}), \\
&\bP_{01+0\mid 1}=\Theta_{n}\Theta_{R^D\mid 10}(1-\Theta_{R^Y\mid n01}), \\
&\bP_{11+0\mid 1}=(1-\Theta_{n})\Theta_{R^D\mid 11}(1-\Theta_{R^Y\mid c11}), \\
&\bP_{+0+0\mid 0}=\Theta_{n}(1-\Theta_{R^D\mid 00})(1-\Theta_{R^Y\mid n00})+(1-\Theta_{n})(1-\Theta_{R^D\mid 00})(1-\Theta_{R^Y\mid c00}),\\
&\bP_{+0+0\mid 1}=\Theta_{n}(1-\Theta_{R^D\mid 10})(1-\Theta_{R^Y\mid n00})+(1-\Theta_{n})(1-\Theta_{R^D\mid 11})(1-\Theta_{R^Y\mid c10}). 
\end{align}
\endgroup

Based on the observable data probabilities, we can have $\Theta_{n}=\frac{1}{4}$, $\Theta_{Y\mid n0}=\frac{1}{2}$, $\Theta_{Y\mid c0}=\frac{1}{3}$, $\Theta_{Y\mid c1}=\frac{1}{2}$, $\Theta_{R^D\mid 00}=\frac{1}{2}$, $\Theta_{R^D\mid 10}=\frac{1}{4}$, $\Theta_{R^D\mid 11}=\frac{1}{3}$, $\Theta_{R^Y\mid n00}=\frac{1}{6}$, $\Theta_{R^Y\mid c00}=\frac{1}{8}$, $\Theta_{R^Y\mid c10}=\frac{1}{4}$, $\Theta_{R^Y\mid n01}=\frac{1}{4}$, $\Theta_{R^Y\mid c01}=\frac{1}{3}$, $\Theta_{R^Y\mid c11}=\frac{1}{2}$, and the CACE $=\frac{1}{6}$. Alternatively, we can have $\Theta_{n}=\frac{23}{60}$, $\Theta_{Y\mid n0}=\frac{1}{2}$, $\Theta_{Y\mid c0}=\frac{4}{13}$, $\Theta_{Y\mid c1}=\frac{1}{2}$, $\Theta_{R^D\mid 00}=\frac{1}{2}$, $\Theta_{R^D\mid 10}=\frac{15}{92}$, $\Theta_{R^D\mid 11}=\frac{15}{37}$, $\Theta_{R^Y\mid n00}=\frac{13}{92}$, $\Theta_{R^Y\mid c00}=\frac{39}{296}$, $\Theta_{R^Y\mid c10}=\frac{2449}{8096}$, $\Theta_{R^Y\mid n01}=\frac{1}{4}$, $\Theta_{R^Y\mid c01}=\frac{13}{37}$, $\Theta_{R^Y\mid c11}=\frac{1}{2}$, and the CACE $=\frac{5}{26}$. Therefore, the CACE cannot be uniquely identified. 

\subsubsection{Assumption 1\textit{UY}$\oplus$2\textit{ZD}}\label{subsec::counterexamples3-1uy+2zd}
For a binary $Y$ with one-sided noncompliance, we consider the following observable data probabilities:
\begin{eqnarray*}
&&(\bP_{0111\mid 0},\bP_{0011\mid 0},\bP_{0111\mid 1},\bP_{0011\mid 1},\bP_{1111\mid 1},\bP_{1011\mid 1},\\&&\bP_{1+01\mid 0},\bP_{0+01\mid 0},\bP_{1+01\mid 1},\bP_{0+01\mid 1},\bP_{01+0\mid 0},\bP_{01+0\mid 1},\bP_{11+0\mid 1},\bP_{+0+0\mid 0},\bP_{+0+0\mid 1})\\&=&\left(\frac{1}{24},\frac{3}{64},\frac{1}{96},\frac{1}{128},\frac{1}{48},\frac{1}{64},\frac{11}{192},\frac{3}{32},\frac{41}{384},\frac{19}{192},\frac{79}{192},\frac{17}{384},\frac{41}{192},\frac{67}{192},\frac{185}{384}\right)\nonumber.
\end{eqnarray*}
Define the parameters:
\begingroup
\allowdisplaybreaks
\begin{align*}
&\Theta_{n}=\bP(U=n),\\
&\Theta_{Y\mid ud}=\bP(Y=1\mid U=u,D=d)~\textup{for}~(u,d)=(n,0),(c,1),(c,0),\\
&\Theta_{R^D\mid zd}=\bP(R^D=1\mid Z=z,D=d)~\textup{for}~(z,d)=(0,0),(1,0),(1,1),\\
&\Theta_{R^Y\mid uyr^D}=\bP(R^Y=1\mid U=u,Y=y,R^D=r^D)~\textup{for}~u=n,c,~y=0,1,~\textup{and}~r^D=0,1.
\end{align*}
\endgroup
The following relationships between the observable data probabilities and the parameters hold,
\begingroup
\allowdisplaybreaks
\begin{align}
&\bP_{0111\mid 0}=\Theta_{Y\mid n0}\Theta_{n}\Theta_{R^D\mid 00}\Theta_{R^Y\mid n11}+\Theta_{Y\mid c0}(1-\Theta_{n})\Theta_{R^D\mid 00}\Theta_{R^Y\mid c11}, \\
&\bP_{0011\mid 0}=(1-\Theta_{Y\mid n0})\Theta_{n}\Theta_{R^D\mid 00}\Theta_{R^Y\mid n01}+(1-\Theta_{Y\mid c0})(1-\Theta_{n})\Theta_{R^D\mid 00}\Theta_{R^Y\mid c01}, \\
&\bP_{0111\mid 1}=\Theta_{Y\mid n0}\Theta_{n}\Theta_{R^D\mid 10}\Theta_{R^Y\mid n11}, \\
&\bP_{0011\mid 1}=(1-\Theta_{Y\mid n0})\Theta_{n}\Theta_{R^D\mid 10}\Theta_{R^Y\mid n01}, \\
&\bP_{1111\mid 1}=\Theta_{Y\mid c1}(1-\Theta_{n})\Theta_{R^D\mid 11}\Theta_{R^Y\mid c11}, \\
&\bP_{1011\mid 1}=(1-\Theta_{Y\mid c1})(1-\Theta_{n})\Theta_{R^D\mid 11}\Theta_{R^Y\mid c01},\\
&\bP_{1+01\mid 0}=\Theta_{Y\mid n0}\Theta_{n}(1-\Theta_{R^D\mid 00})\Theta_{R^Y\mid n10}+\Theta_{Y\mid c0}(1-\Theta_{n})(1-\Theta_{R^D\mid 00})\Theta_{R^Y\mid c10}, \\
&\bP_{0+01\mid 0}=(1-\Theta_{Y\mid n0})\Theta_{n}(1-\Theta_{R^D\mid 00})\Theta_{R^Y\mid n00}\nonumber\\&\hspace{1.5cm}+(1-\Theta_{Y\mid c0})(1-\Theta_{n})(1-\Theta_{R^D\mid 00})\Theta_{R^Y\mid c00}, \\
&\bP_{1+01\mid 1}=\Theta_{Y\mid n0}\Theta_{n}(1-\Theta_{R^D\mid 10})\Theta_{R^Y\mid n10}+\Theta_{Y\mid c1}(1-\Theta_{n})(1-\Theta_{R^D\mid 11})\Theta_{R^Y\mid c10}, \\
&\bP_{0+01\mid 1}=(1-\Theta_{Y\mid n0})\Theta_{n}(1-\Theta_{R^D\mid 10})\Theta_{R^Y\mid n00}\nonumber\\&\hspace{1.5cm}+(1-\Theta_{Y\mid c1})(1-\Theta_{n})(1-\Theta_{R^D\mid 11})\Theta_{R^Y\mid c00},\\
&\bP_{01+0\mid 0}=\Theta_{Y\mid n0}\Theta_{n}\Theta_{R^D\mid 00}(1-\Theta_{R^Y\mid n11})+\Theta_{Y\mid c0}(1-\Theta_{n})\Theta_{R^D\mid 00}(1-\Theta_{R^Y\mid c11})\nonumber\\&\hspace{1.5cm}+(1-\Theta_{Y\mid n0})\Theta_{n}\Theta_{R^D\mid 00}(1-\Theta_{R^Y\mid n01})\nonumber\\&\hspace{1.5cm}+(1-\Theta_{Y\mid c0})(1-\Theta_{n})\Theta_{R^D\mid 00}(1-\Theta_{R^Y\mid c01}), \\
&\bP_{01+0\mid 1}=\Theta_{Y\mid n0}\Theta_{n}\Theta_{R^D\mid 10}(1-\Theta_{R^Y\mid n11})+(1-\Theta_{Y\mid n0})\Theta_{n}\Theta_{R^D\mid 10}(1-\Theta_{R^Y\mid n01}), \\
&\bP_{11+0\mid 1}=\Theta_{Y\mid c1}(1-\Theta_{n})\Theta_{R^D\mid 11}(1-\Theta_{R^Y\mid c11})\nonumber\\&\hspace{1.5cm}+(1-\Theta_{Y\mid c1})(1-\Theta_{n})\Theta_{R^D\mid 11}(1-\Theta_{R^Y\mid c01}), \\
&\bP_{+0+0\mid 0}=\Theta_{Y\mid n0}\Theta_{n}(1-\Theta_{R^D\mid 00})(1-\Theta_{R^Y\mid n10})+\Theta_{Y\mid c0}(1-\Theta_{n})(1-\Theta_{R^D\mid 00})(1-\Theta_{R^Y\mid c10})\nonumber\\&\hspace{1.5cm}+(1-\Theta_{Y\mid n0})\Theta_{n}(1-\Theta_{R^D\mid 00})(1-\Theta_{R^Y\mid n00})\nonumber\\&\hspace{1.5cm}+(1-\Theta_{Y\mid c0})(1-\Theta_{n})(1-\Theta_{R^D\mid 00})(1-\Theta_{R^Y\mid c00}),\\
&\bP_{+0+0\mid 1}=\Theta_{Y\mid n0}\Theta_{n}(1-\Theta_{R^D\mid 10})(1-\Theta_{R^Y\mid n10})+(1-\Theta_{Y\mid n0})\Theta_{n}(1-\Theta_{R^D\mid 10})(1-\Theta_{R^Y\mid n00})\nonumber\\&\hspace{1.5cm}+\Theta_{Y\mid c1}(1-\Theta_{n})(1-\Theta_{R^D\mid 11})(1-\Theta_{R^Y\mid c10})\nonumber\\&\hspace{1.5cm}+(1-\Theta_{Y\mid c1})(1-\Theta_{n})(1-\Theta_{R^D\mid 11})(1-\Theta_{R^Y\mid c00}). 
\end{align}
\endgroup

Based on the observable data probabilities, we can have $\Theta_{n}=\frac{1}{4}$, $\Theta_{Y\mid n0}=\frac{1}{2}$, $\Theta_{Y\mid c0}=\frac{1}{3}$, $\Theta_{Y\mid c1}=\frac{1}{2}$, $\Theta_{R^D\mid 00}=\frac{1}{2}$, $\Theta_{R^D\mid 10}=\frac{1}{4}$, $\Theta_{R^D\mid 11}=\frac{1}{3}$, $\Theta_{R^Y\mid n00}=\frac{1}{6}$, $\Theta_{R^Y\mid n10}=\frac{1}{4}$, $\Theta_{R^Y\mid c00}=\frac{1}{3}$, $\Theta_{R^Y\mid c10}=\frac{1}{3}$, $\Theta_{R^Y\mid n01}=\frac{1}{4}$, $\Theta_{R^Y\mid n11}=\frac{1}{3}$, $\Theta_{R^Y\mid c01}=\frac{1}{8}$, $\Theta_{R^Y\mid c11}=\frac{1}{6}$, and the CACE $=\frac{1}{6}$. Alternatively, we can have $\Theta_{n}=\frac{3}{8}$, $\Theta_{Y\mid n0}=\frac{1}{2}$, $\Theta_{Y\mid c0}=\frac{1}{10}$, $\Theta_{Y\mid c1}=\frac{1}{4}$, $\Theta_{R^D\mid 00}=\frac{1}{2}$, $\Theta_{R^D\mid 10}=\frac{1}{6}$, $\Theta_{R^D\mid 11}=\frac{2}{5}$, $\Theta_{R^Y\mid n00}=\frac{1}{12}$, $\Theta_{R^Y\mid n10}=\frac{25}{48}$, $\Theta_{R^Y\mid c00}=\frac{11}{36}$, $\Theta_{R^Y\mid c10}=\frac{13}{48}$, $\Theta_{R^Y\mid n01}=\frac{1}{4}$, $\Theta_{R^Y\mid n11}=\frac{1}{3}$, $\Theta_{R^Y\mid c01}=\frac{1}{12}$, $\Theta_{R^Y\mid c11}=\frac{1}{3}$, and the CACE $=\frac{3}{20}$. Therefore, the CACE cannot be uniquely identified. 

\subsubsection{Assumption 1\textit{DY}$\oplus$2\textit{ZD}}\label{subsec::counterexamples3-1dy+2zd}

For a binary $Y$ with two-sided noncompliance, we consider the following observable data probabilities:
\begin{eqnarray*}
&&(\bP_{0111\mid 0},\bP_{0011\mid 0},\bP_{1111\mid 0},\bP_{1011\mid 0},\bP_{0111\mid 1},\bP_{0011\mid 1},\bP_{1111\mid 1},\bP_{1011\mid 1},\\&&\bP_{1+01\mid 0},\bP_{0+01\mid 0},\bP_{1+01\mid 1},\bP_{0+01\mid 1},\bP_{01+0\mid 0},\bP_{11+0\mid 0},\bP_{01+0\mid 1},\bP_{11+0\mid 1},\bP_{+0+0\mid 0},\bP_{+0+0\mid 1})\\&=&\left(\frac{1}{32},\frac{1}{16},\frac{1}{64},\frac{1}{64},\frac{1}{32},\frac{1}{128},\frac{1}{96},\frac{1}{12},\frac{13}{384},\frac{7}{64},\frac{1}{24},\frac{73}{384},\frac{7}{32},\frac{1}{16},\frac{7}{128},\frac{11}{96},\frac{173}{384},\frac{179}{384}\right)\nonumber.
\end{eqnarray*}
Define the parameters:
\begingroup
\allowdisplaybreaks
\begin{align*}
&\Theta_{dy\mid z}=\bP(D=d,Y=y\mid Z=z)~\textup{for}~(z,d,y)=(0,0,0),(0,1,1),(0,1,0),(1,0,1),(1,0,0),(1,1,0),\\
&\Theta_{R^D\mid zd}=\bP(R^D=1\mid Z=z,D=d)~\textup{for}~z=0,1~\textup{and}~d=0,1,\\
&\Theta_{R^Y\mid dyr^D}=\bP(R^Y=1\mid D=d,Y=y,R^D=r^D)~\textup{for}~d=0,1,~y=0,1,~\textup{and}~r^D=0,1.
\end{align*}
\endgroup
The following relationships between the observable data probabilities and the parameters hold,
\begingroup
\allowdisplaybreaks
\begin{align}
&\bP_{0111\mid 0}=(1-\Theta_{00\mid 0}-\Theta_{11\mid 0}-\Theta_{10\mid 0})\Theta_{R^D\mid 00}\Theta_{R^Y\mid 011}, \\
&\bP_{0011\mid 0}=\Theta_{00\mid 0}\Theta_{R^D\mid 00}\Theta_{R^Y\mid 001}, \\
&\bP_{1111\mid 0}=\Theta_{11\mid 0}\Theta_{R^D\mid 01}\Theta_{R^Y\mid 111}, \\
&\bP_{1011\mid 0}=\Theta_{10\mid 0}\Theta_{R^D\mid 01}\Theta_{R^Y\mid 101}, \\
&\bP_{0111\mid 1}=\Theta_{01\mid 1}\Theta_{R^D\mid 10}\Theta_{R^Y\mid 011}, \\
&\bP_{0011\mid 1}=\Theta_{00\mid 1}\Theta_{R^D\mid 10}\Theta_{R^Y\mid 001}, \\
&\bP_{1111\mid 1}=(1-\Theta_{01\mid 1}-\Theta_{00\mid 1}-\Theta_{10\mid 1})\Theta_{R^D\mid 11}\Theta_{R^Y\mid 111},\\
&\bP_{1011\mid 1}=\Theta_{10\mid 1}\Theta_{R^D\mid 11}\Theta_{R^Y\mid 101}, \\
&\bP_{1+01\mid 0}=(1-\Theta_{00\mid 0}-\Theta_{11\mid 0}-\Theta_{10\mid 0})(1-\Theta_{R^D\mid 00})\Theta_{R^Y\mid 010}\nonumber\\&\hspace{1.5cm}+\Theta_{11\mid 0}(1-\Theta_{R^D\mid 01})\Theta_{R^Y\mid 110},\\
&\bP_{0+01\mid 0}=\Theta_{00\mid 0}(1-\Theta_{R^D\mid 00})\Theta_{R^Y\mid 000}+\Theta_{10\mid 0}(1-\Theta_{R^D\mid 01})\Theta_{R^Y\mid 100},\\
&\bP_{1+01\mid 1}=\Theta_{01\mid 1}(1-\Theta_{R^D\mid 10})\Theta_{R^Y\mid 010}\nonumber\\&\hspace{1.5cm}+(1-\Theta_{01\mid 1}-\Theta_{00\mid 1}-\Theta_{10\mid 1})(1-\Theta_{R^D\mid 11})\Theta_{R^Y\mid 110},\\
&\bP_{0+01\mid 1}=\Theta_{00\mid 1}(1-\Theta_{R^D\mid 10})\Theta_{R^Y\mid 000}+\Theta_{10\mid 1}(1-\Theta_{R^D\mid 11})\Theta_{R^Y\mid 100},\\
&\bP_{01+0\mid 0}=\Theta_{00\mid 0}\Theta_{R^D\mid 00}(1-\Theta_{R^Y\mid 001})\nonumber\\&\hspace{1.5cm}+(1-\Theta_{00\mid 0}-\Theta_{11\mid 0}-\Theta_{10\mid 0})\Theta_{R^D\mid 00}(1-\Theta_{R^Y\mid 011}),\\
&\bP_{11+0\mid 0}=\Theta_{10\mid 0}\Theta_{R^D\mid 01}(1-\Theta_{R^Y\mid 101})+\Theta_{11\mid 0}\Theta_{R^D\mid 01}(1-\Theta_{R^Y\mid 111}),\\
&\bP_{01+0\mid 1}=\Theta_{00\mid 1}\Theta_{R^D\mid 10}(1-\Theta_{R^Y\mid 001})+\Theta_{01\mid 1}\Theta_{R^D\mid 10}(1-\Theta_{R^Y\mid 011}),\\
&\bP_{11+0\mid 1}=\Theta_{10\mid 1}\Theta_{R^D\mid 11}(1-\Theta_{R^Y\mid 101})\nonumber\\&\hspace{1.5cm}+(1-\Theta_{01\mid 1}-\Theta_{00\mid 1}-\Theta_{10\mid 1})\Theta_{R^D\mid 11}(1-\Theta_{R^Y\mid 111}),\\
&\bP_{+0+0\mid 0}=\Theta_{00\mid 0}(1-\Theta_{R^D\mid 00})(1-\Theta_{R^Y\mid 000})\nonumber\\&\hspace{1.5cm}+(1-\Theta_{00\mid 0}-\Theta_{11\mid 0}-\Theta_{10\mid 0})(1-\Theta_{R^D\mid 00})(1-\Theta_{R^Y\mid 010})\nonumber\\&\hspace{1.5cm}+\Theta_{10\mid 0}(1-\Theta_{R^D\mid 01})(1-\Theta_{R^Y\mid 100})+\Theta_{11\mid 0}(1-\Theta_{R^D\mid 01})(1-\Theta_{R^Y\mid 110}),\\
&\bP_{+0+0\mid 1}=\Theta_{00\mid 1}(1-\Theta_{R^D\mid 10})(1-\Theta_{R^Y\mid 000})+\Theta_{01\mid 1}(1-\Theta_{R^D\mid 10})(1-\Theta_{R^Y\mid 010})\nonumber\\&\hspace{1.5cm}+\Theta_{10\mid 1}(1-\Theta_{R^D\mid 11})(1-\Theta_{R^Y\mid 100})\nonumber\\&\hspace{1.5cm}+(1-\Theta_{01\mid 1}-\Theta_{00\mid 1}-\Theta_{10\mid 1})(1-\Theta_{R^D\mid 11})(1-\Theta_{R^Y\mid 110}).
\end{align}
\endgroup

Based on the observable data probabilities, we can have $\Theta_{10\mid 0}=\frac{1}{8}$, $\Theta_{11\mid 0}=\frac{1}{4}$, $\Theta_{00\mid 0}=\frac{1}{2}$, $\Theta_{00\mid 1}=\frac{1}{8}$, $\Theta_{01\mid 1}=\frac{1}{4}$, $\Theta_{10\mid 1}=\frac{1}{2}$, $\Theta_{R^D\mid 00}=\frac{1}{2}$, $\Theta_{R^D\mid 01}=\frac{1}{4}$, $\Theta_{R^D\mid 10}=\frac{1}{4}$, $\Theta_{R^D\mid 11}=\frac{1}{3}$, $\Theta_{R^Y\mid 001}=\frac{1}{4}$, $\Theta_{R^Y\mid 011}=\frac{1}{2}$, $\Theta_{R^Y\mid 101}=\frac{1}{2}$, $\Theta_{R^Y\mid 111}=\frac{1}{4}$, $\Theta_{R^Y\mid 000}=\frac{1}{4}$, $\Theta_{R^Y\mid 010}=\frac{1}{6}$, $\Theta_{R^Y\mid 100}=\frac{1}{2}$, $\Theta_{R^Y\mid 110}=\frac{1}{8}$, and the CACE $=0$. Alternatively, we can have $\Theta_{10\mid 0}=\frac{1}{6}$, $\Theta_{11\mid 0}=\frac{1}{3}$, $\Theta_{00\mid 0}=\frac{2}{5}$, $\Theta_{00\mid 1}=\frac{1}{12}$, $\Theta_{01\mid 1}=\frac{1}{6}$, $\Theta_{10\mid 1}=\frac{3}{5}$, $\Theta_{R^D\mid 00}=\frac{5}{8}$, $\Theta_{R^D\mid 01}=\frac{3}{16}$, $\Theta_{R^D\mid 10}=\frac{3}{8}$, $\Theta_{R^D\mid 11}=\frac{5}{18}$, $\Theta_{R^Y\mid 001}=\frac{1}{4}$, $\Theta_{R^Y\mid 011}=\frac{1}{2}$, $\Theta_{R^Y\mid 101}=\frac{1}{2}$, $\Theta_{R^Y\mid 111}=\frac{1}{4}$, $\Theta_{R^Y\mid 000}=\frac{1535}{4108}$, $\Theta_{R^Y\mid 010}=\frac{135}{428}$, $\Theta_{R^Y\mid 100}=\frac{10515}{26702}$, $\Theta_{R^Y\mid 110}=\frac{905}{11128}$, and the CACE $=-\frac{7}{15}$. Therefore, the CACE cannot be uniquely identified.

\subsubsection{Assumption 1\textit{ZY}$\oplus$2\textit{ZD}}\label{subsec::counterexamples3-1zy+2zd}

For a binary $Y$ with two-sided noncompliance, we consider the following observable data probabilities:
\begin{eqnarray*}
&&(\bP_{0111\mid 0},\bP_{0011\mid 0},\bP_{1111\mid 0},\bP_{1011\mid 0},\bP_{0111\mid 1},\bP_{0011\mid 1},\bP_{1111\mid 1},\bP_{1011\mid 1},\\&&\bP_{1+01\mid 0},\bP_{0+01\mid 0},\bP_{1+01\mid 1},\bP_{0+01\mid 1},\bP_{01+0\mid 0},\bP_{11+0\mid 0},\bP_{01+0\mid 1},\bP_{11+0\mid 1},\bP_{+0+0\mid 0},\bP_{+0+0\mid 1})\\&=&\left(\frac{1}{32},\frac{1}{16},\frac{1}{32},\frac{1}{128},\frac{1}{64},\frac{1}{64},\frac{1}{96},\frac{1}{12},\frac{1}{24},\frac{11}{128},\frac{13}{384},\frac{41}{192},\frac{7}{32},\frac{7}{128},\frac{1}{16},\frac{11}{96},\frac{179}{384},\frac{173}{384}\right)\nonumber.
\end{eqnarray*}
Define the parameters:
\begingroup
\allowdisplaybreaks
\begin{align*}
&\Theta_{dy\mid z}=\bP(D=d,Y=y\mid Z=z)~\textup{for}~(z,d,y)=(0,0,0),(0,1,1),(0,1,0),(1,0,1),(1,0,0),(1,1,0),\\
&\Theta_{R^D\mid zd}=\bP(R^D=1\mid Z=z,D=d)~\textup{for}~z=0,1~\textup{and}~d=0,1,\\
&\Theta_{R^Y\mid zyr^D}=\bP(R^Y=1\mid Z=z,Y=y,R^D=r^D)~\textup{for}~z=0,1,~y=0,1,~\textup{and}~r^D=0,1.
\end{align*}
\endgroup
The following relationships between the observable data probabilities and the parameters hold,
\begin{align}
&\bP_{0111\mid 0}=(1-\Theta_{00\mid 0}-\Theta_{11\mid 0}-\Theta_{10\mid 0})\Theta_{R^D\mid 00}\Theta_{R^Y\mid 011}, \\
&\bP_{0011\mid 0}=\Theta_{00\mid 0}\Theta_{R^D\mid 00}\Theta_{R^Y\mid 001}, \\
&\bP_{1111\mid 0}=\Theta_{11\mid 0}\Theta_{R^D\mid 01}\Theta_{R^Y\mid 011}, \\
&\bP_{1011\mid 0}=\Theta_{10\mid 0}\Theta_{R^D\mid 01}\Theta_{R^Y\mid 001}, \\
&\bP_{0111\mid 1}=\Theta_{01\mid 1}\Theta_{R^D\mid 10}\Theta_{R^Y\mid 111}, \\
&\bP_{0011\mid 1}=\Theta_{00\mid 1}\Theta_{R^D\mid 10}\Theta_{R^Y\mid 101}, \\
&\bP_{1111\mid 1}=(1-\Theta_{01\mid 1}-\Theta_{00\mid 1}-\Theta_{10\mid 1})\Theta_{R^D\mid 11}\Theta_{R^Y\mid 111},\\
&\bP_{1011\mid 1}=\Theta_{10\mid 1}\Theta_{R^D\mid 11}\Theta_{R^Y\mid 101}, \\
&\bP_{1+01\mid 0}=(1-\Theta_{00\mid 0}-\Theta_{11\mid 0}-\Theta_{10\mid 0})(1-\Theta_{R^D\mid 00})\Theta_{R^Y\mid 010}\nonumber\\&\hspace{1.5cm}+\Theta_{11\mid 0}(1-\Theta_{R^D\mid 01})\Theta_{R^Y\mid 010},\\
&\bP_{0+01\mid 0}=\Theta_{00\mid 0}(1-\Theta_{R^D\mid 00})\Theta_{R^Y\mid 000}+\Theta_{10\mid 0}(1-\Theta_{R^D\mid 01})\Theta_{R^Y\mid 000},\\
&\bP_{1+01\mid 1}=\Theta_{01\mid 1}(1-\Theta_{R^D\mid 10})\Theta_{R^Y\mid 110}\nonumber\\&\hspace{1.5cm}+(1-\Theta_{01\mid 1}-\Theta_{00\mid 1}-\Theta_{10\mid 1})(1-\Theta_{R^D\mid 11})\Theta_{R^Y\mid 110},\\
&\bP_{0+01\mid 1}=\Theta_{00\mid 1}(1-\Theta_{R^D\mid 10})\Theta_{R^Y\mid 100}+\Theta_{10\mid 1}(1-\Theta_{R^D\mid 11})\Theta_{R^Y\mid 100},\\
&\bP_{01+0\mid 0}=\Theta_{00\mid 0}\Theta_{R^D\mid 00}(1-\Theta_{R^Y\mid 001})\nonumber\\&\hspace{1.5cm}+(1-\Theta_{00\mid 0}-\Theta_{11\mid 0}-\Theta_{10\mid 0})\Theta_{R^D\mid 00}(1-\Theta_{R^Y\mid 011}),\\
&\bP_{11+0\mid 0}=\Theta_{10\mid 0}\Theta_{R^D\mid 01}(1-\Theta_{R^Y\mid 001})+\Theta_{11\mid 0}\Theta_{R^D\mid 01}(1-\Theta_{R^Y\mid 011}),\\
&\bP_{01+0\mid 1}=\Theta_{00\mid 1}\Theta_{R^D\mid 10}(1-\Theta_{R^Y\mid 101})+\Theta_{01\mid 1}\Theta_{R^D\mid 10}(1-\Theta_{R^Y\mid 111}),\\
&\bP_{11+0\mid 1}=\Theta_{10\mid 1}\Theta_{R^D\mid 11}(1-\Theta_{R^Y\mid 101})\nonumber\\&\hspace{1.5cm}+(1-\Theta_{01\mid 1}-\Theta_{00\mid 1}-\Theta_{10\mid 1})\Theta_{R^D\mid 11}(1-\Theta_{R^Y\mid 111}),\\
&\bP_{+0+0\mid 0}=\Theta_{00\mid 0}(1-\Theta_{R^D\mid 00})(1-\Theta_{R^Y\mid 000})\nonumber\\&\hspace{1.5cm}+(1-\Theta_{00\mid 0}-\Theta_{11\mid 0}-\Theta_{10\mid 0})(1-\Theta_{R^D\mid 00})(1-\Theta_{R^Y\mid 010})\nonumber\\&\hspace{1.5cm}+\Theta_{10\mid 0}(1-\Theta_{R^D\mid 01})(1-\Theta_{R^Y\mid 000})+\Theta_{11\mid 0}(1-\Theta_{R^D\mid 01})(1-\Theta_{R^Y\mid 010}),\\
&\bP_{+0+0\mid 1}=\Theta_{00\mid 1}(1-\Theta_{R^D\mid 10})(1-\Theta_{R^Y\mid 100})+\Theta_{01\mid 1}(1-\Theta_{R^D\mid 10})(1-\Theta_{R^Y\mid 110})\nonumber\\&\hspace{1.5cm}+\Theta_{10\mid 1}(1-\Theta_{R^D\mid 11})(1-\Theta_{R^Y\mid 100})\nonumber\\&\hspace{1.5cm}+(1-\Theta_{01\mid 1}-\Theta_{00\mid 1}-\Theta_{10\mid 1})(1-\Theta_{R^D\mid 11})(1-\Theta_{R^Y\mid 110}).
\end{align}
Based on the observable data probabilities, we can have $\Theta_{10\mid 0}=\frac{1}{8}$, $\Theta_{11\mid 0}=\frac{1}{4}$, $\Theta_{00\mid 0}=\frac{1}{2}$, $\Theta_{00\mid 1}=\frac{1}{8}$, $\Theta_{01\mid 1}=\frac{1}{4}$, $\Theta_{10\mid 1}=\frac{1}{2}$, $\Theta_{R^D\mid 00}=\frac{1}{2}$, $\Theta_{R^D\mid 01}=\frac{1}{4}$, $\Theta_{R^D\mid 10}=\frac{1}{4}$, $\Theta_{R^D\mid 11}=\frac{1}{3}$, $\Theta_{R^Y\mid 001}=\frac{1}{4}$, $\Theta_{R^Y\mid 011}=\frac{1}{2}$, $\Theta_{R^Y\mid 101}=\frac{1}{2}$, $\Theta_{R^Y\mid 111}=\frac{1}{4}$, $\Theta_{R^Y\mid 000}=\frac{1}{4}$, $\Theta_{R^Y\mid 010}=\frac{1}{6}$, $\Theta_{R^Y\mid 100}=\frac{1}{2}$, $\Theta_{R^Y\mid 110}=\frac{1}{8}$, and the CACE $=0$. Alternatively, we can have $\Theta_{10\mid 0}=\frac{1}{6}$, $\Theta_{11\mid 0}=\frac{1}{3}$, $\Theta_{00\mid 0}=\frac{2}{5}$, $\Theta_{00\mid 1}=\frac{1}{12}$, $\Theta_{01\mid 1}=\frac{1}{6}$, $\Theta_{10\mid 1}=\frac{3}{5}$, $\Theta_{R^D\mid 00}=\frac{5}{8}$, $\Theta_{R^D\mid 01}=\frac{3}{16}$, $\Theta_{R^D\mid 10}=\frac{3}{8}$, $\Theta_{R^D\mid 11}=\frac{5}{18}$, $\Theta_{R^Y\mid 001}=\frac{1}{4}$, $\Theta_{R^Y\mid 011}=\frac{1}{2}$, $\Theta_{R^Y\mid 101}=\frac{1}{2}$, $\Theta_{R^Y\mid 111}=\frac{1}{4}$, $\Theta_{R^Y\mid 000}=\frac{165}{548}$, $\Theta_{R^Y\mid 010}=\frac{5}{37}$, $\Theta_{R^Y\mid 100}=\frac{205}{466}$, $\Theta_{R^Y\mid 110}=\frac{65}{408}$, and the CACE $=-\frac{7}{15}$. Therefore, the CACE cannot be uniquely identified. 

\subsubsection{Assumption 1\textit{U}$\oplus$2\textit{ZD}}\label{subsec::counterexamples3-1u+2zd}

For a binary $Y$ with two-sided noncompliance, we consider the following observable data probabilities:
\begin{eqnarray*}
&&(\bP_{0111\mid 0},\bP_{0011\mid 0},\bP_{1111\mid 0},\bP_{1011\mid 0},\bP_{0111\mid 1},\bP_{0011\mid 1},\bP_{1111\mid 1},\bP_{1011\mid 1},\\&&\bP_{1+01\mid 0},\bP_{0+01\mid 0},\bP_{1+01\mid 1},\bP_{0+01\mid 1},\bP_{01+0\mid 0},\bP_{11+0\mid 0},\bP_{01+0\mid 1},\bP_{11+0\mid 1},\bP_{+0+0\mid 0},\bP_{+0+0\mid 1})\\&=&\left(\frac{1}{32},\frac{5}{96},\frac{1}{96},\frac{1}{96},\frac{1}{32},\frac{1}{32},\frac{5}{288},\frac{5}{288},\frac{47}{576},\frac{7}{64},\frac{59}{576},\frac{59}{576},\frac{1}{6},\frac{1}{24},\frac{1}{16},\frac{13}{144},\frac{143}{288},\frac{157}{288}\right)\nonumber.
\end{eqnarray*}
Define the parameters:
\begingroup
\allowdisplaybreaks
\begin{align*}
&\Theta_{u}=\bP(U=u)~\textup{for}~u=a,n,\\
&\Theta_{Y\mid ud}=\bP(Y=1\mid U=u,D=d)~\textup{for}~(u,d)=(a,1),(n,0),(c,1),(c,0),\\
&\Theta_{R^D\mid zd}=\bP(R^D=1\mid Z=z,D=d)~\textup{for}~z=0,1~\textup{and}~d=0,1,\\
&\Theta_{R^Y\mid ur^D}=\bP(R^Y=1\mid U=u,R^D=r^D)~\textup{for}~u=a,n,c~\textup{and}~r^D=0,1.
\end{align*}
\endgroup
The following relationships between the observable data probabilities and the parameters hold,
\begingroup
\allowdisplaybreaks
\begin{align}
&\bP_{0111\mid 0}=\Theta_{Y\mid n0}\Theta_{n}\Theta_{R^D\mid 00}\Theta_{R^Y\mid n1}+\Theta_{Y\mid c0}(1-\Theta_{n}-\Theta_{a})\Theta_{R^D\mid 00}\Theta_{R^Y\mid c1},\\
&\bP_{0011\mid 0}=(1-\Theta_{Y\mid n0})\Theta_{n}\Theta_{R^D\mid 00}\Theta_{R^Y\mid n1}+(1-\Theta_{Y\mid c0})(1-\Theta_{n}-\Theta_{a})\Theta_{R^D\mid 00}\Theta_{R^Y\mid c1},\\
&\bP_{1111\mid 0}=\Theta_{Y\mid a1}\Theta_{a}\Theta_{R^D\mid 01}\Theta_{R^Y\mid a1},\\
&\bP_{1011\mid 0}=(1-\Theta_{Y\mid a1})\Theta_{a}\Theta_{R^D\mid 01}\Theta_{R^Y\mid a1},\\
&\bP_{0111\mid 1}=\Theta_{Y\mid n0}\Theta_{n}\Theta_{R^D\mid 10}\Theta_{R^Y\mid n1},\\
&\bP_{0011\mid 1}=(1-\Theta_{Y\mid n0})\Theta_{n}\Theta_{R^D\mid 10}\Theta_{R^Y\mid n1},\\
&\bP_{1111\mid 1}=\Theta_{Y\mid a1}\Theta_{a}\Theta_{R^D\mid 11}\Theta_{R^Y\mid a1}+\Theta_{Y\mid c1}(1-\Theta_{n}-\Theta_{a})\Theta_{R^D\mid 11}\Theta_{R^Y\mid c1},\\
&\bP_{1011\mid 1}=(1-\Theta_{Y\mid a1})\Theta_{a}\Theta_{R^D\mid 11}\Theta_{R^Y\mid a1}+(1-\Theta_{Y\mid c1})(1-\Theta_{n}-\Theta_{a})\Theta_{R^D\mid 11}\Theta_{R^Y\mid c1},\\
&\bP_{1+01\mid 0}=\Theta_{Y\mid n0}\Theta_{n}(1-\Theta_{R^D\mid 00})\Theta_{R^Y\mid n0}\nonumber+\Theta_{Y\mid a1}\Theta_{a}(1-\Theta_{R^D\mid 01})\Theta_{R^Y\mid a0}\\&\hspace{1.5cm}+\Theta_{Y\mid c0}(1-\Theta_{n}-\Theta_{a})(1-\Theta_{R^D\mid 00})\Theta_{R^Y\mid c0},\\
&\bP_{0+01\mid 0}=(1-\Theta_{Y\mid n0})\Theta_{n}(1-\Theta_{R^D\mid 00})\Theta_{R^Y\mid n0}\nonumber+(1-\Theta_{Y\mid a1})\Theta_{a}(1-\Theta_{R^D\mid 01})\Theta_{R^Y\mid a0}\\&\hspace{1.5cm}+(1-\Theta_{Y\mid c0})(1-\Theta_{n}-\Theta_{a})(1-\Theta_{R^D\mid 00})\Theta_{R^Y\mid c0},\\
&\bP_{1+01\mid 1}=\Theta_{Y\mid n0}\Theta_{n}(1-\Theta_{R^D\mid 10})\Theta_{R^Y\mid n0}+\Theta_{Y\mid a1}\Theta_{a}(1-\Theta_{R^D\mid 11})\Theta_{R^Y\mid a0}\nonumber\\&\hspace{1.5cm}+\Theta_{Y\mid c1}(1-\Theta_{n}-\Theta_{a})(1-\Theta_{R^D\mid 11})\Theta_{R^Y\mid c0},\\
&\bP_{0+01\mid 1}=(1-\Theta_{Y\mid n0})\Theta_{n}(1-\Theta_{R^D\mid 10})\Theta_{R^Y\mid n0}+(1-\Theta_{Y\mid a1})\Theta_{a}(1-\Theta_{R^D\mid 11})\Theta_{R^Y\mid a0}\nonumber\\&\hspace{1.5cm}+(1-\Theta_{Y\mid c1})(1-\Theta_{n}-\Theta_{a})(1-\Theta_{R^D\mid 11})\Theta_{R^Y\mid c0},\\
&\bP_{01+0\mid 0}=\Theta_{n}\Theta_{R^D\mid 00}(1-\Theta_{R^Y\mid n1})+(1-\Theta_{n}-\Theta_{a})\Theta_{R^D\mid 00}(1-\Theta_{R^Y\mid c1}),\\
&\bP_{11+0\mid 0}=\Theta_{a}\Theta_{R^D\mid 01}(1-\Theta_{R^Y\mid a1}),\\
&\bP_{01+0\mid 1}=\Theta_{n}\Theta_{R^D\mid 10}(1-\Theta_{R^Y\mid n1}),\\
&\bP_{11+0\mid 1}=\Theta_{a}\Theta_{R^D\mid 11}(1-\Theta_{R^Y\mid a1})+(1-\Theta_{n}-\Theta_{a})\Theta_{R^D\mid 11}(1-\Theta_{R^Y\mid c1}),\\
&\bP_{+0+0\mid 0}=\Theta_{n}(1-\Theta_{R^D\mid 00})(1-\Theta_{R^Y\mid n0})\nonumber+\Theta_{a}(1-\Theta_{R^D\mid 01})(1-\Theta_{R^Y\mid a0})\nonumber\\&\hspace{1.5cm}+(1-\Theta_{n}-\Theta_{a})(1-\Theta_{R^D\mid 00})(1-\Theta_{R^Y\mid c0}),\\
&\bP_{+0+0\mid 1}=\Theta_{n}(1-\Theta_{R^D\mid 10})(1-\Theta_{R^Y\mid n0})\nonumber+\Theta_{a}(1-\Theta_{R^D\mid 11})(1-\Theta_{R^Y\mid a0})\\&\hspace{1.5cm}+(1-\Theta_{n}-\Theta_{a})(1-\Theta_{R^D\mid 11})(1-\Theta_{R^Y\mid c0}).
\end{align}
\endgroup

Based on the observable data probabilities, we can have $\Theta_{n}=\frac{1}{4}$, $\Theta_{a}=\frac{1}{4}$, $\Theta_{Y\mid n0}=\frac{1}{2}$, $\Theta_{Y\mid c0}=\frac{1}{4}$, $\Theta_{Y\mid a1}=\frac{1}{2}$, $\Theta_{Y\mid c1}=\frac{1}{2}$, $\Theta_{R^D\mid 00}=\frac{1}{3}$, $\Theta_{R^D\mid 01}=\frac{1}{4}$, $\Theta_{R^D\mid 10}=\frac{1}{2}$, $\Theta_{R^D\mid 11}=\frac{1}{6}$, $\Theta_{R^Y\mid n0}=\frac{1}{4}$, $\Theta_{R^Y\mid a0}=\frac{1}{2}$, $\Theta_{R^Y\mid c0}=\frac{1}{6}$, $\Theta_{R^Y\mid n1}=\frac{1}{2}$, $\Theta_{R^Y\mid a1}=\frac{1}{3}$, $\Theta_{R^Y\mid c1}=\frac{1}{4}$, and the CACE $=\frac{1}{4}$. Alternatively, we can have $\Theta_{n}=\frac{1}{4}$, $\Theta_{a}=\frac{7}{16}$, $\Theta_{Y\mid n0}=\frac{1}{2}$, $\Theta_{Y\mid c0}=\frac{1}{8}$, $\Theta_{Y\mid a1}=\frac{1}{2}$, $\Theta_{Y\mid c1}=\frac{1}{2}$, $\Theta_{R^D\mid 00}=\frac{4}{9}$, $\Theta_{R^D\mid 01}=\frac{1}{7}$, $\Theta_{R^D\mid 10}=\frac{1}{2}$, $\Theta_{R^D\mid 11}=\frac{1}{6}$, $\Theta_{R^Y\mid n0}=\frac{11}{312}$, $\Theta_{R^Y\mid a0}=\frac{31}{78}$, $\Theta_{R^Y\mid c0}=\frac{16}{75}$, $\Theta_{R^Y\mid n1}=\frac{1}{2}$, $\Theta_{R^Y\mid a1}=\frac{1}{3}$, $\Theta_{R^Y\mid c1}=\frac{1}{5}$, and the CACE $=\frac{3}{8}$. Therefore, the CACE cannot be uniquely identified. 

\vskip 0.2in
\bibliography{sample}

\end{document}